\documentclass[10pt, oneside]{article}
\usepackage{hyperref}
\usepackage{enumitem} %so the whole list is not indented
\usepackage{easylist}  	
\usepackage{graphicx}							
\usepackage{amssymb}
\usepackage[usenames]{color}
\usepackage{epsf}
\usepackage[usenames,dvipsnames]{pstricks}
\usepackage{epsfig}
\usepackage{pst-grad} % For gradients
\usepackage{url}
\usepackage{wrapfig}
\usepackage{float} %to force table/figure with [H]
\newlength{\bibitemsep}
\newlength{\bibparskip}
\let\oldthebibliography\thebibliography
\renewcommand\thebibliography[1]{%
  \oldthebibliography{#1}%
  \setlength{\parskip}{\bibitemsep}%
  \setlength{\itemsep}{\bibparskip}%
}

\title{The X-rule: universal computation in a non-isotropic Life-like Cellular Automaton}
\author{Jos\'e Manuel G\'omez Soto%
\thanks{jmgomezuam@gmail.com, http://matematicas.reduaz.mx/$\sim$jmgomez}%
\hspace{2ex}{\it \small Universidad Aut\'onoma de Zacatecas.}\\
\hspace{2ex}{\it \small  Unidad Acad\'emica de Matem\'aticas. Zacatecas, Zac. M\'exico.}\\
Andrew Wuensche%
\thanks{andy@ddlab.org,  http://www.ddlab.org}%
\hspace{2ex}{\it \small Discrete Dynamics Lab, London, UK}}

\date{\small 15 March 2015 (to appear in Journal of Cellular Automata)}	% Activate to display a given date or no date

\begin{document}

\maketitle

\vspace{-3ex}
%^^^^^^^^^^^^^^^^^^^^^^^^^^^^^^^^^^^^^^^^^^^^^^^^^^^^^^^^^^^^^^^^^^^^
\begin{abstract}
\noindent 
We present a new Life-like cellular automaton (CA) capable of logic
universality -- the X-rule. The CA is 2D, binary, with a Moore
neighborhood and $\lambda$ parameter similar to the game-of-Life, but
is not based on birth/survival and is non-isotropic. We outline the
search method. Several glider types and stable structures emerge
spontaneously within \mbox{X-rule dynamics}. We construct glider-guns
based on periodic oscillations between stable barriers, and
interactions to create logical gates.
\end{abstract}

\begin{center}
{\it keywords: universality, cellular automata, glider-gun, logical gates.}
\end{center}

%^^^^^^^^^^^^^^^^^^^^^^^^^^^^^^^^^^^^^^^^^^^^^^^^^^^^^^^^^^^^^^^^^^^^^^^^^^^^
\section{Introduction} 
\label{Introduction}% A tag for referring to this section (optional).
\begin{wrapfigure}{L}{0.42\textwidth}
%\vspace{-3ex}
\vspace{-0.036\textwidth} %note 1ex=0.012\\textwidth
\includegraphics[width=0.40\textwidth,bb=32 18 336 323, clip=]{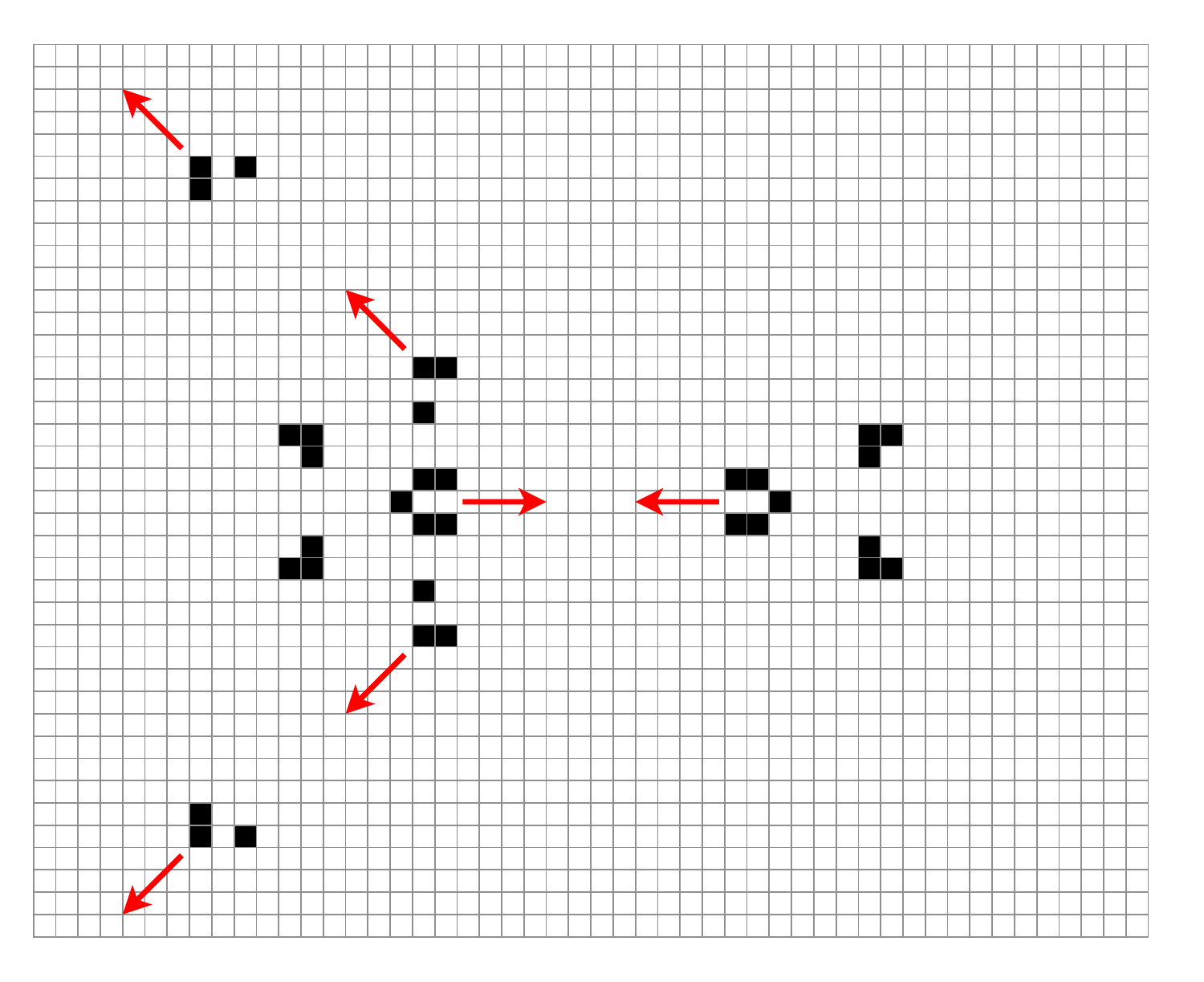}
%\vspace{-2.5ex}
\vspace{-0.03\textwidth}
\caption[X-rule glider-gun GGa]
{\textsf{\small X-rule glider-gun GGa}}\label{gga-g1.pdf}
\label{X-rule}
\color{red}
\setlength{\unitlength}{0.001\textwidth}
\thicklines
\begin{picture}(400,400)(0,-400)
\end{picture}
\color{black}
\vspace{-0.42\textwidth}
\end{wrapfigure}

Ever since the game-of-Life cellular automaton (CA) was created by
Conway in 1970, and its first glider-gun constructed subsequently by
Gosper\cite{Gardner1970,Berlekamp1982}, the search has
continued for alternative rules capable of universal
computation\cite{Adamatzky2002,Eppstein2010}, where logical gates can
be constructed from mobile and static configurations, and where a
glider-gun, a logical information mechanism for generating gliders at
regular intervals, is a key requirement.

\begin{figure}[htb]
\raisebox{3ex}{(a)}\includegraphics[width=.38\linewidth]{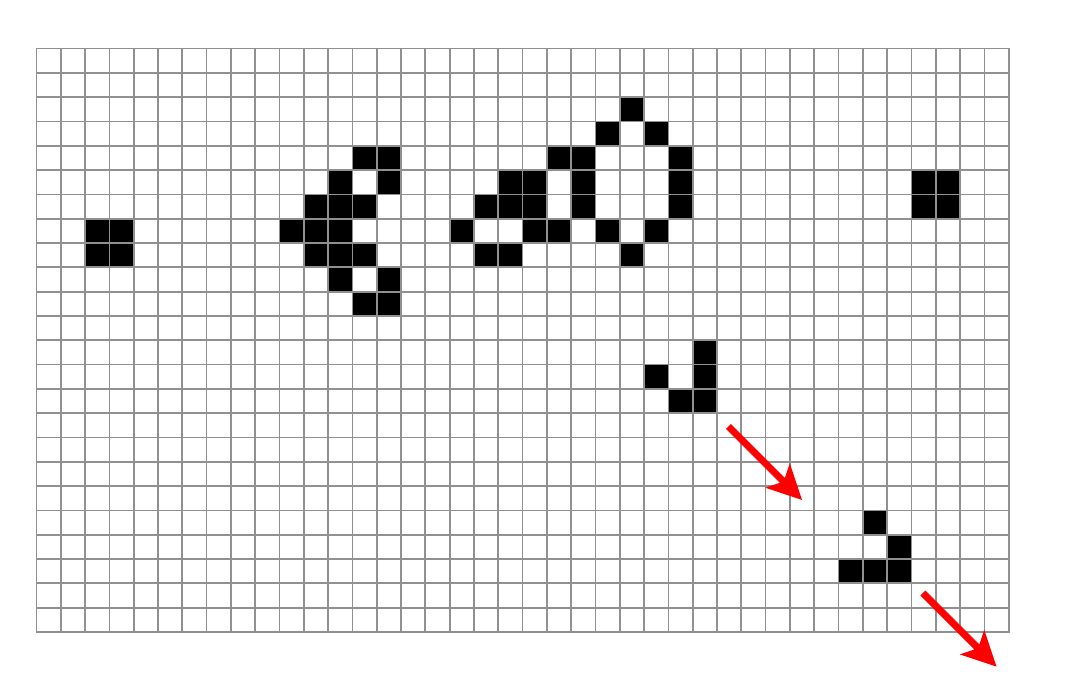}
\hfill
\raisebox{3ex}{(b)}\includegraphics[width=.55\linewidth]{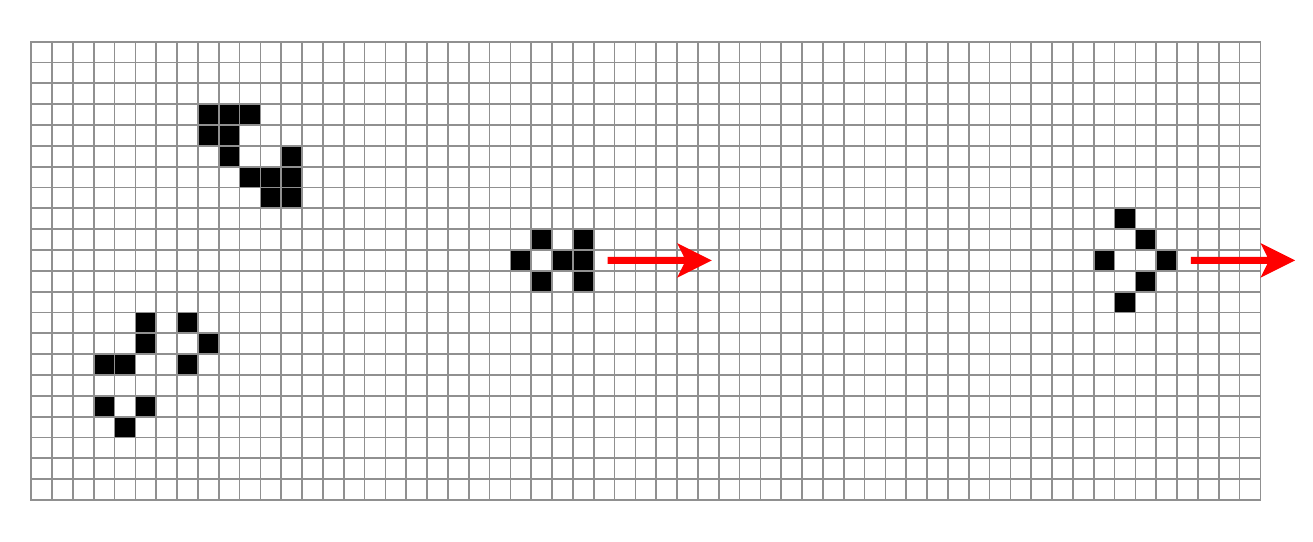}\\%[1ex] 
\raisebox{3ex}{(c)}\includegraphics[width=.44\linewidth]{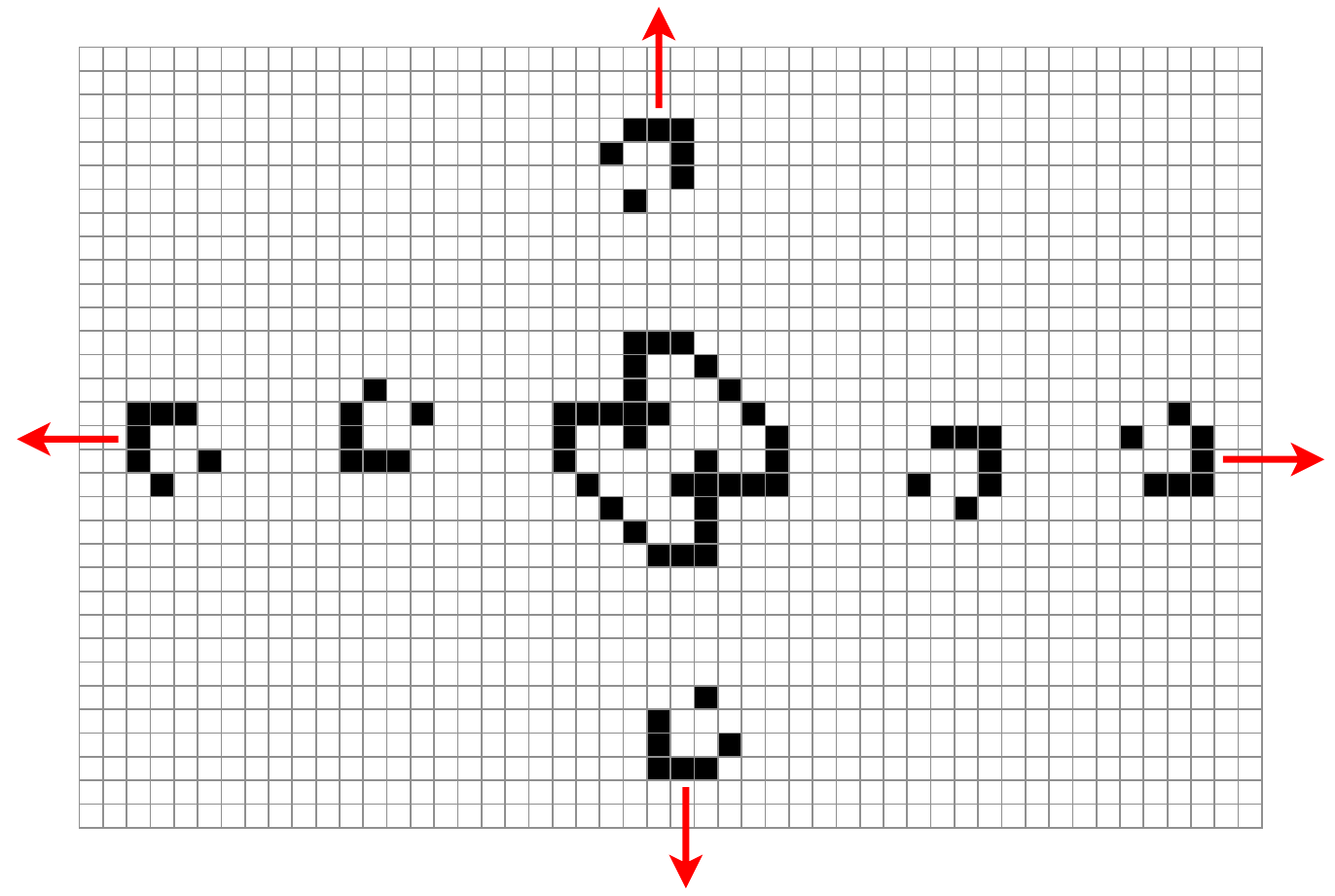} 
\hfill
\raisebox{3ex}{(d)}\includegraphics[width=.39\linewidth]{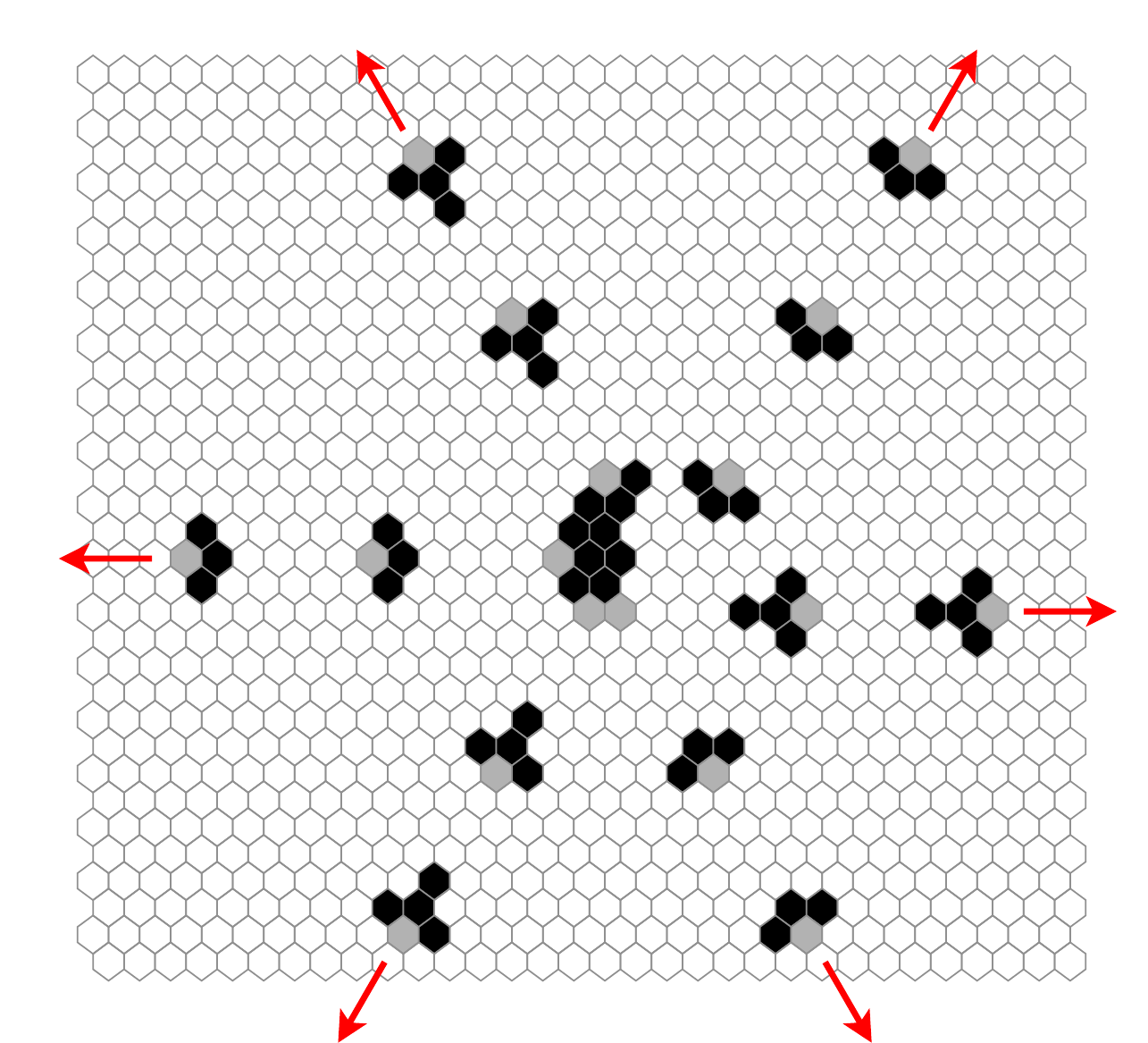}
\vspace{-2ex} 
\caption[Other glider-guns]{
{\textsf{Snapshots of glider-guns in other 2D CA: 
(a)~Conway's game-of-Life~b3/s23 showing Gosper's glider-gun\cite{Berlekamp1982},
(b)~Eppstein~b35/s236\cite{Eppstein2010}, 
(c)~Sapin\cite{Sapin2004}, (d)~Adamatsky-Wuensche~spiral-rule\cite{Wuensche2006}.
}}}
\vspace{-2ex}
\label{Other glider-guns}
\end{figure}

This paper presents a new Life-like CA, the X-rule (figure~\ref{XruleT.pdf}) 
-- a 2d binary CA
with a Moore neighborhood and a $\lambda$ parameter\cite{Langton90} 
similar to the
game-of-Life, but which is not based on birth/survival and is
non-isotropic, and where glider-guns based on periodic oscillations
between stable barriers are constructed (figures \ref{X-rule},
\ref{glider-guns GGa and GGb}), and interactions between gliders and
stable structures (eaters) are arranged to create the logical gates
required for logic universality -- implementing any logic circuit,
and potentially universality in the Turing sense, though this is
reserved for a later investigation.

Other 2D CA with aspects of universality have been presented in addition
to the game-of-Life involving both glider-guns and stable structures
(figure~\ref{Other glider-guns}).  Eppstein\cite{Eppstein2010} has
come up with several birth/survival CA, notably b35/s236 with a
constructed glider-gun.  Sapin's isotropic outer-totalstic
rule\cite{Sapin2004}, not based on birth/survival, features a
spontaneously emergent glider-gun.  In 3-value systems there is the
Adamatzky-Wuensche 2D hexagonal 7-neighbor totalistic ``spiral''
rule\cite{Wuensche2006} with spontaneously emergent glider-guns.

\begin{figure}[h]
\begin{center}
\includegraphics[width=.8\linewidth]{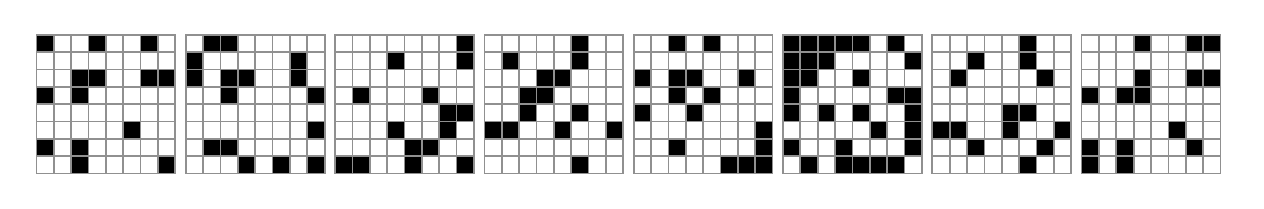}   
\end{center} %vector PostScript file
\vspace{-5ex}
\caption[X-rule precursor]
{\textsf{The rule-table of the X-rule precursor,
512 neighborhood outputs are shown
in descending order of neighborhood values\cite{Wolfram83}, 
from left to right, then in successive rows from the top.}}
\label{precur.pdf}
\end{figure}

A remarkable quality of the X-rule is that it achieves logical
universality, and potentially also in the Turing sense, by means of
constructed glider-guns, analogous to Gosper's (figure~\ref{Other glider-guns}a)
but different in that
they are made from a kit of parts, gliders and reflectors, that can be
put together in many combinations to produce periodic oscillators
based on bouncing/reflecting behaviour -- pairs of gliders bouncing
against each other and trapped between reflectors from which other
glider types are ejected at periodic intervals -- this is achieved by
introducing specific non-isotropic
outputs (figure~\ref{X-rule nhoods-mutated}) within an 
isotropic\footnote{Isotropic in the sense we use it
means that given a neighborhood pattern, any orientation of the pattern 
(spin, reflection, vertical flip) has the same output in the rule-table,
so the resulting CA dynamics will be equivalent whatever the orientation
of an initial state.}
precursor rule~(figure~\ref{precur.pdf}).
Increasing the gap between reflectors increases the
glider-gun period and reduces glider ejection frequency.  These
properties provide the \mbox{X-rule} with flexible and versatile
computational dynamics.

\enlargethispage{3ex}
\clearpage

Succeeding sections describe the following:\\[-4ex]
\begin{list}{$\Box$}{\parsep 0ex \itemsep 0ex 
          \leftmargin 2ex \rightmargin 0ex \labelwidth 4ex \labelsep 1ex}
\item[\ref{Searching rule-space for emergent gliders}.] The search
of isotropic rule-space for emergent gliders coexisting with stable structures.

\item[\ref{Constructing glider-guns}.] The pivotal section in the paper --
the construction of glider-guns based on simpler periodic oscillators,
and the derivation of the \mbox{X-rule} itself.

\item[\ref{Emergent structures in the X-rule universe}.] 
The \mbox{X-rule} universe, including its gliders, eaters and collisions.

\item[\ref{Glider-guns}.] A detailed description of two types of basic glider-guns.

\item[\ref{Compound glider-guns}.] Compound glider-guns, combining basic
glider-guns to enhance the diversity of gun dynamics.

\item[\ref{Logical gates -- logic universality}.] Examples of the logical gates,
NOT, AND, OR and NAND, to achieve logic universality.
\end{list}

%^^^^^^^^^^^^^^^^^^^^^^^^^^^^^^^^^^^^^^^^^^^^^^^^^^^^^^^^^^^^^^^^^^^^^
\section{Searching rule-space for emergent gliders}  
\label{Searching rule-space for emergent gliders}

\begin{figure}[H]
\begin{center}
\includegraphics[width=.9\linewidth]{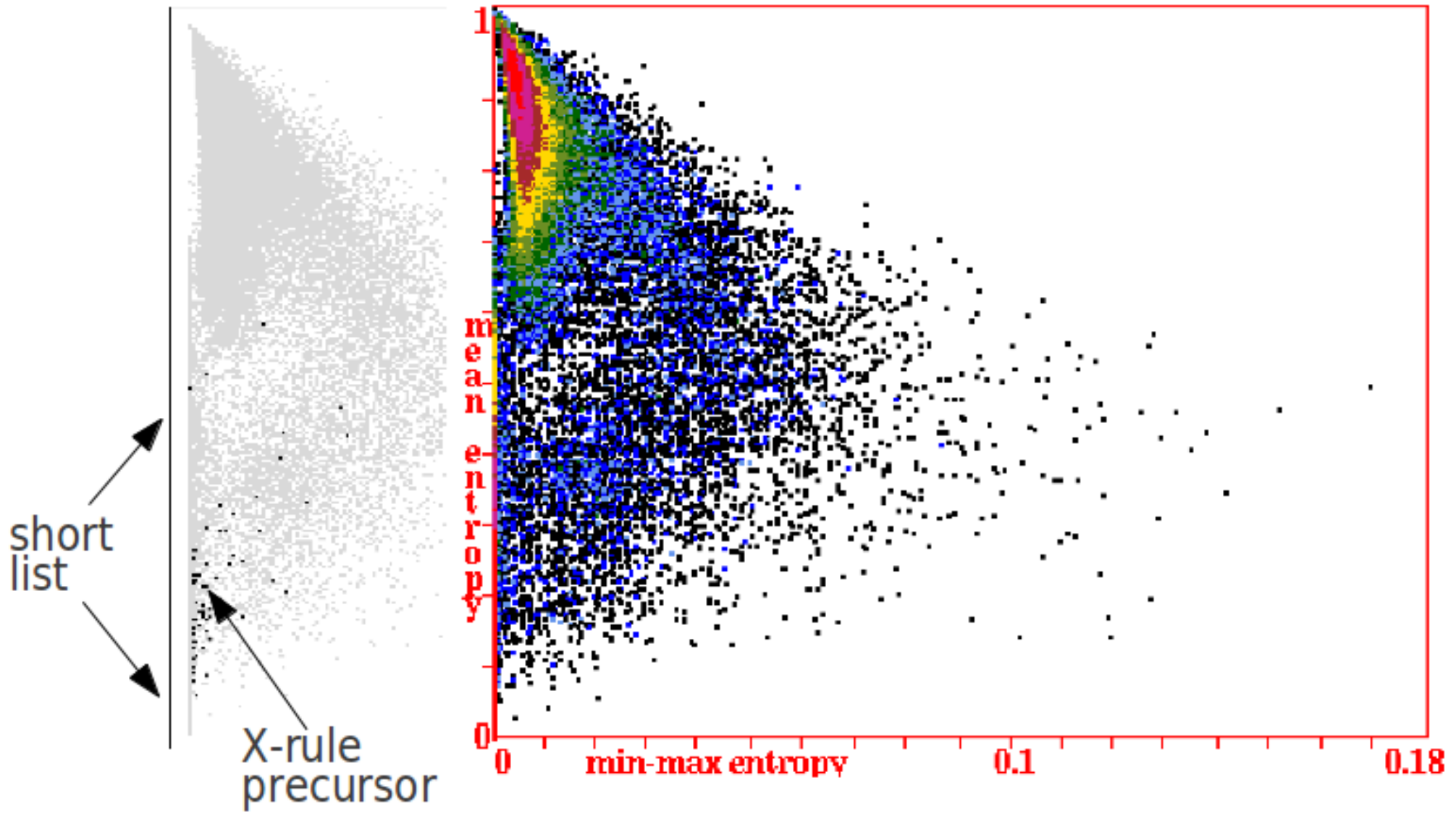}   
\end{center} %vector PostScript file
\vspace{-3ex}
\caption[Input-entropy scatter-plot]
{\textsf{The scatter-plot of a  sample of 93000+ rules,
plotting min-max entropy variability against mean entropy.
The left panel shows the location of the shortlist of about
70 rules, and the precursor to the X-rule.
}}
\label{scatter-plot}
\end{figure}

Definitions of Cellular Automata (CA) can be found from many sources,
so will not be repeated here, other than to note that this paper is
dealing with binary 2D classical synchronous CA, comparable to the
Game-of-Life (GoL) with a Moore neighborhood
but not based on birth/survival, and with periodic (or null) boundary
conditions.  

The Moore neighborhood
\hspace{-2ex}
\raisebox{-2.5ex}{\includegraphics[width=7ex]{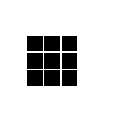}}
\hspace{-3ex} has 3x3=9 neighbors giving a full
lookup-table with $2^9$=512 outputs, a rule-space of $2^{512}$, but we
started off with isotropic rules only -- equal outputs for any
neighborhood rotation, reflection, or vertical flip -- where the number of effective
outputs reduces to 102, thus rule-space = $2^{102}$\cite{Sapin2010}.

\begin{figure}[H]
\begin{center}
\includegraphics[width=1\linewidth]{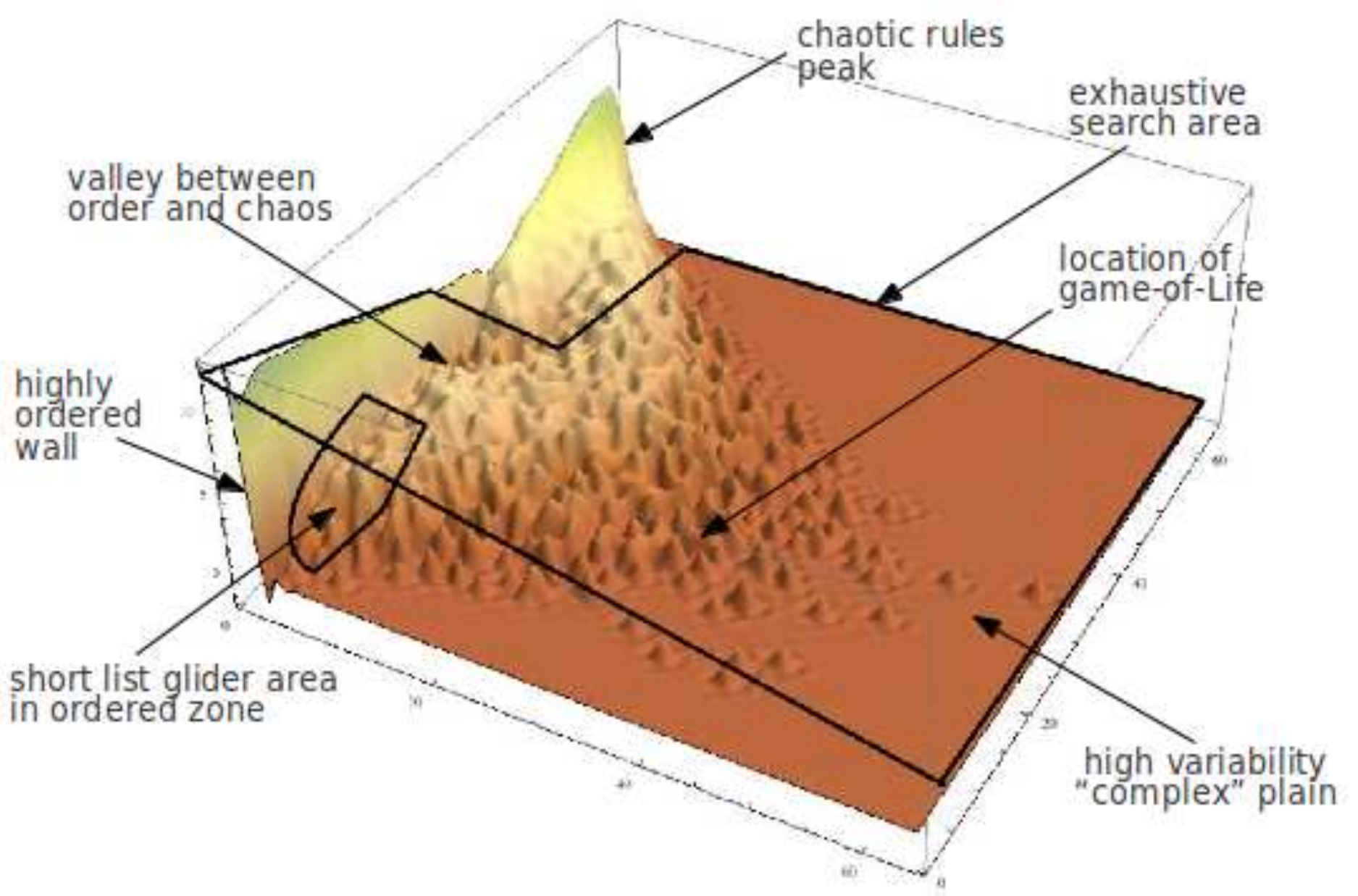}   
\end{center} %vector PostScript file
\vspace{-6ex}
\caption[Scatter-plot landscape]
{\textsf{The scatter-plot in Fig.\ref{scatter-plot}
showing rule frequency (log $y$) of rules on a 256x256 grid,
showing the area of exhaustive search and the short list of glider rules.
Characteristic dynamical behaviour is found in different parts of the
landscape. Rules in the high variability ``complex'' sector
of the scatter plot were found unsuitable because of over-active
dynamics, including unabated glider collisions and the lack of
stable structures.}}
\label{rule frequency}
\end{figure}

We searched in a sample of isotropic rule-space where look-up tables
were selected at random but biased probabilistically to have a similar
density of 1s as GoL ($\lambda$ parameter = 0.273 $\pm\approx$ 0.06).
This rule-space is more general than GoL -- not based on
birth/survival, and not totalistic.  Because of the $\lambda$ bias,
the isotropic rule-space searched was smaller than 
$2^{102}$ in~\cite{Sapin2010}. In addition, the all-0 neighborhood
was made to output 0.

Within these constraints, a sample of 93000+ rules was generated in
DDLab\cite{ddlab, Wuensche2011} using Wuensche's input-entropy
method\cite{Wuensche99,Wuensche05}, which creates a scatter-plot of
min-max input-entropy variability against mean~entropy, and which
separates rule-space into fuzzy sectors of chaos, order, and
complexity (figures~\ref{scatter-plot} and \ref{rule frequency}).  To
generate the scatter-plot we track how frequently the different
entries in the lookup-table are actually looked up in a moving window
of time-steps, started once the CA has settled into its typical
behaviour.  The Shannon entropy of this frequency distribution, the
input-entropy $S$, at time-step $t$, for one time-step ($w$=1), is
given by $S^t = -\sum_{i=0}^{L-1} \left( \frac{Q_{i}^{t}}{n} \times
log\left(\frac{Q_{i}^{t}}{n}\right)\right)$, where $Q_{i}^{t}$ is the
lookup frequency of neighborhood $i$ at time $t$. $L$ is the
rule-table size, and $n$ is the size of the CA.

A number of parameters need to be defined. For this experiment they
were as follows.  The measures were smoothed by being averaged over a
moving window of $w=10$ time-steps. The measures were started after 30
time-steps, and then taken for a further 400 time-steps.  The 2d CA
was $40\times40$ -- the lattice should be big enough but not too big
for effective tracking of entropy variability\cite{Wuensche05}.  Each
rule was run from 5 random initial states and average measures were
plotted -- course grained onto a 256$\times$256 grid -- the entropy
variability ($x$-axis) against the mean entropy
($y$-axis). Variability was measured according to min-max, meaning the
maximum increase in entropy from any dynamical minimum. The assembled
sample was sorted by both decreasing $x$ and $y$, and the data plotted
in figures~\ref{scatter-plot} and \ref{rule frequency}.  The plot
roughly classifies rule-space between chaos, order and complexity.
Individual rules from the plot can be selected and visually scanned by
efficient methods in DDLab\cite{ddlab,Wuensche2011}.

%----------------------------------------------------------------------
\subsection{Shortlist of glider rules in the ordered sector}

Avoiding the densely populated chaotic sector, an exhaustive search
was made of CA dynamics in the scatter-plot looking for spontaneously
emerging gliders and stable structures.  Although gliders were
frequent in the complex sector -- the sector with high entropy
variability -- this also turned out to be too unstable, although this
is the sector were GoL, Eppstien's and Sapin's rule would occur.

The exhaustive search thus concentrated within the ordered sector of
the scatter plot (figure \ref{scatter-plot}), where a shortlist of
about 70 rules with both gliders and stable structures were
identified.  From this list we selected just 5 rules with gliders travelling
both orthogonally and diagonally.

%^^^^^^^^^^^^^^^^^^^^^^^^^^^^^^^^^^^^^^^^^^^^^^^^^^^^^^^^^^^^^^^^
\section{Constructing glider-guns}
\label{Constructing glider-guns}

%----------------------------------------------------------------------
\subsection{Periodic oscillators -- the X-rule precursor}
\label{Periodic oscillators -- the X-rule precursor}

\vspace{-3ex}

\noindent Having found gliders the next step was to build a glider-gun.  
To do this, the main idea was to build a periodic bouncing-colliding structure,
a dynamical oscillator driving periodic collisions, which eventually, with some
modification to the pattern or rule, might eject gliders.

With this in mind, from the final short-list of 5 isotropic rules, 
we selected a rule, the X-rule precursor,
where bouncing from collisions was observed.
This bouncing behaviour was promising because it could
provide components for a periodic bouncing-colliding structure.

\begin{figure}[h]
\textsf{\small
\begin{center}
\begin{tabular}[t]{ @{}c@{} @{}c@{}  @{}c@{}  @{}c@{}  @{}c@{}   @{}c@{}   }
& \multicolumn{4}{c}{$\leftarrow$---------------- Ga ----------------$\rightarrow$} &\\[1ex]
  \includegraphics[height=.08\linewidth,bb= 0 0 30 30, clip=]{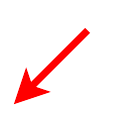}
&  \includegraphics[height=.08\linewidth,bb=-1 -2  42 35, clip=]{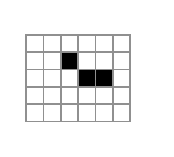}\color{white}-\color{black}%
& \includegraphics[height=.08\linewidth,bb=-3 -2  42 35,  clip=]{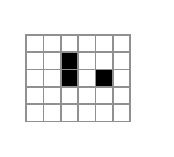}\color{white}-\color{black}%
& \includegraphics[height=.08\linewidth,bb=-3 -2  42 35,  clip=]{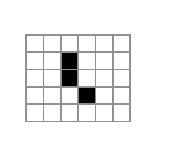}%
& \includegraphics[height=.08\linewidth,bb=-3 -2  42 35,  clip=]{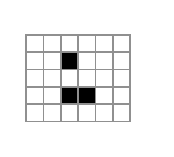}%
& \includegraphics[height=.08\linewidth,bb=-3 -2  42 35,  clip=]{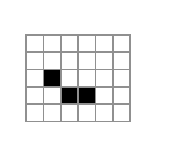}\\[-2ex]  
& 1 & 2 & 3 & 4 & 5 
\end{tabular}
\end{center}
}
\vspace{-4ex}
\caption[glider Ga]%
{\textsf{Glider Ga, moving SouthWest, time-steps shown. 
}}
\label{glider-Ga}
\vspace{-4ex}
\end{figure}

\begin{figure}[h]
\textsf{\small
\begin{center}
\begin{tabular}[t]{ @{}c@{} @{}c@{}  @{}c@{}  @{}c@{}  @{}c@{}   @{}c@{}   }
 \multicolumn{4}{c}{$\leftarrow$-------------------------- Gc --------------------------$\rightarrow$} & &\\[1ex]
  \includegraphics[height=.1\linewidth,bb=-3 -2  62 45, clip=]{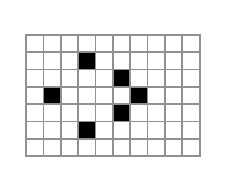}\color{white}-\color{black}%
& \includegraphics[height=.1\linewidth,bb=-3 -2  62 45,  clip=]{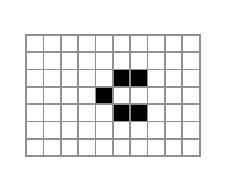}\color{white}-\color{black}%
& \includegraphics[height=.1\linewidth,bb=-3 -2  62 45,  clip=]{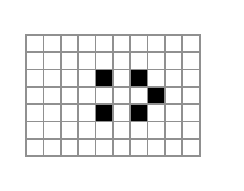}%
& \includegraphics[height=.1\linewidth,bb=-3 -2  62 45,  clip=]{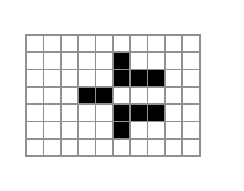}%
& \includegraphics[height=.1\linewidth,bb=-3 -2  62 45,  clip=]{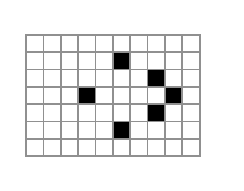}%
& \includegraphics[width=.09\linewidth,bb= 5 -3 32 22, clip=]{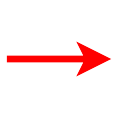}\\[-2ex]  
1 & 2 & 3 & 4 & 5 & 
\end{tabular}
\end{center}
}
\vspace{-4ex}
\caption[glider Gc]%
{\textsf{Glider Gc, moving East, time-steps shown. 
}}
\label{glider-Gc}
\end{figure}

\enlargethispage{4ex}
Two spontaneously emergent gliders\footnote{A third glider,
Gb (figure~\ref{glider-Gb}) is also supported but less likely to emerge than Ga or Gc,
because its phase patterns are more complex and Gb lacks a simple
predecessor.} in the X-rule precursor are
\mbox{glider~Ga
\hspace{-2ex}
\raisebox{-2ex}{\includegraphics[width=7ex]{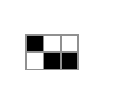}}
\hspace{-2ex}}
free to move in all diagonal directions (fig~\ref{glider-Ga}) and 
glider\footnote{In game-of-Life terminology a structure moving
  diagonally is a ``glider'' -- moving orthogonally its a
  ``space-ship''. In this paper we use ``glider'' for both types.}
Gc 
\hspace{-2ex}
\raisebox{-2ex}{\includegraphics[width=7ex]{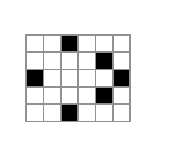}}
\hspace{-2ex}   
free to move in all orthogonal directions (fig~\ref{glider-Gc}). The glider speed
of Ga=$c$/4 and of Gc=$c$/2, where $c$ 
is the speed of light\footnote{A Moore neighborhood limits the maximum displacement
of any pattern to one cell per time-step either orthogonally or diagonally.
This is the CA's speed of light, $c$.}.
%\clearpage

\vspace{-3ex}

\vspace{-2ex}

\begin{figure}[h]
\begin{center}
\begin{minipage}[b]{.9\linewidth}
\raisebox{4ex}{1}\includegraphics[height=.12\linewidth]{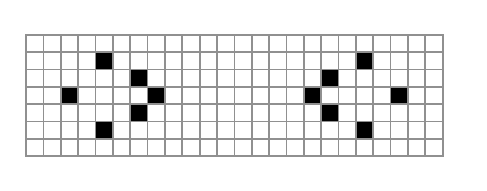}\hfill
\raisebox{4ex}{9}\includegraphics[height=.12\linewidth]{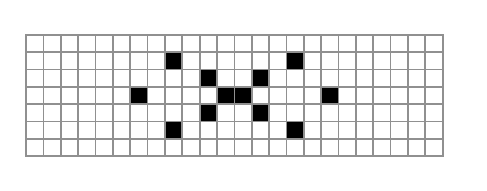}\hfill
\raisebox{4ex}{24}\includegraphics[height=.12\linewidth]{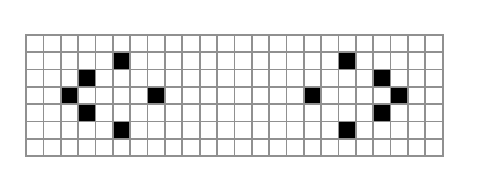}
\end{minipage}
\vspace{-3ex}
\caption[glider bouncing] %checked another word for bounce is ``rebound''
{\textsf{Head-on collision of two Gc gliders which bounce back, time-steps shown.}}
\label{gliderbounce}
\end{center}
\vspace{-2ex}
\end{figure}

An example of gliders ``bouncing'' is when two Gc gliders 
collide head-on and reverse direction. The initial separation
must be an even number including zero, as in figure \ref{gliderbounce}
where the initial separation is 8 cells. 

\begin{figure}[h]
\begin{center}
\begin{minipage}[b]{.63\linewidth}
\raisebox{4ex}{1}\includegraphics[height=.19\linewidth]{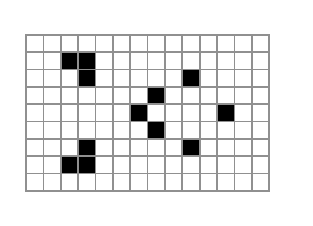}\hfill
\raisebox{4ex}{12}\includegraphics[height=.19\linewidth]{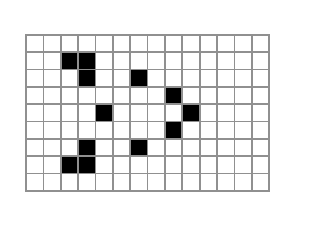}\hfill
\raisebox{4ex}{16}\includegraphics[height=.19\linewidth]{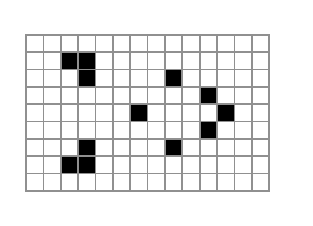}
\end{minipage}
\vspace{-3ex}
\caption[Reflect bouncing] %checked
{\textsf{Glider Gc reflecting off the reflector Rc, time-steps shown.}}
\label{reflectorbounce}
\end{center}
\label{Reflect bouncing}
\vspace{-5ex}
\end{figure}

Also observed in the X-rule precursor were small stable emergent 
configurations (eaters Ea)
\raisebox{-2ex}{
\includegraphics[height=4ex,bb=-1 -1 20 20, clip=]{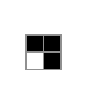}
\includegraphics[height=4ex,bb=-1 -1 20 20, clip=]{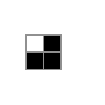}
\includegraphics[height=4ex,bb=-1 -1 20 20, clip=]{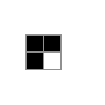}
\includegraphics[height=4ex,bb=-1 -1 20 20, clip=]{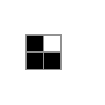}
\includegraphics[height=4ex,bb=-1 -1 20 20, clip=]{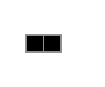}
\includegraphics[height=4ex,bb=-1 -1 20 20, clip=]{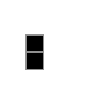}}.
With two Ea eaters we built a reflector (Rc) that ``reflects'' 
glider Gc back after a collision,
where Rc itself remains stable (figure~\ref{reflectorbounce}).
This reflecting behaviour occurs when the initial gap between glider Gc and
Rc is an even number including zero.

For clarity we will reserve the word
``bounce'' for glider-glider interactions as in figure~\ref{gliderbounce} and
``reflect'' for interactions between a glider and any stable configuration
such as Ea or Rc, as in figure~\ref{reflectorbounce}

\begin{figure}[h]
\begin{center}
\begin{minipage}[b]{.67\linewidth}
\color{white}-\color{black}\raisebox{4ex}{1}\includegraphics[height=.18\linewidth]{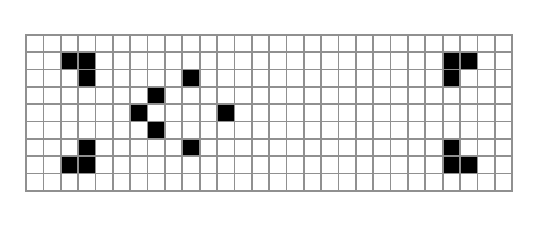}\hfill
\raisebox{4ex}{16}\includegraphics[height=.18\linewidth]{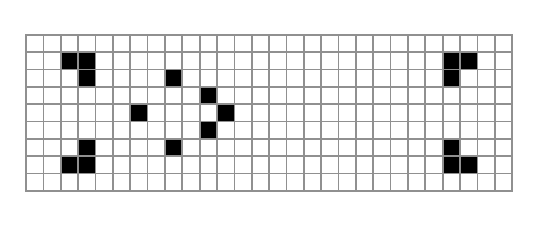}\\[-1ex]
\raisebox{4ex}{36}\includegraphics[height=.18\linewidth]{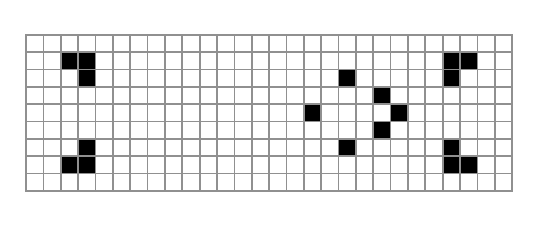}\hfill
\raisebox{4ex}{51}\includegraphics[height=.18\linewidth]{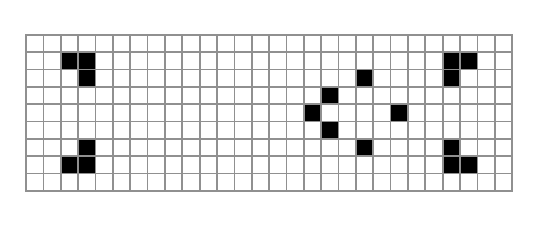}
\end{minipage}
\end{center}
\vspace{-5ex}
\caption[Simple reflecting oscillator] %checked
{\textsf{Glider Gc reflecting continuously between two stable Rc reflectors, 
gap=20, period=70, time-steps shown. We call this periodic structure
a ``simple reflecting oscillator'', SRO.}}
\label{pre-periodic-oscillator}
\vspace{-3ex}
\end{figure}

As the rule is isotropic, the mirror image of the reflecting interaction
would also be is valid, so we could construct a periodic oscillator with one Gc glider
reflecting back and forth between two Rc reflectors,
called a ``simple reflecting oscillator'', SRO
 (Figure \ref{pre-periodic-oscillator}).

\begin{figure}[h]
\begin{center}
\begin{minipage}[b]{.67\linewidth}
\color{white}-\color{black}\raisebox{4ex}{1}\includegraphics[height=.18\linewidth]{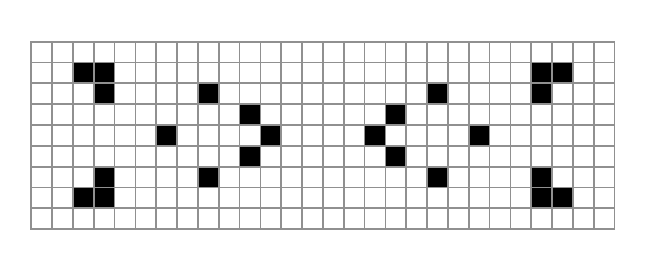}\hfill
\raisebox{4ex}{5}\includegraphics[height=.18\linewidth]{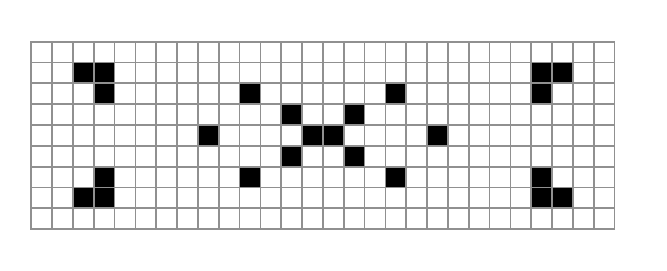}\\[-1ex]
\raisebox{4ex}{20}\includegraphics[height=.18\linewidth]{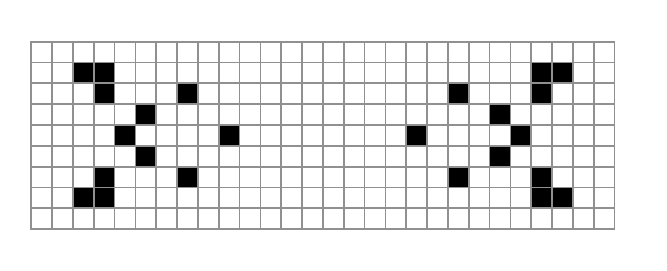}\hfill
\raisebox{4ex}{31}\includegraphics[height=.18\linewidth]{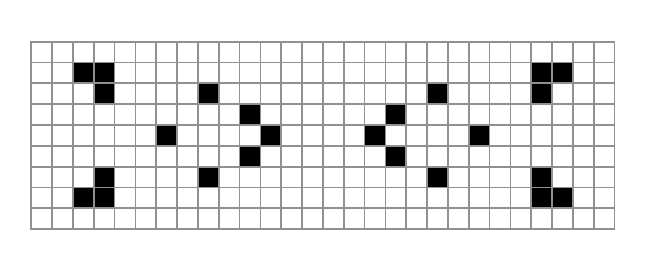}
\end{minipage}
\hfill
\begin{minipage}[b]{.3\linewidth}
\includegraphics[height=.75\linewidth]{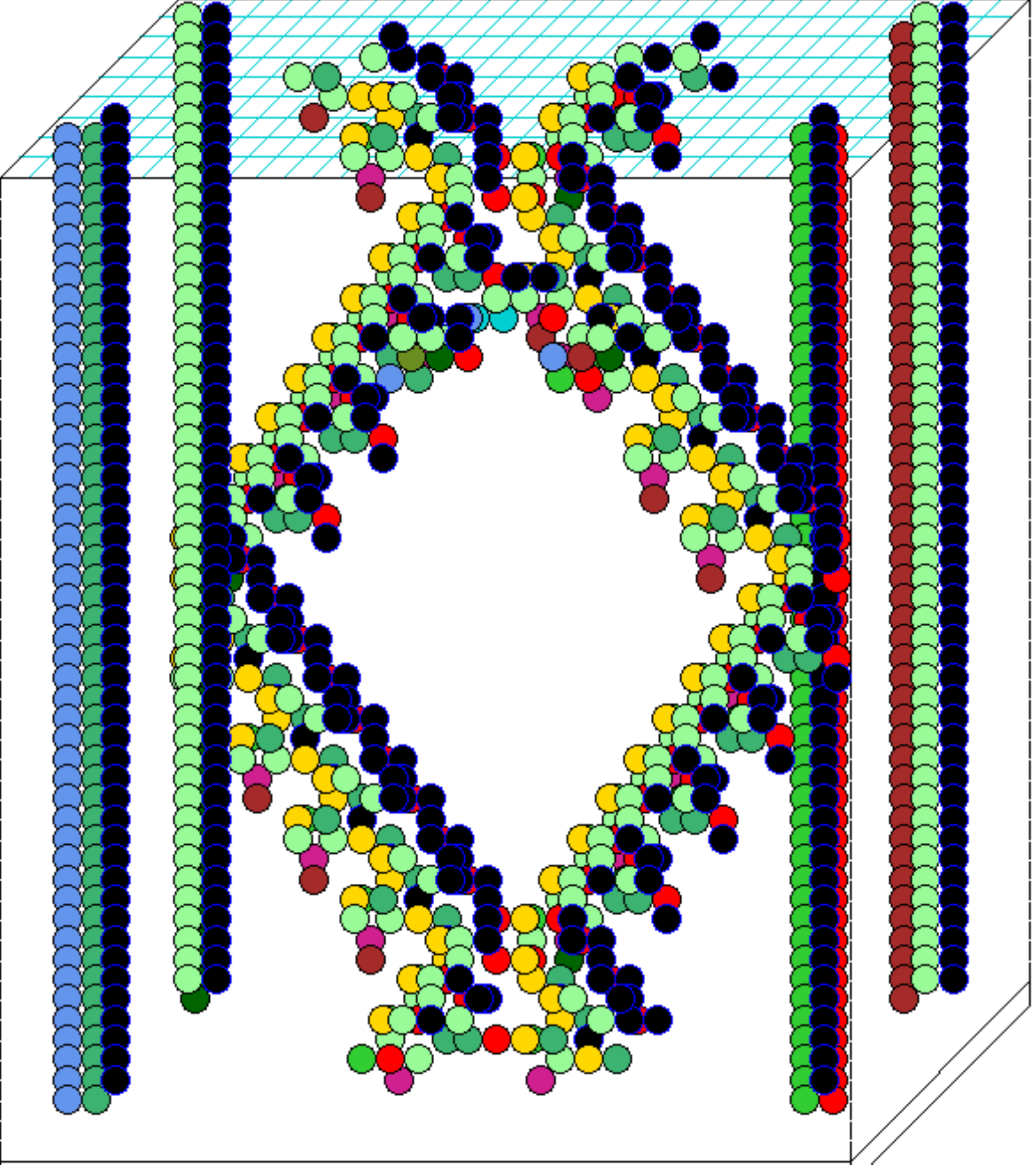}
\end{minipage}
\end{center}
\vspace{-4ex}
\caption[A periodic reflecting/bouncing structure] %checked
{\textsf{Two Gc gilders reflecting/bouncing between two reflectors, 
    gap=20, period=30, time-steps shown. We call this periodic structure
    a ``reflecting/bouncing oscillator'', RBO.
    {\it(right)} a representation showing 2D time-steps.}}
\label{pcs}
\vspace{-1ex}
\end{figure}

Finally we obtained what we were looking for, 
a more complex periodic collision structure (figure~\ref{pcs})
formed from two Gc gliders continually colliding and bouncing
between two reflectors, which we call a ``reflecting/bouncing oscillator'' RBO.   
The RBO is an assembly of several components, so is
flexible in that the distance between reflectors (figure~\ref{pcs})
and the phase of the various collisions could be varied.  We ran experiments
to test these variations, hoping to obtain gliders, but without
success. Although the RBO we constructed did
not eject gliders, the rule itself seemed a promising candidate
whereby mutations to its lookup-table might achieve a glider-gun, so
we kept this rule as the precursor to the X-rule.

\vspace{-1ex}

%----------------------------------------------------------------------
\subsection{Creating glider-guns -- the X-rule}
\label{Creating glider-guns -- the X-rule}

\noindent Our strategy for creating glider-guns was to mutate outputs in
the \mbox{X-rule}~precursor's rule-table with no restriction on 
preserving isotropy. These mutations had to fulfil two objectives: 
firstly, to preserve the essence of the periodic oscillating 
behaviour SRO and RBO, and the gliders 
Ga~\raisebox{-1.5ex}{\includegraphics[height=2ex,bb=0 -1 19 12 clip=]{mga1}}
\hspace{.5ex}
and Gc~\raisebox{-2.5ex}{\includegraphics[height=3.5ex,bb=0 0 35 27 clip=]{miniglider_c1}}
\hspace{.3ex},
from the \mbox{X-rule precursor} 
described in section~\ref{Periodic oscillators -- the X-rule precursor},
and secondly to produce ejected gliders from the periodic oscillators.
The methodology for achieving this was as follows:
\clearpage

\begin{figure}[h]
\begin{center}
\includegraphics[width=.8\linewidth]{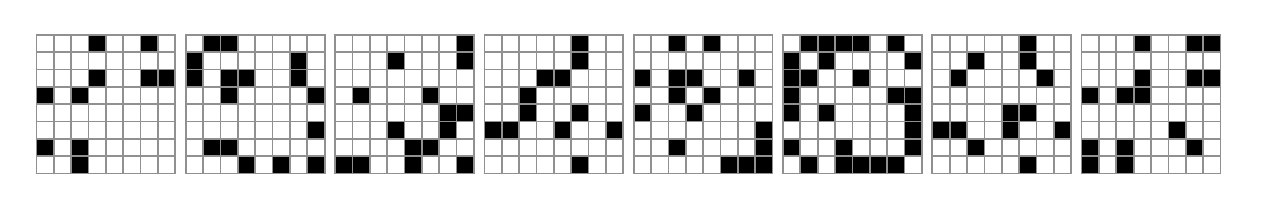}   
\end{center}
\vspace{-5ex}
\caption[X-rule]
{\textsf{The rule-table of the X-rule --
512 neighborhood outputs are shown
in descending order of neighborhood values\cite{Wolfram83}, 
from left to right, then in successive rows from the top.
}}
\label{XruleT.pdf}
\end{figure}

\vspace{-3ex}
\begin{figure}[h]
\begin{center}
\includegraphics[width=.8\linewidth]{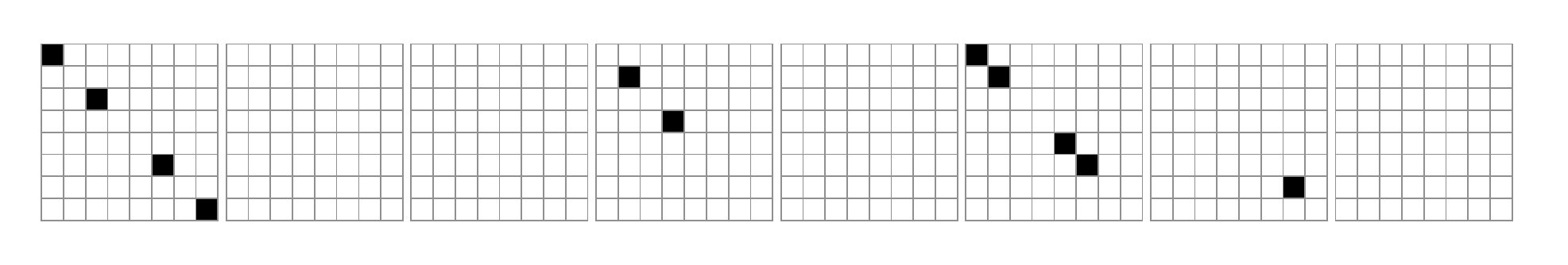} 
\end{center}
\vspace{-5ex}
\caption[X-rule mutations]
{\textsf{The position of 11 mutated outputs -- the difference between
the X-rule and the X-rule precursor. All the mutations are changes from 1 to 0.
}}
\label{X-rule mutations}
\end{figure}

\begin{enumerate}[leftmargin=*]

\item We identified the 36 neighborhood outputs that had no effect on
the oscillating behaviour or on the gliders, giving $2^{36}$
possible combinatorial mutations.  However, for ease of computation
we considered just a random subset of $2^{17}$.\\[-3ex]

\item  We automatically tested rules in the subset sample with
a ``reflecting/bouncing oscillator'', RBO, varying the gaps between
the reflectors, looking for periodic glider ejection.\\[-3ex]

\item  In these experiments we obtained two different glider-guns
in a rule later named the X-rule.
One glider-gun GGa ejected 
%Ga~\epsfig{file=mga1.pdf, height=2ex,bb=0 -1 19 12, clip=}
Ga~\raisebox{-1.5ex}{\includegraphics[height=2ex,bb=0 -1 19 12 clip=]{mga1}}
\hspace{.5ex}
gliders SouthWest and NorthWest, 
the same glider that had previously been observed as freely
emergent in Section~\ref{Periodic oscillators -- the X-rule precursor}.
The other glider-gun GGb ejected a new type of glider
Gb~\raisebox{-1.2ex}{\includegraphics[height=5ex,bb=0 -1 25 40, clip=]{miniglider_b1}} 
South and North.
These glider-guns are described in detail in section~\ref{Two basic glider-guns}.

\end{enumerate}

In the testing sequence before and after the X-rule, GGa (but not GGb) was also detected
in rules that were close variations of the X-rule, but we decided to focus on the X-rule itself
because it supported two glider-guns, reasoning that two glider-guns are better than one.

\begin{figure}[h]
\begin{center}
{\textsf{\small
\begin{minipage}[b]{.9\linewidth}
\raisebox{4ex}{iso$\to$}\hspace{-2ex}
\includegraphics[width=6ex,bb=-2 -2 25 25, clip=]{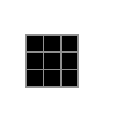}\hfill
\includegraphics[width=6ex,bb=-2 -2 25 25,clip=]{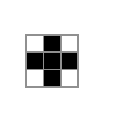}\hfill
\includegraphics[width=6ex,bb=-2 -2 25 25,clip=]{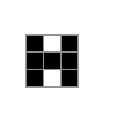}\hspace{-2ex}
\includegraphics[width=6ex,bb=-2 -2 25 25,clip=]{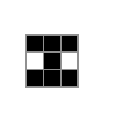}\hfill
\includegraphics[width=6ex,bb=-2 -2 25 25,clip=]{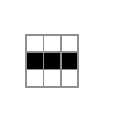}\hspace{-2ex}
\includegraphics[width=6ex,bb=-2 -2 25 25,clip=]{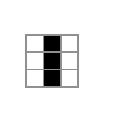}\hspace{3ex}
\raisebox{4ex}{non-iso$\to$}\hspace{-2ex}
\includegraphics[width=6ex,bb=-2 -2 25 25, clip=]{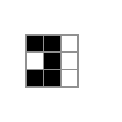}\hspace{-2ex}
\includegraphics[width=6ex,bb=-2 -2 25 25,clip=]{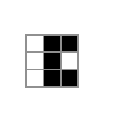}\hfill
\includegraphics[width=6ex,bb=-2 -2 25 25,clip=]{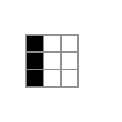}\hspace{-2ex}
\includegraphics[width=6ex,bb=-2 -2 25 25,clip=]{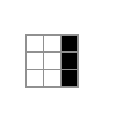}\hfill
\includegraphics[width=6ex,bb=-2 -2 25 25,clip=]{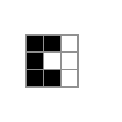}
\end{minipage}
}}
\vspace{-4ex}
\caption[X-rule mutations]
{\textsf{11 neighborhoods where the precursor output was mutated 
to arrive at the X-rule. The first 6 from the left retain isotropy, 
the final 5 and their spins/flips give 12 non-isotropic neighborhoods, about 2.23\% of the total.}}
\label{X-rule nhoods-mutated}
\end{center}
\end{figure}

The X-rule differs from its precursor by 11 out of 512 neighborhood outputs.
The mutations were all from 1 to 0 -- their positions in the rule-table are 
shown in figure~\ref{X-rule mutations} and the actual neighborhoods in 
figure~\ref{X-rule nhoods-mutated}. 
Six of these mutations retain isotropy, 5 are non-isotropic and their spins/flips 
give 12 non-isotropic neighborhoods. The remaining 500 isotropic neighborhoods allow
most but not all gliders to move in reflected and rotated directions.
\enlargethispage{4ex}

\clearpage

%^^^^^^^^^^^^^^^^^^^^^^^^^^^^^^^^^^^^^^^^^^^^^^^^^^^^^^^^^^^^^^^^^^^^
\section{Emergent structures in the X-rule universe}
\label{Emergent structures in the X-rule universe}

%----------------------------------------------------------------------
\subsection{Gliders}
\label{Gliders}

The X-rule conserves two emergent glider types from its precursor,
glider~Ga~\raisebox{-1.5ex}{\includegraphics[height=2ex,bb=0 -1 19 12 clip=]{mga1}}
able to move in all diagonal directions (figure~\ref{glider-Ga}), 
and glider~Gc~\raisebox{-2.3ex}{\includegraphics[height=3.5ex,bb=0 0 35 27 clip=]{miniglider_c1}}
\hspace{.5ex} able to move in all orthogonal directions
(figure~\ref{glider-Gc}). Both these gliders emerge easily from a random initial seed
because the phase patterns of Ga are very simple, and Gc has a simple predecessor -- 
the pattern Gc-p~\raisebox{-2ex}{\includegraphics[height=3ex,bb=0 -1 15 18 clip=]{intW}}
\hspace{.5ex} and its rotations.\\[-2ex] 

The X-rule presents two further gliders
Gd~\raisebox{-1.5ex}{\epsfig{file=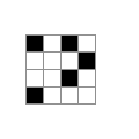, height=3.5ex,bb=0 0 25 25 clip=}}\hspace{.5ex}
and Gb~\raisebox{-1.5ex}{\epsfig{file=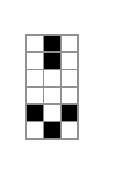, height=5ex,bb=0 -1 23 45, clip=}}\hspace{.5ex}.
Glider Gd, not present in the precursor, is an emergent orthogonal asymmetric glider
moving only West and East with speed $c$/2, in 4 phases where the asymmetry alternates 
about a horizontal axis.

\vspace{-3ex}
\begin{figure}[H]
\textsf{\small
\begin{center}
\begin{tabular}[t]{ @{}c@{} @{}c@{}  @{}c@{}  @{}c@{}  @{}c@{}   @{}c@{}   }
 \multicolumn{4}{c}{\hspace{1.5ex}$\leftarrow$---------------- Gd ----------------$\rightarrow$} & &\\[-2ex]
  \includegraphics[height=.09\linewidth,bb=-3 2  48 45, clip=]{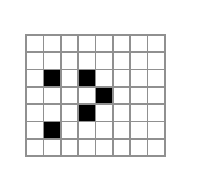}%
& \includegraphics[height=.09\linewidth,bb=-3 2  48 45,  clip=]{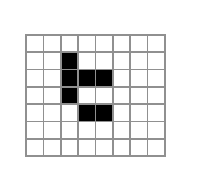}%
& \includegraphics[height=.09\linewidth,bb=-3 2  48 45,  clip=]{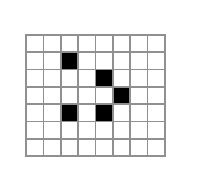}%
& \includegraphics[height=.09\linewidth,bb=-3 2  48 45,  clip=]{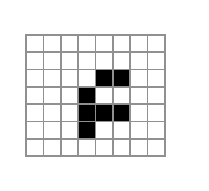}%
& \includegraphics[height=.09\linewidth,bb=-3 2  48 45,  clip=]{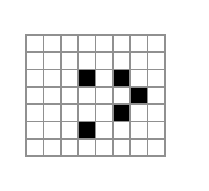}%
& \includegraphics[width=.08\linewidth,bb= -3 -7 32 45, clip=]{ArrowE.pdf}\\[-1.5ex]   
1 & 2 & 3 & 4 & 5 & 
\end{tabular}
\end{center}
}
\vspace{-4ex}
\caption[glider Gd]%
{\textsf{Asymmetric glider Gd shown moving East. The reverse moves West. 
}}
\label{glider-Gd}
\vspace{-4ex}
\end{figure} 

\vspace{-2ex}
\begin{figure}[H]
\textsf{\small
\begin{center}
\begin{tabular}[t]{ @{}c@{} @{}c@{}  @{}c@{}  @{}c@{}  @{}c@{}   @{}c@{}}
& \multicolumn{4}{c}{$\leftarrow$------------- Gb -------------$\rightarrow$} &\\
\includegraphics[height=.12\linewidth,bb= 0 -5  30 34, clip=]{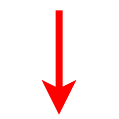}
& \includegraphics[height=.13\linewidth,bb= 0 -1  45 60, clip=]{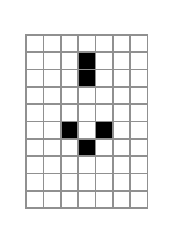}%
& \includegraphics[height=.13\linewidth,bb= 0 -1  45 60,  clip=]{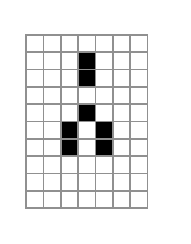}%
& \includegraphics[height=.13\linewidth,bb= 0 -1  45 60,  clip=]{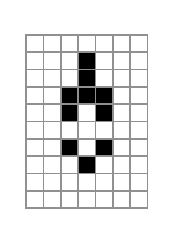}%
& \includegraphics[height=.13\linewidth,bb= 0 -1  45 60,  clip=]{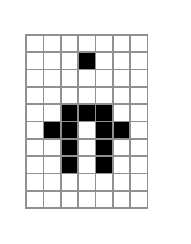}%
& \includegraphics[height=.13\linewidth,bb= 0 -1  45 60,  clip=]{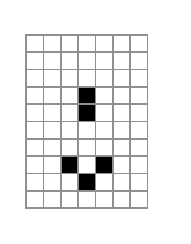}\\[-1.5ex]  
& 1 & 2 & 3 & 4 & 5 
\end{tabular}
\end{center}
}
\vspace{-4ex}
\caption[glider Gb] %
{\textsf{Glider Gb shown moving South. The reverse moves North. 
}}
\label{glider-Gb}
\end{figure}

In the X-rule, Gb is an orthogonal glider moving only North and South
with speed $c$/2, in 4 phases. In the X-rule precursor, Gb also
functions -- in all directions, but was not initially observed because
each of its phases has a moderately intricate pattern with a low
probability of emerging and surviving from a random seed, and Gb's
phases also lack a simple predecessor.

\begin{figure}[h]
\textsf{\small
\begin{tabular}[t]{ @{}c@{} @{}c@{}  @{}c@{}  @{}c@{}  @{}c@{}  @{}c@{}  @{}c@{}  @{}c@{}  @{}c@{}  @{}c@{} }
& \multicolumn{4}{c}{$\leftarrow$--------------- Gc ---------------$\rightarrow$} & Gc-p &% 
\multicolumn{4}{c}{$\leftarrow$--------------- Gb ---------------$\rightarrow$}\\
\includegraphics[width=.07\linewidth,bb=-2 7 23 28, clip=]{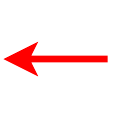}%   
& \includegraphics[width=.11\linewidth,bb=-2 -1 65 40, clip=]{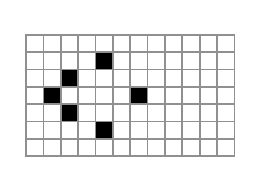}
& \includegraphics[width=.11\linewidth,bb=-2 -1 65 40, clip=]{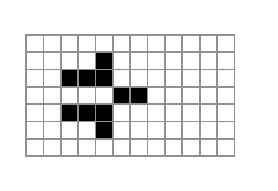}%
& \includegraphics[width=.11\linewidth,bb=-2 -1 65 40, clip=]{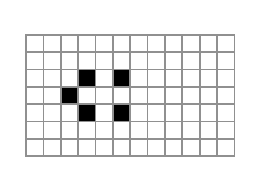}%
& \includegraphics[width=.11\linewidth,bb=-2 -1 65 40, clip=]{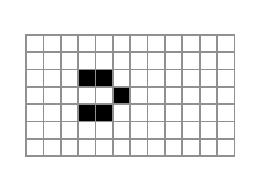}%
& \includegraphics[width=.11\linewidth,bb=-2 -1 65 40, clip=]{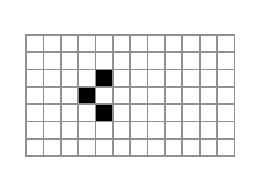}%
& \includegraphics[width=.11\linewidth,bb=-2 -1 65 40, clip=]{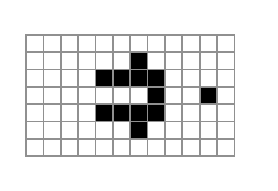}%
& \includegraphics[width=.11\linewidth,bb=-2 -1 65 40, clip=]{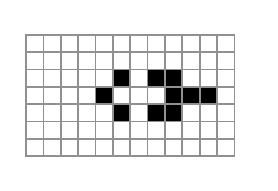}%
& \includegraphics[width=.11\linewidth,bb=-2 -1 65 40, clip=]{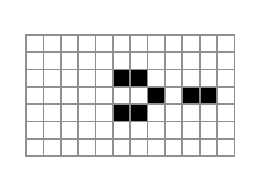}%
& \includegraphics[width=.11\linewidth,bb=-2 -1 65 40, clip=]{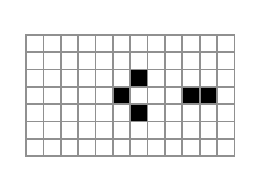}\\[-1.5ex]  
& 9 & 8 & 7 & 6 & 5 & 4 & 3 & 2 & 1
\end{tabular}
}
\vspace{-3ex}
\caption[glider Gb West to Gc] %
{\textsf{Starting with glider Gb (phase 1) pointing West, the pattern
    transits all of its 4 phases, then transforms to Gc-p which is the predecessor
    of glider Gc.
}}
\label{glider Gb West to Gc}
\end{figure}

In the X-rule glider Gb is not free to move
East and West, only North and South. If Gb (phase 1 in figure~\ref{glider-Gb}) is
pointed towards the 
West~\raisebox{-1.5ex}{\epsfig{file=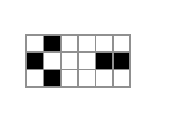, height=2.5ex,bb=2 0 34 17 clip=}}\hspace{.5ex}
it transits into glider Gc moving West via Gc-p 
\raisebox{-2ex}{\epsfig{file=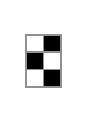, height=3ex,bb=0 -1 15 18 clip=}}\hspace{.5ex}
as in figure~\ref{glider Gb West to Gc}.\\

\vspace{-2ex}

Things become more complicated  when glider Gb is pointed East. If Gb phase 4 
in figure~\ref{glider Gb West to Gc}
is reversed towards the East it will transit 
via Gc-p~\raisebox{-2ex}{\epsfig{file=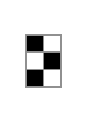, height=3ex,bb=0 -1 15 18 clip=}}\hspace{.5ex} 
into glider Gc moving East -- a reversed scenario to figure~\ref{glider Gb West to Gc}.
Gc-p pointing West 
\hspace{-.5ex} \raisebox{-2ex}{\epsfig{file=intW.pdf, height=3ex,bb=0 -1 15 18 clip=}}\hspace{.5ex} 
or East \hspace{-.5ex}\raisebox{-2ex}{\epsfig{file=intE.pdf, height=3ex,bb=0 -1 15 18 clip=}}\hspace{.5ex} 
provides the predecessors of Gc.

\vspace{-2ex}
\begin{figure}[h]
\textsf{\small
\begin{center}
\begin{tabular}[t]{ @{}c@{} @{}c@{}  @{}c@{}  @{}c@{}  @{}c@{}  @{}c@{}  @{}c@{}  }
{$\leftarrow$ Gb $\rightarrow$} & Gc-p &
\multicolumn{4}{c}{$\leftarrow$--------------- Gb ---------------$\rightarrow$} & \\[-1ex]
  \includegraphics[width=.11\linewidth,bb=5 -1 78 45, clip=]{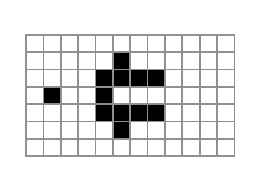}%
& \includegraphics[width=.11\linewidth,bb=5 -1 78 45, clip=]{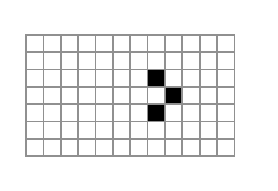}%
& \includegraphics[width=.11\linewidth,bb=5 -1 78 45, clip=]{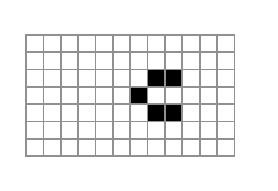}%
& \includegraphics[width=.11\linewidth,bb=5 -1 78 45, clip=]{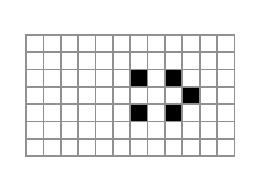}%
& \includegraphics[width=.11\linewidth,bb=5 -1 78 45, clip=]{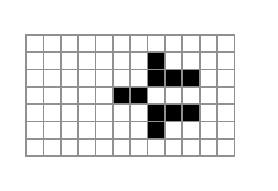}%
& \includegraphics[width=.11\linewidth,bb=5 -1 78 45, clip=]{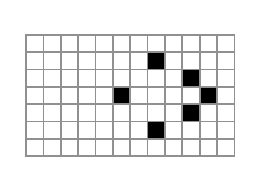}%
& \includegraphics[width=.07\linewidth,bb= 0 -3 32 35, clip=]{ArrowE.pdf}\\[-1.5ex]   
4 & 5 & 6 & 7 & 8 & 9 & 
\end{tabular}
\end{center}
}
\vspace{-3ex}
\caption[glider Gb phase 4 East to Gc] %
{\textsf{Starting with glider Gb (phase 4) pointing East, 
the pattern changes to Gc-p,
then transits to glider Gc moving East.
}}
\label{[glider Gb phase 4 East to Gc}
\end{figure}

However, if Gb phase 1 in figure~\ref{glider Gb West to Gc}
is turned towards the East, it will transit Gb phase 2 and 3 pointing East but not reach phase 4,
instead it will transform
in stages to the final result of a pair of Ga gliders moving SW and NW, and a pair of
Gc gliders moving West and East.

\begin{figure}[h]
\textsf{\small
\begin{center}
%\setlength{\fboxsep}{0pt}
%\hspace{-4ex}
\begin{tabular}[t]{ @{}c@{} @{}c@{}  @{}c@{}  @{}c@{}  @{}c@{} @{}c@{}}
\multicolumn{3}{c}{$\leftarrow$------- Gb -------$\rightarrow$} & & & \\[-5ex]
  \begin{minipage}[b]{.09\linewidth}
  \includegraphics[width=1\linewidth,bb=-2 -6 48 40, clip=]{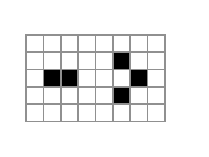}\\[-6.7ex]
  \begin{center} 1 \end{center}
   \end{minipage}%
& \begin{minipage}[b]{.09\linewidth}
  \includegraphics[width=1\linewidth,bb=-2 -6 48 40, clip=]{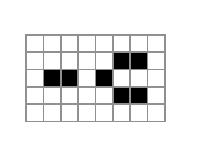}\\[-6.7ex]
  \begin{center} 2 \end{center}
   \end{minipage}%
& \begin{minipage}[b]{.09\linewidth} 
  \includegraphics[width=1\linewidth,bb=-2 -6 48 40, clip=]{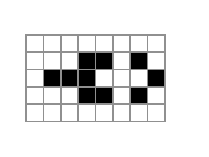}\\[-6.7ex]
  \begin{center} 3 \end{center}
   \end{minipage}%
& \begin{minipage}[b]{.07\linewidth} 
  \includegraphics[width=1\linewidth,bb= -3 -9 32 28, clip=]{ArrowE.pdf}\\[-6.7ex]
  \begin{center}  \end{center}
  \end{minipage}% 
& \begin{minipage}[b]{.16\linewidth} 
  \includegraphics[width=1\linewidth,bb=-2 -2 88 50, clip=]{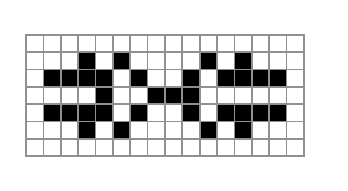}\\[-6.7ex]
  \begin{center} 14 \end{center}
  \end{minipage}% 
& \includegraphics[width=.37\linewidth,bb=-40 0 200 90, clip=]{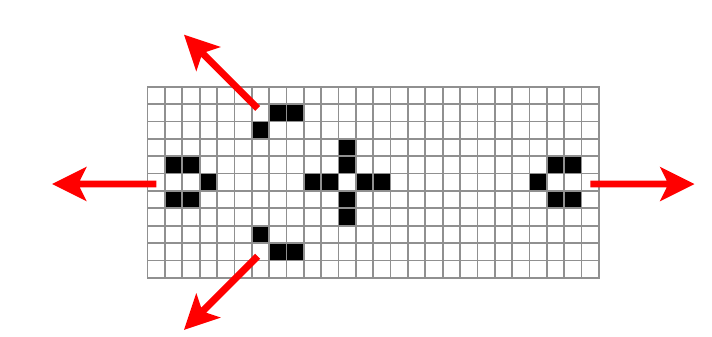}\\[-6.5ex]
\end{tabular}
\end{center}
}
%\vspace{.5ex}
\caption[glider Gb phase 1] %
{\textsf{Starting with glider Gb (phase 1) pointing East 
the \raisebox{3.5ex}{24} \hspace{-3ex} pattern transits the 
first 3 Gb phases,
then transforms according the central part of glider-gun GGa. For example
steps 14 and 24 above are the same as the central part of steps 27 and 37 in figure~\ref{glider-gun GGa}.
At step 24 above, a pair of Ga gliders move SW and NW, and a pair of Gc gliders move West and East.
The remaining patterns in the central area will disappear in 2 more time-steps.
}}
\label{[glider Gb phase 1 East to Ga}
\end{figure}

%----------------------------------------------------------------------
\subsection{Combined gliders}
\label{Combined gliders}

\begin{figure}[h]
\textsf{\small
\begin{center}
\begin{minipage}[b]{.7\linewidth}
\includegraphics[width=1\linewidth,bb=-2 3 373 70, clip=]{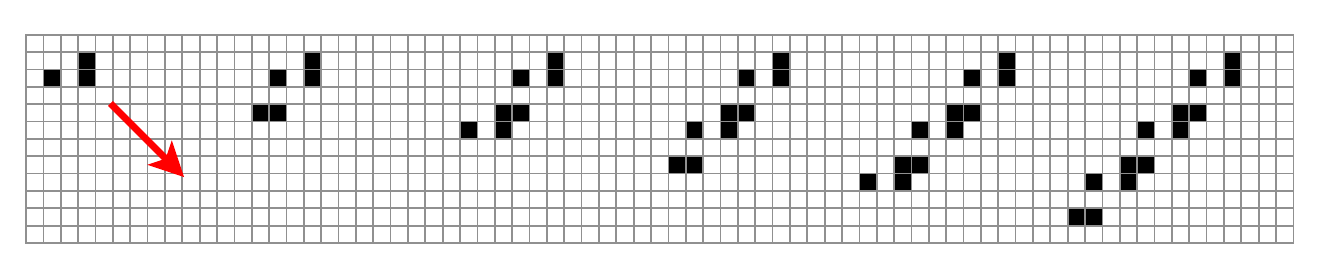}
\end{minipage} 
\end{center}
}
\vspace{-3ex}
\caption[Ga combined gliders SE]
{\textsf{Glider Ga2 (left) and combined Ga2 gliders of increasing size 2 -- 6,
moving SE in 4 phases at a speed of $c/4$. There is a one cell 
overlap between adjacent gliders in the combination. Ga1 combines with Ga3, and Ga2 with Ga4.
%5 time-steps are shown. 
}}
\label{Ga compound gliders SE}
\end{figure}

Ga gliders can be joined together to created combined Ga gliders of arbitrary size
with a one cell overlap between adjacent gliders, as illustrated in 
figure~\ref{Ga compound gliders SE} for combined Ga moving SE.
Combined Ga gliders move in any diagonal direction cycling through the usual 4 phases. 

\clearpage

%----------------------------------------------------------------------
\subsection{Gliders colliding with gliders}
\label{Gliders colliding with gliders gliders}

\begin{figure}[h]
\textsf{\small
\begin{center}
\begin{minipage}[b]{1\linewidth}
\includegraphics[width=1\linewidth]{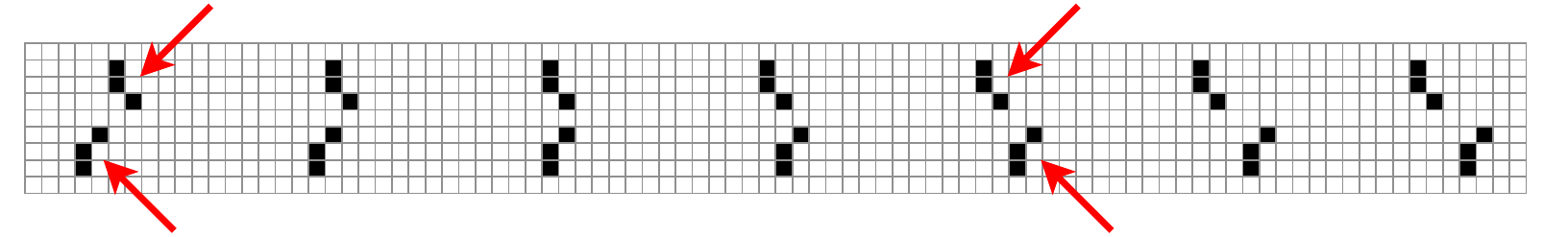}\\[-2ex]%lattice=90x9
\color{white}-----\color{black} P1 \hspace{7.5ex} P2 \hspace{8ex} P3 \hspace{9.5ex}%
P4\hspace{9.5ex} P5  \hspace{9.5ex} P6 \hspace{9.5ex} P7\\[2ex]
\includegraphics[width=1\linewidth]{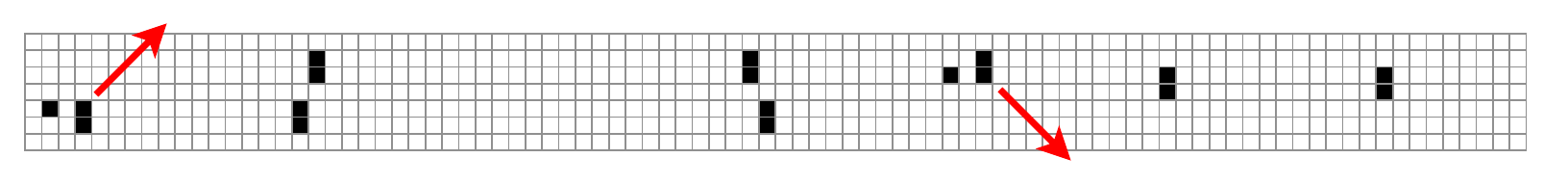}%lattice=90x7
\end{minipage} 
\end{center}
}
\vspace{-5ex}
\caption[Ga1SW collisions with Ga1NW]
{\textsf{Examples of glider Ga moving SW and colliding with Ga moving NW
with impact points P1 to P6 resulting in the following at time-step 9:\\
P1:~GaSW$\Leftrightarrow$GaNW $\to$ GaNE.
P2:~GaSW$\Leftrightarrow$GaNW $\to$ two Ea6.
P3:~GaSW$\Leftrightarrow$GaNW destroyed.
P4:~GaSW$\Leftrightarrow$GaNW $\to$ two Ea6.
P5:~GaSW$\Leftrightarrow$GaNW $\to$ Ea6.
P6:~GaSW$\Leftrightarrow$GaNW $\to$ Ea6.
}}
\label{Ga1SW collisions with Ga1NW}
\vspace{-5ex}
\end{figure}

\begin{figure}[h]
\textsf{\small
\begin{center}
\begin{minipage}[b]{1\linewidth}
\includegraphics[width=1\linewidth]{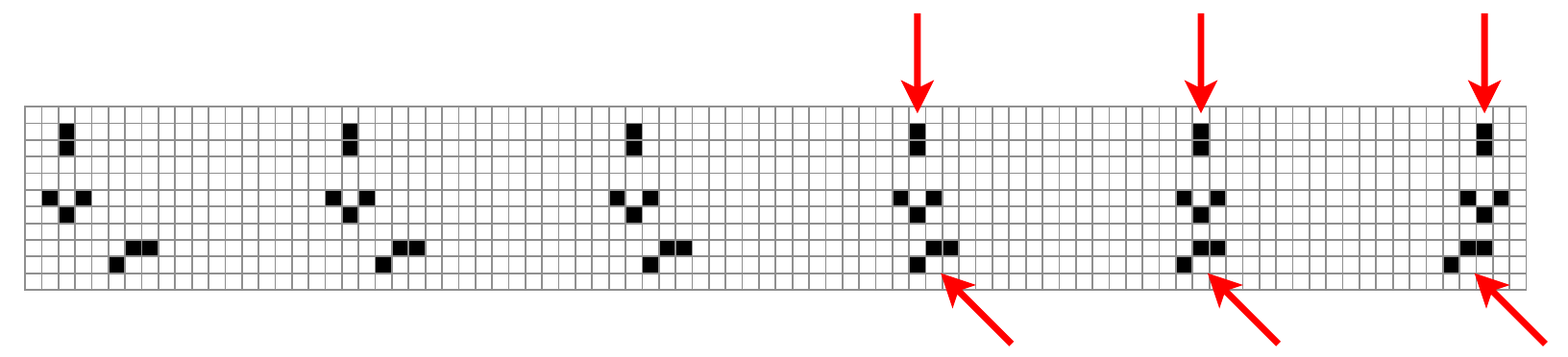}\\[-4ex]%lattice=90x11
\color{white}-----\color{black} P1 \hspace{11ex} P2 \hspace{11ex} P3 \hspace{11.5ex}%
P4\hspace{11.5ex} P5  \hspace{11.5ex} P6\\[2ex]
\includegraphics[width=1\linewidth]{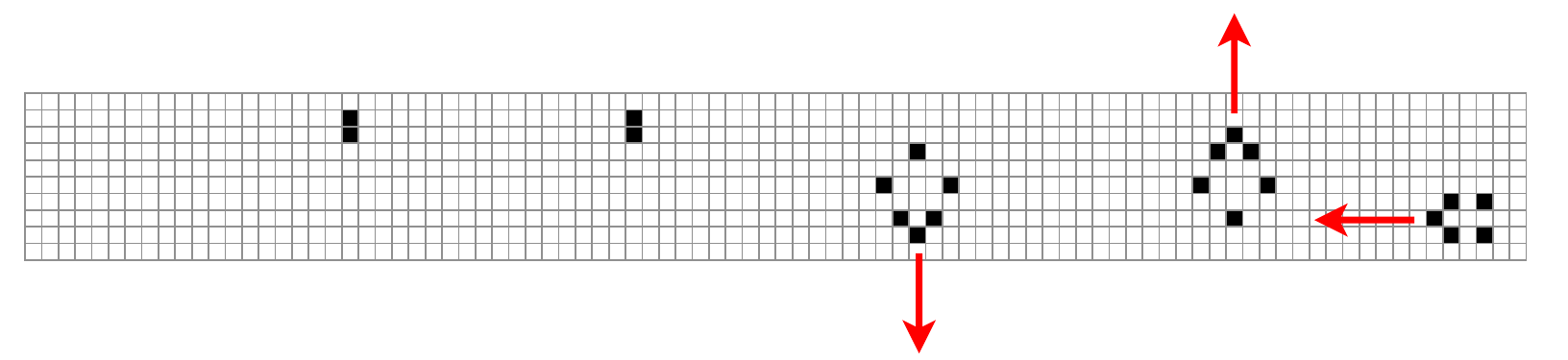}%lattice=90x10
\end{minipage}
\vspace{-8ex} 
\end{center}
}
\caption[Ga1NW collisions with Gb1 South]
{\textsf{Examples of glider Ga1 moving NW and colliding with Gb1 moving South.
with impact points P1 to P6 resulting in the following at time-step 12:\\
P1:~GaSW$\Leftrightarrow$GbS destroyed.
P2,P3:~GaSW$\Leftrightarrow$Gb~South $\to$ Ea6.
P4:~GaSW$\Leftrightarrow$Gb~South $\to$ Gc South.
P5:~GaSW$\Leftrightarrow$Gb~South $\to$ Gc North.
P6:~GaSW$\Leftrightarrow$Gb~South $\to$ Gc West.
}}
\label{Ga1NW collisions with Gb1 South}
\end{figure}

The outcome of glider collisions is highly sensitive to the collision phases,
and the point and angle of impact. Gliders can self destruct, form a stable structure,
transform and bounce off at different angles.
Figures~\ref{Ga1SW collisions with Ga1NW} and \ref{Ga1NW collisions with Gb1 South}
give just a flavour of the diversity of behaviour.

\clearpage

%----------------------------------------------------------------------
\subsection{Gliders colliding with eaters}
\label{Gliders colliding with eaters}

\begin{figure}[h]
\textsf{\small
\begin{center}
\begin{minipage}[b]{1\linewidth}
\includegraphics[width=1\linewidth]{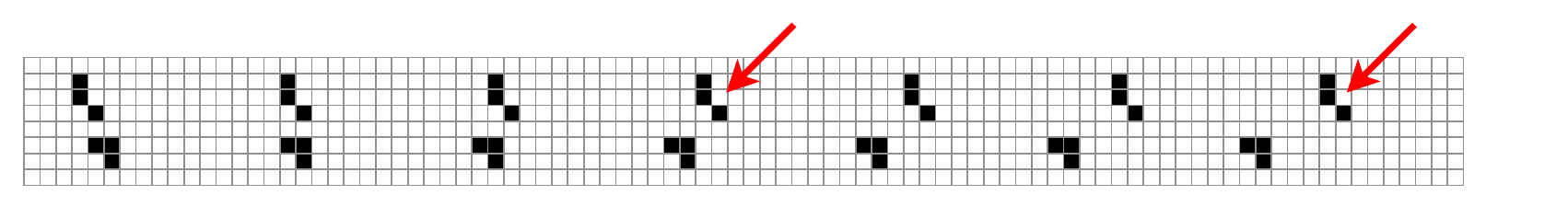}\\[-2ex] %lattice=90x8
\color{white}-----\color{black} P1 \hspace{7ex} P2 \hspace{6ex} P3 \hspace{7ex} P4\hspace{7ex} P5  \hspace{7ex} P6 \hspace{7ex} P7\\[1ex]
\includegraphics[width=1\linewidth]{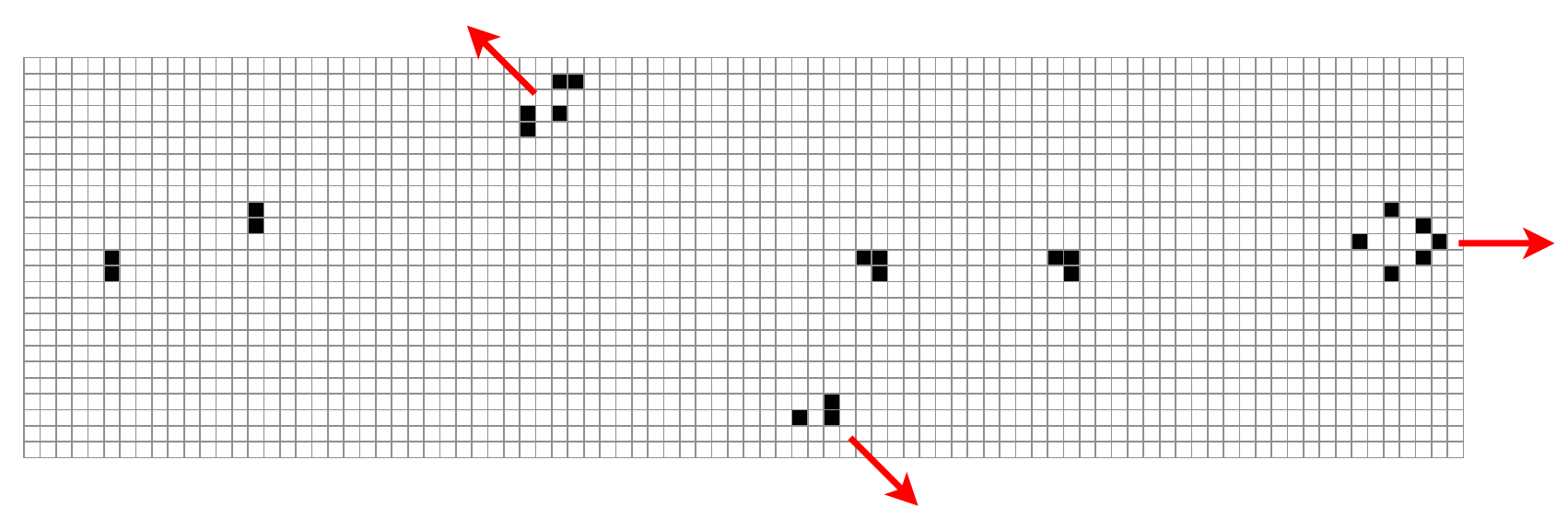}%lattice=90x25
\end{minipage} 
\end{center}
}
\vspace{-5ex}
\caption[Ea1 collisions with GaSW]
{\textsf{Examples of glider Ga moving SW colliding into eater Ea1
\hspace{-1ex}\raisebox{-1ex}{\includegraphics[height=2.8ex,bb=-1 5 20 20, clip=]{Ea1.pdf}}
with impact points P1 to P7 resulting in the following
at time-step 44:\\
P1,P2:~Ea1 $\to$ Ea6, Ga destroyed.
P3:~Ea1 destroyed, Ga destroyed.
P4:~Ea1 destroyed, Ga $\to$ combined$\times$2-Ga NW and Ga SE.
P5,P6:~Ea1 stable, Ga destroyed.  
P7:~Ea1 destroyed, Ga $\to$ Gc East.
}}
\label{Ea1 collisions with GaSW}
\end{figure}

\begin{figure}[h]
\textsf{\small
\begin{center}
\begin{minipage}[b]{1\linewidth}
\includegraphics[width=1\linewidth]{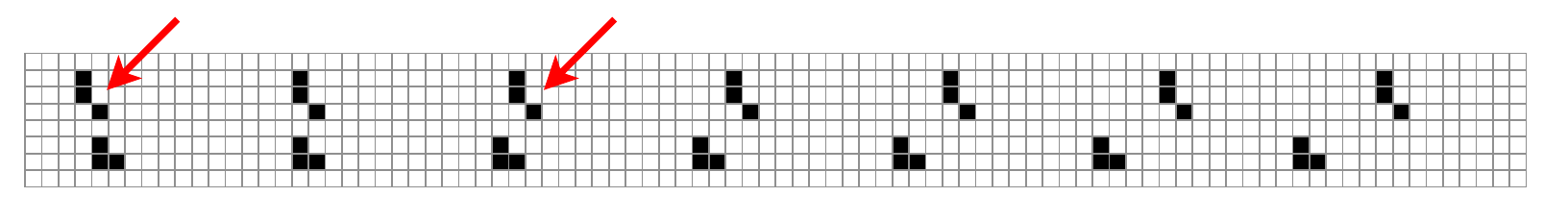}\\[-2ex]%lattice=90x7
\color{white}--------\color{black} P1 \hspace{7ex} P2 \hspace{6ex} P3 \hspace{7ex} P4\hspace{7ex} P5  \hspace{7ex} P6 \hspace{7ex} P7\\[1ex]
\includegraphics[width=1\linewidth]{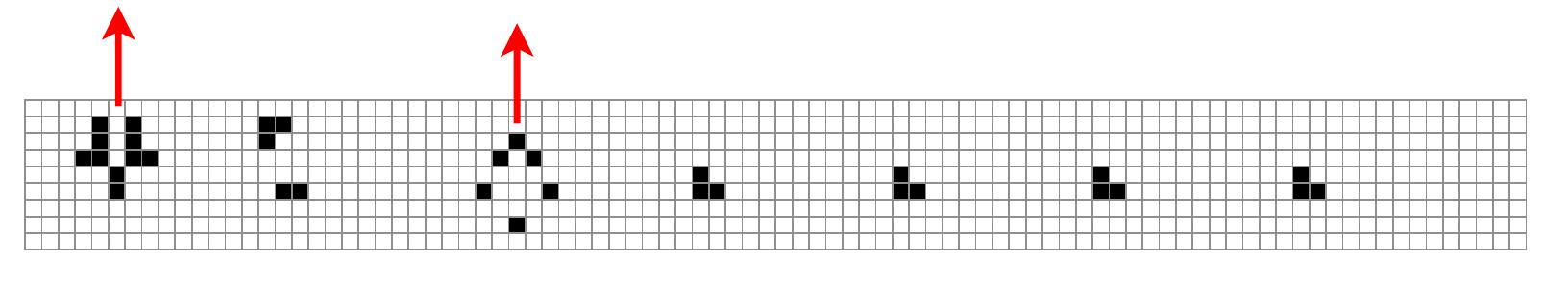}%lattice=90x11
\end{minipage} 
\end{center}
}
\vspace{-4ex}
\caption[Ea4 collisions with GaSW]
{\textsf{Examples of glider Ga moving SW colliding into eater Ea4
\hspace{-1ex}\raisebox{-1ex}{\includegraphics[height=2.8ex,bb=-1 5 20 20, clip=]{Ea4.pdf}}
with impact points P1 to P7 resulting in the following at time-step 13:\\
P1:~Ea4 destroyed, Ga $\to$ Gc North. 
P2:~Ea4  $\to$ Ea3 and Ea5, Ga destroyed.
P3:~Ea4 destroyed, Ga $\to$ Gc North.
P4~--~P7:~Ea4 conserved, Ga destroyed.
}}
\label{Ea4 collisions with GaSW}
\end{figure}

\begin{figure}[t]
\vspace{-2ex}
\textsf{\small
\begin{center}
\begin{minipage}[b]{1\linewidth}
\includegraphics[width=1\linewidth]{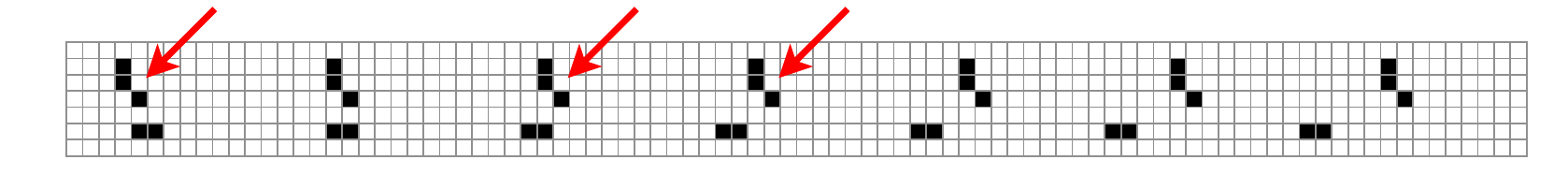}\\[-2ex]%lattice=90x7
\color{white}--------\color{black} P1 \hspace{7ex} P2 \hspace{6ex} P3 \hspace{7ex} P4\hspace{7ex} P5  \hspace{7ex} P6 \hspace{7ex} P7\\[1ex]
\includegraphics[width=1\linewidth]{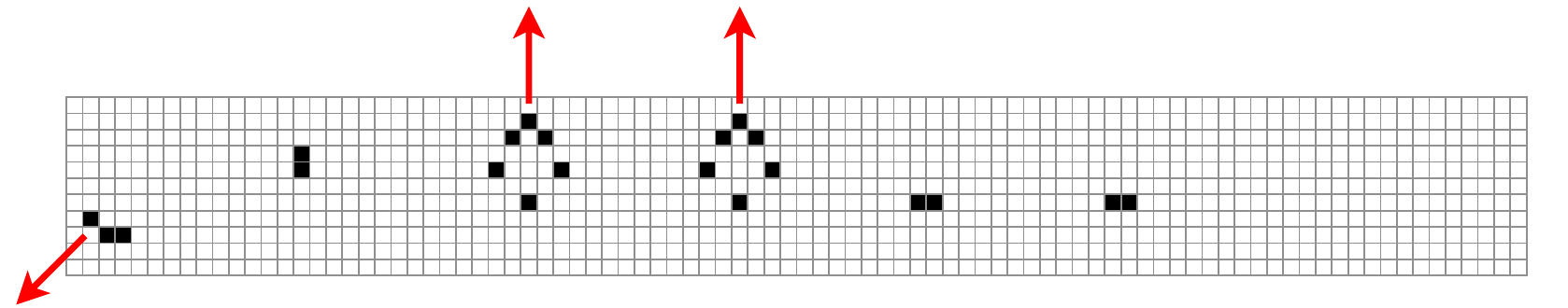}%lattice=90x11
\end{minipage} 
\end{center}
}
\vspace{-4ex}
\caption[Ea5 collisions with GaSW]
{\textsf{Examples of glider Ga moving SW colliding into eater Ea5
\hspace{-1ex}\raisebox{-1ex}{\includegraphics[height=2.8ex,bb=-1 5 20 20, clip=]{Ea5.pdf}}
with impact points P1 to P7 resulting in the following at time-step 14:\\
P1:~Ea5 destroyed, Ga conserved.
P2:~Ea5~$\to$~Ea6, Ga destroyed.
P3,~P4:~Ea5 destroyed, Ga $\to$ Gc North.
P5,~P6:~Ea5 conserved, Ga destroyed.
P7:~Ea destroyed, Ga destroyed.
}}
\label{Ea5 collisions with GaSW}
\vspace{-4ex}
\end{figure}

\begin{figure}[h]
\textsf{\small
\begin{center}
\begin{minipage}[b]{1\linewidth}
\includegraphics[width=1\linewidth]{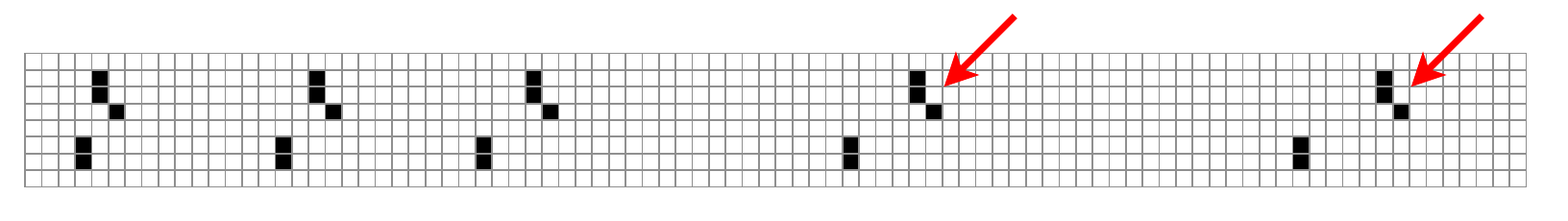}\\[-2ex]%lattice=90x7
\color{white}------\color{black} P1 \hspace{7ex} P2 \hspace{6ex} P3 \hspace{15ex} P4\hspace{21ex} P5 \\[1ex]
\includegraphics[width=1\linewidth]{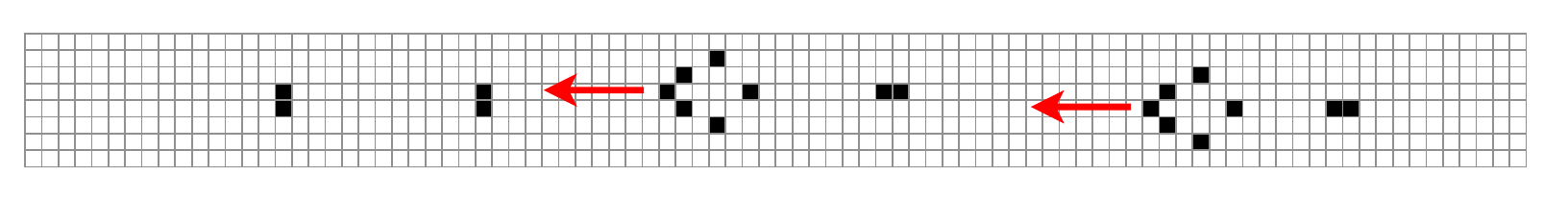}%lattice=90x11
\end{minipage} 
\end{center}
}
\vspace{-4ex}
\caption[Ea6 collisions with GaSW]
{\textsf{Examples of glider Ga moving SW and colliding into eater Ea6
\hspace{-1ex}\raisebox{-1ex}{\includegraphics[height=2.8ex,bb=-1 5 20 20, clip=]{Ea6.pdf}}
with impact points P1 to P5 resulting in the following at time-step 34:\\
P1:~Ea6 destroyed, Ga destroyed.
P2:~Ea6 conserved, Ga destroyed.
P3:~Ea6 conserved, Ga destroyed.
P4:~Ea6 $\to$ Ea5, Ga $\to$ Gc West.
P5:~Ea6 $\to$ Ea5, Ga $\to$ Gc West.
}}
\label{Ea6 collisions with GaSW}
\vspace{-20ex}
\end{figure}

\begin{figure}[b]
\textsf{\small
\begin{center}
\begin{minipage}[b]{1\linewidth}
\includegraphics[width=1\linewidth]{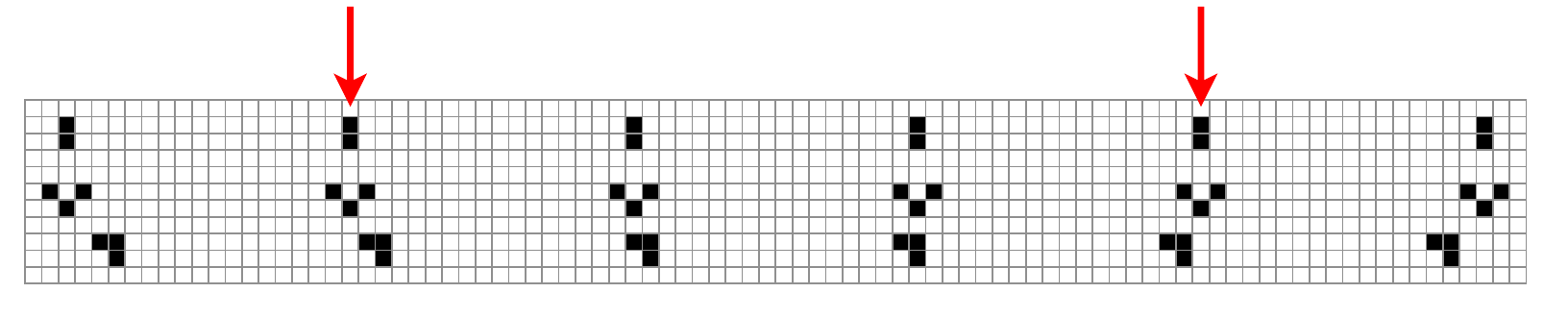}\\[-2ex] %lattice=90x8
\color{white}-----\color{black} P1 \hspace{10.5ex} P2 \hspace{10.8ex}% 
P3 \hspace{10.8ex} P4\hspace{10.8ex} P5  \hspace{10.8ex} P6\\[1ex]
\includegraphics[width=1\linewidth]{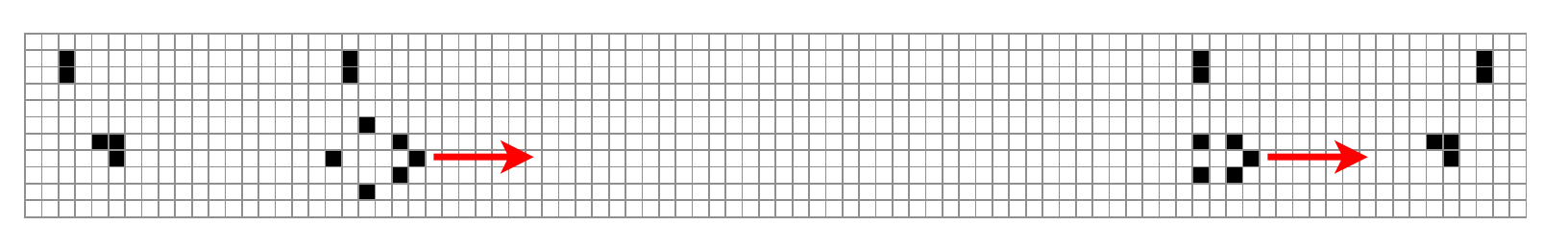}%lattice=90x25
\end{minipage} 
\end{center}
}
\vspace{-4ex}
\caption[Ea1 collisions with Gb South]
{\textsf{Examples of glider Gb moving South colliding into eater Ea1
\hspace{-1ex}\raisebox{-1ex}{\includegraphics[height=2.8ex,bb=-1 5 20 20, clip=]{Ea1.pdf}}
with impact points P1 to P6 resulting in the following
at time-step 44:\\
P1:~Ea1 $\to$ Ea6 and Ea1, Gb destroyed.
P2:~Ea1 $\to$ Ea6, Gb $\to$ Gc1 East.
P3,P4:~Ea1 destroyed, Gb destroyed.
P5:~Ea1 $\to$ Ea6, Gb $\to$ Gc3 East.
P6:~Ea1 $\to$ Ea6 and Ea1, Gb destroyed.
}}
\label{Ea1 collisions with Gb South}
\end{figure}

Just like its precursor, the X-rule presents small stable emergent configurations Ea
of 6 types, Ea1 -- Ea6, which include all rotations
1\hspace{-1.2ex}\raisebox{-1ex}{\includegraphics[height=2.8ex,bb=-1 5 20 20, clip=]{Ea1.pdf}}
2\hspace{-1.2ex}\raisebox{-1ex}{\includegraphics[height=2.8ex,bb=-1 5 20 20, clip=]{Ea2.pdf}}
3\hspace{-1.2ex}\raisebox{-1ex}{\includegraphics[height=2.8ex,bb=-1 5 20 20, clip=]{Ea3.pdf}}
4\hspace{-1.2ex}\raisebox{-1ex}{\includegraphics[height=2.8ex,bb=-1 5 20 20, clip=]{Ea4.pdf}}
5\hspace{-1.2ex}\raisebox{-1ex}{\includegraphics[height=2.8ex,bb=-1 5 20 20, clip=]{Ea5.pdf}}
6\hspace{-1.2ex}\raisebox{-1ex}{\includegraphics[height=2.8ex,bb=-1 5 20 20, clip=]{Ea6.pdf}}\hspace{-1ex}.
They are know as eaters because in most cases they destroy colliding gliders,
but eaters Ea may themselves be destroyed or transformed in the process, or
instead of being destroyed a glider can be transformed and/or reflected --
the exact behaviour is highly sensitive to the collision phase, point and angle of
impact,  and the Ea type.
As in collisions between gliders in section~\ref{Gliders colliding with gliders gliders},
the examples in figures~\ref{Ea1 collisions with GaSW} -- \ref{Ea1 collisions with Gb South}
give just a flavour of the diversity of behaviour.

\vspace{-1ex}

\vspace{-5ex}
\clearpage

\vspace{-4ex}

\clearpage

%----------------------------------------------------------------------
\subsection{Combined eaters make reflectors}
\label{Combined eaters make reflectors}

\begin{figure}[H]
\textsf{\small
\begin{center}
\begin{minipage}[b]{1\linewidth}
\includegraphics[height=.129\linewidth,bb= -4 -3 583 77, clip=]{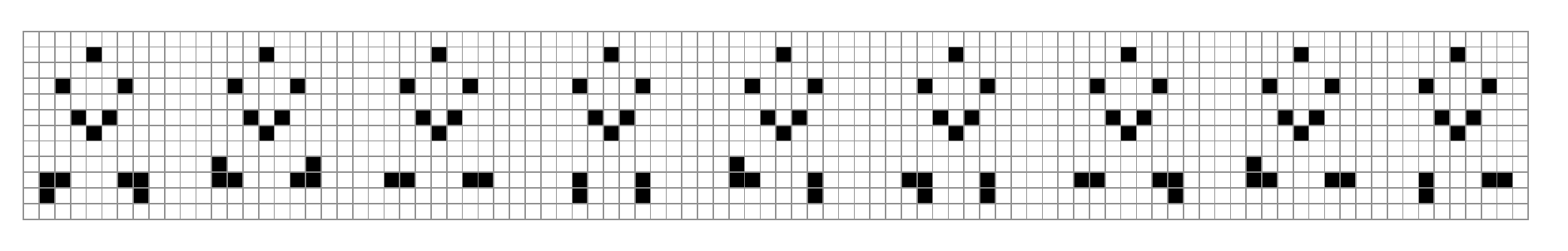}% 
\hfill%
\raisebox{2.5ex}{\includegraphics[height=.09\linewidth,bb= 6 -4  27 34, clip=]{ArrowS.pdf}}\\[-5ex]
\begin{center}Gc1 heading South approaches variations of Rc at a distance of 2 cells.\end{center} 
\includegraphics[height=.129\linewidth,bb= -4 -3 583 77, clip=]{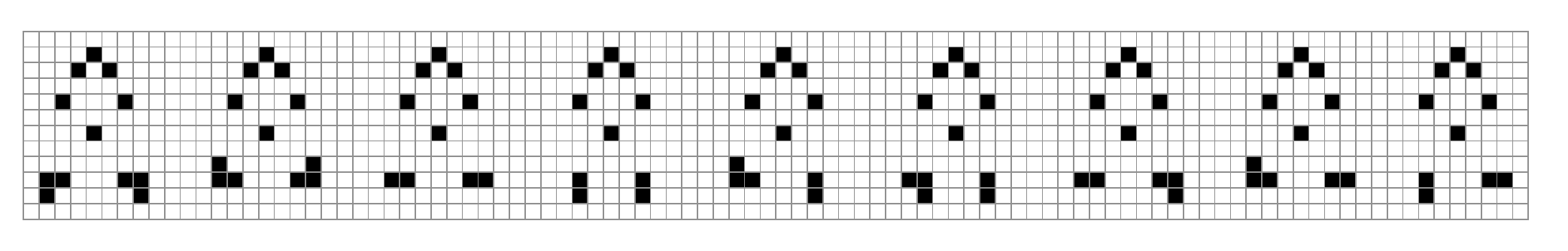}%
\hfill%
\raisebox{2.5ex}{\includegraphics[height=.09\linewidth,bb= 6 -4  27 34, clip=]{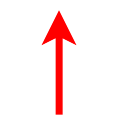}}\\[-5ex]
\begin{center}Gc bounces off Rc shown at time-step 16 with Gc1 heading North.\end{center} 
\end{minipage} 
\end{center}
}
\vspace{-2ex}
\caption[Gc1 bouncing off reflector Rc]
{\textsf{Examples of glider Gc1 bouncing off various combinations of reflector Rc.
The behaviour is valid in any orthogonal direction provided that the distance between
Gc1 and Rc is an even number, otherwise glider Gc is destroyed.
}}
\label{Gc1 bouncing off reflector Rc}
\end{figure}

As in its precursor, the simple Ea eaters in the X-rule can be combined
with a three cells gap to make a stable reflector Rc,
which can either destroy or reflect back Gc gliders.
As shown in figure~\ref{Reflect bouncing}, if the distance between
an incoming glider Gc1 and the reflector Rc is an odd number, Gc1 is destroyed --
for an even number including zero, Gc1 is reflected.
This behaviour is preserved in the X-rule and
occurs with many combinations of Rc as shown 
in figure~\ref{Gc1 bouncing off reflector Rc}.

%----------------------------------------------------------------------
\subsection{Simple reflecting oscillators}
\label{Simple periodic oscillators}

In the X-rule precursor, figure~\ref{pre-periodic-oscillator}, 
we described a simple reflecting oscillator, SRO,
made up of two opposing Rc reflectors with a Gc glider reflected back and forth in between.
The SRO is conserved in the X-rule in any orthogonal orientation.

\begin{figure}[H]
\textsf{\small
\begin{center}
\begin{minipage}[b]{1\linewidth}
\includegraphics[height=.1\linewidth,bb= 0 8 575 67, clip=]{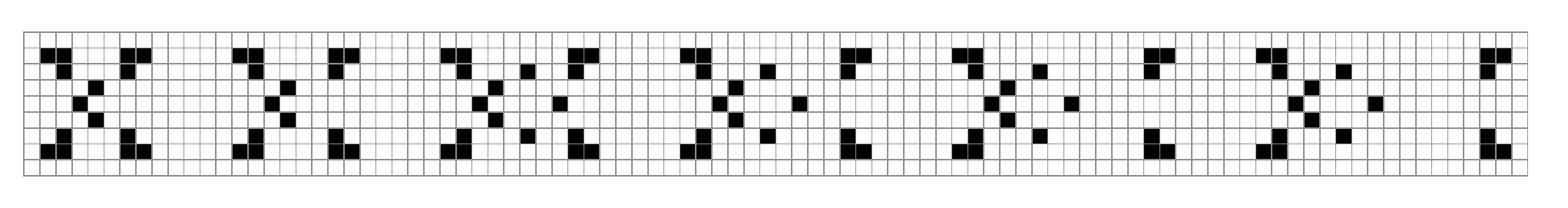}\\[-5ex]
\begin{center}\color{white}-x-\color{black}$g$=3 \color{white}---x----\color{black} $g$=4 \color{white}---xx----\color{black} $g$=6%
\color{white}---xx-------\color{black} $g$=8 \color{white}------x-------\color{black}%
$g$=10 \color{white}----x--------\color{black} $g$=12 \color{white}---------\color{black}\end{center}
\includegraphics[height=.1\linewidth,bb= 0 8 575 67, clip=]{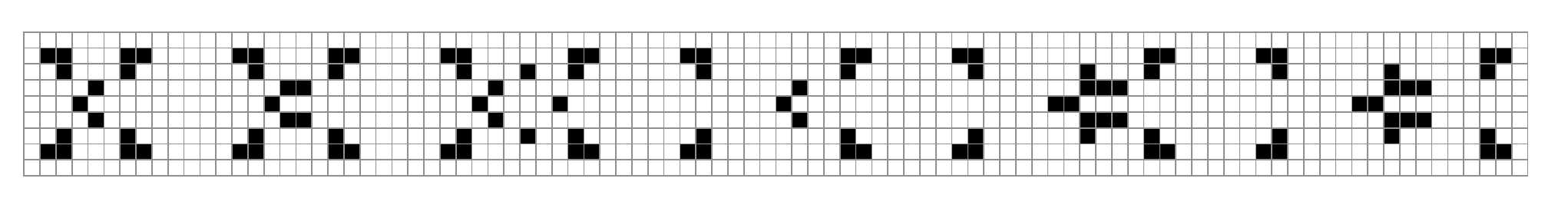}\\[-5ex]
\begin{center}\color{white}--\color{black}$p$=1 \color{white}---x---\color{black} $p$=6 \color{white}--xxx-----\color{black} $p$=14%
\color{white}-----xx-----\color{black} $g$=22 \color{white}------------\color{black}%
$g$=30 \color{white}-----xx-------\color{black} $g$=38 \color{white}----------\color{black}\end{center}
\end{minipage} 
\end{center}
}
\vspace{-5ex}
\caption[oscillators]
{\textsf{Examples simple reflecting oscillators, SRO. 
{\it Top:} Initial oscillator states with gaps $g$ shown.
 {\it Bottom:} The oscillator states after 15 time-steps. 
The complete periods $p$ are shown below. 
For increasing even gaps $g$, period $p$ increases by +8 time-steps. 
}}
\label{oscillators}
\end{figure}

In figure~\ref{oscillators}, as well as glider Gc, we observe the
small pattern \mbox{Gc-p}, rotations of
\raisebox{-2ex}{\epsfig{file=intW.pdf, height=3ex,bb=0 -1 15 18 clip=}}\hspace{.5ex}
-- predecessors of Gc, which are able to oscillate
between two Rc reflectors with a minimal gap of 3 or 4 cells,
thereafter \mbox{Gc-p} still appears in the oscillation sequence with
the next gap of 6 big enough to accommodate a complete Gc glider, and
thereafter any even number gap can continue to increase provided Gc or
\mbox{Gc-a} is started at an appropriate position.

\section{Glider-guns}  
\label{Glider-guns}

\subsection{Two basic glider-guns}  
\label{Two basic glider-guns}

\vspace{-12ex}
\begin{figure}[htb]
\textsf{\small
\begin{center}
\begin{minipage}[b]{.9\linewidth}
   \begin{minipage}[b]{.45\linewidth}
\includegraphics[width=1\linewidth]{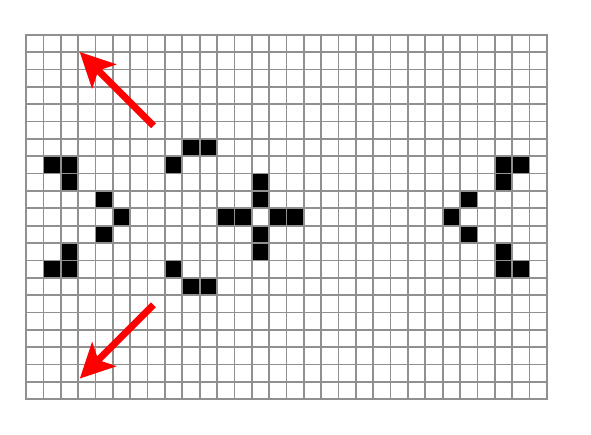}\\[-1.5ex]
glider-gun GGa shooting gliders Ga\\[2ex]
\color{white}-\color{black}
\includegraphics[width=.16\linewidth,bb=5 5 40 37 clip=]{gaSW1.pdf}
\includegraphics[width=.16\linewidth,bb=5 5 40 37, clip=]{gaSW2.pdf}
\includegraphics[width=.16\linewidth,bb=5 5 40 37, clip=]{gaSW3.pdf}
\includegraphics[width=.16\linewidth,bb=5 5 40 37, clip=]{gaSW4.pdf}
\includegraphics[width=.16\linewidth,bb=5 5 40 37, clip=]{gaSW5.pdf}\\[-1ex]
\color{white}---\color{black}1\hspace{5.3ex}2\hspace{5.3ex}3\hspace{5.3ex}4\hspace{5.3ex}5
   \end{minipage}
\hfill
   \begin{minipage}[b]{.45\linewidth}
\includegraphics[width=1\linewidth]{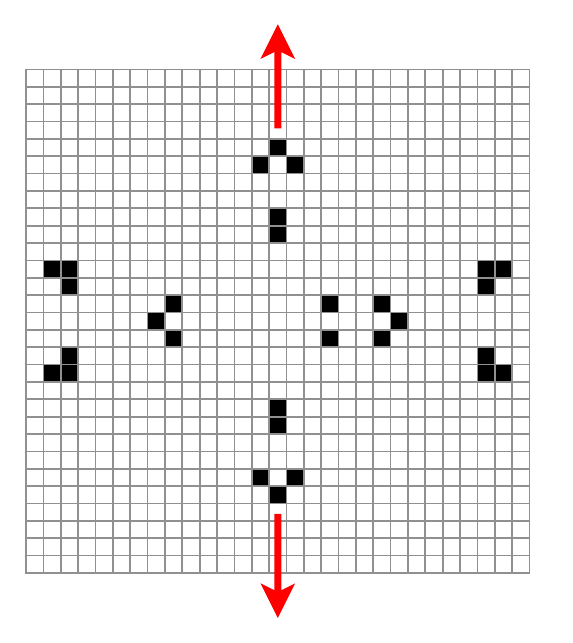}\\[-1.5ex]
glider-gun GGb shooting gliders Gb\\[2ex]
\color{white}-\color{black}
\includegraphics[width=.18\linewidth,bb=5 5 45 60, clip=]{gbS1.pdf}
\includegraphics[width=.18\linewidth,bb=5 5 45 60, clip=]{gbS2.pdf}
\includegraphics[width=.18\linewidth,bb=5 5 45 60, clip=]{gbS3.pdf}
\includegraphics[width=.18\linewidth,bb=5 5 45 60, clip=]{gbS4.pdf}
\includegraphics[width=.18\linewidth,bb=5 5 45 60, clip=]{gbS5.pdf}\\[-1ex]
\color{white}---\color{black}1\hspace{5.3ex}2\hspace{5.3ex}3\hspace{5.3ex}4\hspace{5.3ex}5
   \end{minipage}
\end{minipage}
\end{center}
}
\vspace{-3ex}
\caption[The 2 basic glider-guns]
{\textsf{Snapshots of the two basic glider-guns and glider evolution.\\
{\it (left)} Diagonal glider-gun GGa shoots Ga gliders NW an SW
with speed=$c$/4, time-step 37 in figure~\ref{glider-gun GGa}.
{\it (right)} Orthogonal glider-gun GGb shoots gliders Gb North and South
with speed=$c$/2, time-step 14 in figure~\ref{glider-gun GGb}.
}}
\label{glider-guns GGa and GGb}
\end{figure}

\begin{figure}[h]
\textsf{\small
\begin{minipage}[b]{.24\linewidth}
\includegraphics[width=1\linewidth,bb=4 1 162 117,clip=]{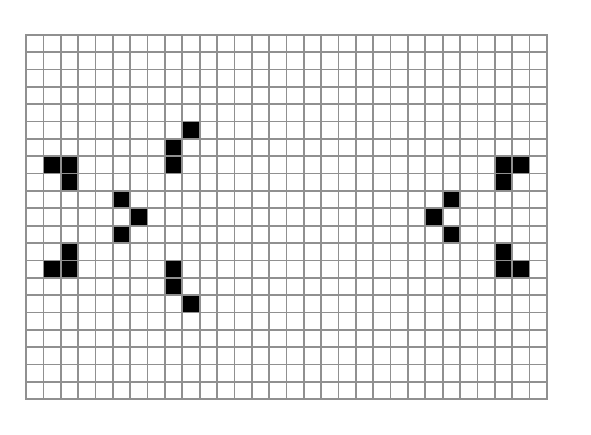}\\[-1ex]
1 sym \end{minipage}
\begin{minipage}[b]{.24\linewidth}
\includegraphics[width=1\linewidth,bb=4 1 162 117,clip=]{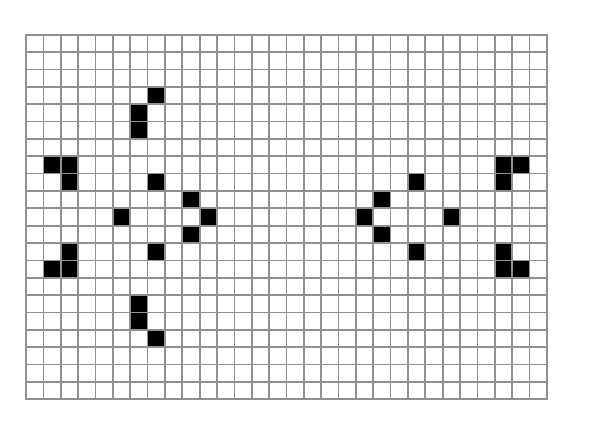}\\[-1ex]
9 sym \end{minipage}
\begin{minipage}[b]{.24\linewidth}
\includegraphics[width=1\linewidth,bb=4 1 162 117,clip=]{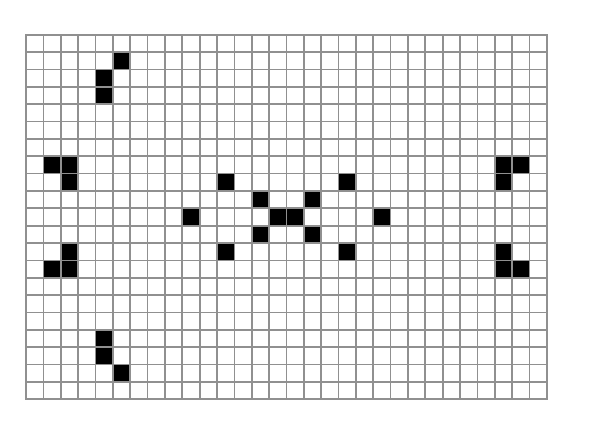}\\[-1ex]
17 sym \end{minipage}
\begin{minipage}[b]{.24\linewidth}
\includegraphics[width=1\linewidth,bb=4 1 162 117,clip=]{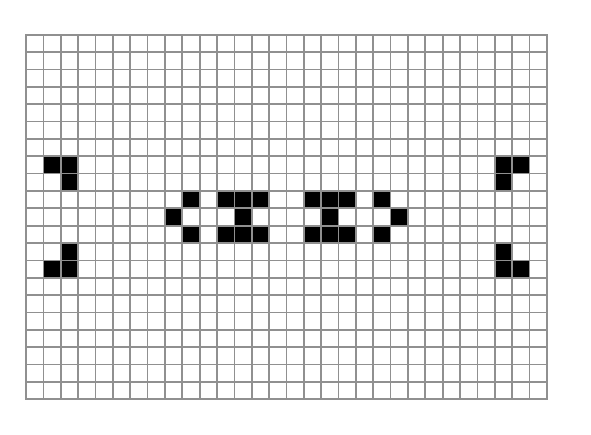}\\[-1ex]
26 sym \end{minipage}\\[2ex]
%------------------------------------------------
\begin{minipage}[b]{.24\linewidth}
\includegraphics[width=1\linewidth,bb=4 1 162 117,clip=]{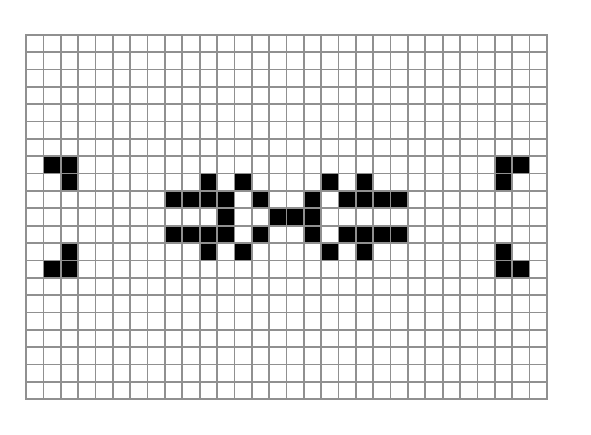}\\[-1ex]
27 asym \end{minipage}
\begin{minipage}[b]{.24\linewidth}
\includegraphics[width=1\linewidth,bb=4 1 162 117,clip=]{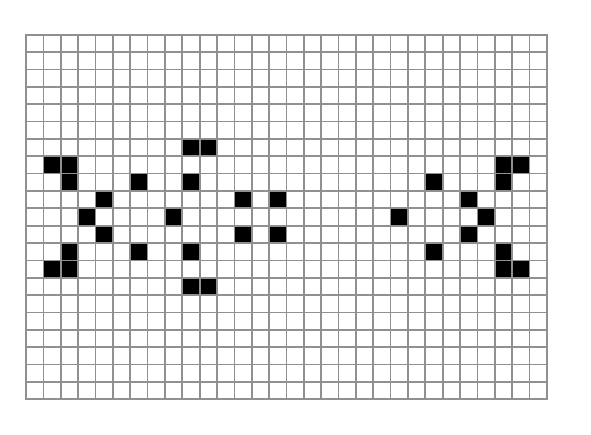}\\[-1ex]
36 asym glider\end{minipage}
\begin{minipage}[b]{.24\linewidth}
\includegraphics[width=1\linewidth,bb=4 1 162 117,clip=]{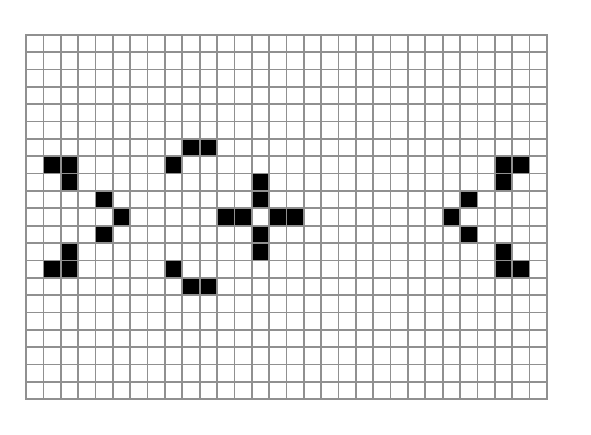}\\[-1ex]
37 asym glide\end{minipage}
\begin{minipage}[b]{.24\linewidth}
\includegraphics[width=1\linewidth,bb=4 1 162 117,clip=]{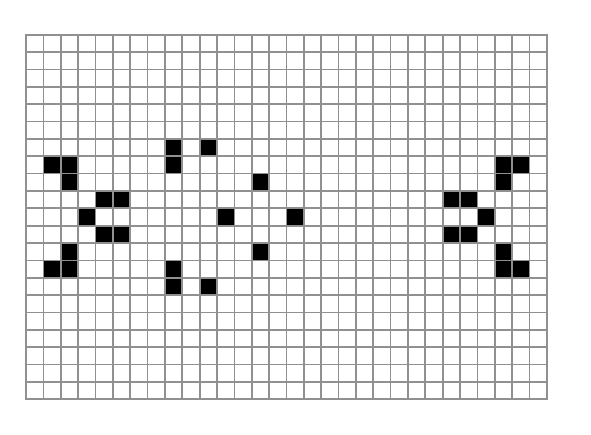}\\[-1ex]
38 asym glide\end{minipage}\\[-4ex]
%---------------------this grid is 30x21
}
\caption[Basic glider-guns GGa]
{\textsf{The basic diagonal glider-gun GGa has a minimum reflector gap of 24 cells,
    an oscillator period 38 time-steps.
    The snapshots (on a 30$\times$x21 lattice) show spans of symmetry (sym) 1--26,
    and asymmetry (asym) 27--38 
    of the oscillating structures.
    Gliders Ga detach at time-steps 36 in the asymmetric span, moving NW and SW.
}}
\label{glider-gun GGa}
\end{figure}

\begin{figure}[h]
\textsf{\small
\begin{minipage}[b]{.24\linewidth}
\includegraphics[width=1\linewidth,bb=2 8 148 155,clip=]{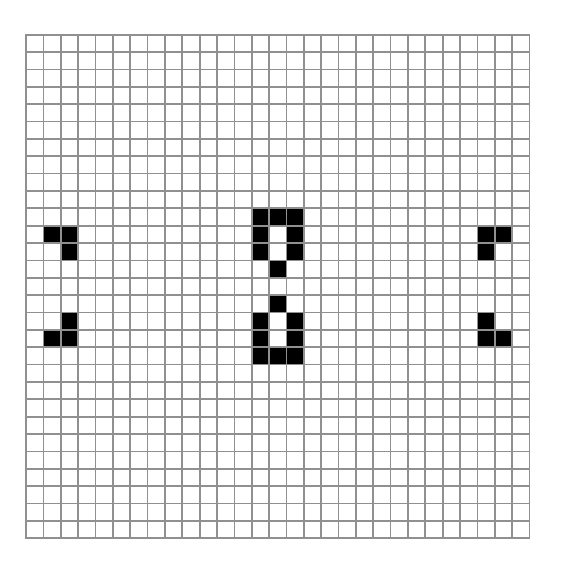}\\[-1ex]
1 sym \end{minipage}
\begin{minipage}[b]{.24\linewidth}
\includegraphics[width=1\linewidth,bb=2 8 148 155,clip=]{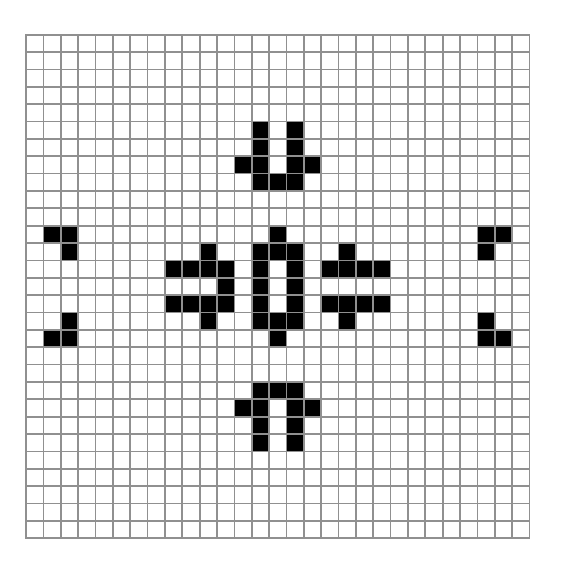}\\[-1ex]
13 sym glider \end{minipage}
\begin{minipage}[b]{.24\linewidth}
\includegraphics[width=1\linewidth,bb=2 8 148 155,clip=]{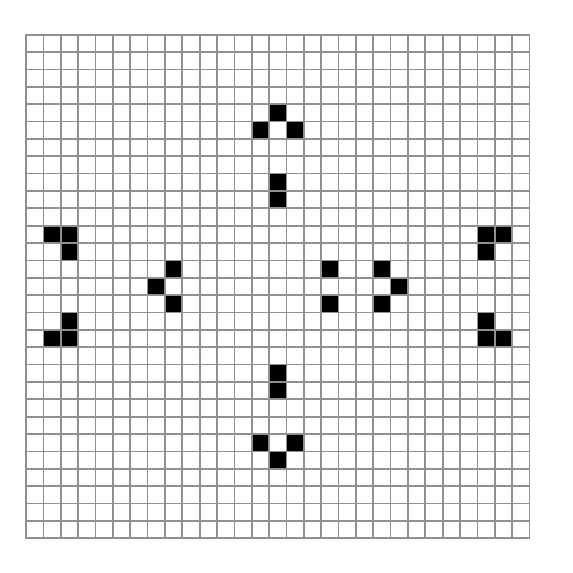}\\[-1ex]
14 asym glider \end{minipage}
\begin{minipage}[b]{.24\linewidth}
\includegraphics[width=1\linewidth,bb=2 8 148 155,clip=]{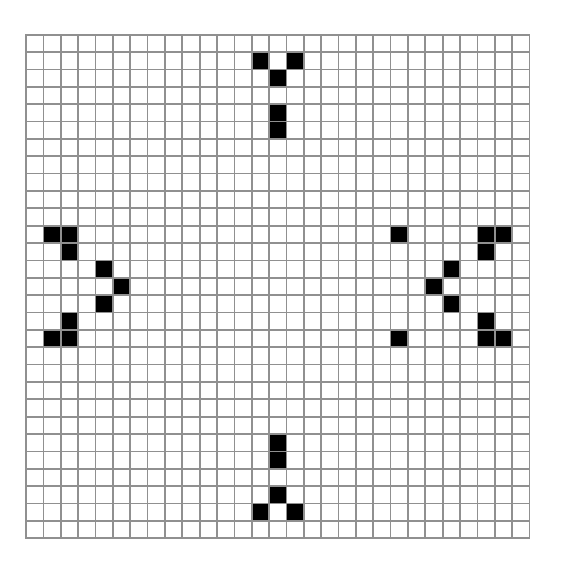}\\[-1ex]
23 asym \end{minipage}\\[2ex]
%------------------------------------------------
\begin{minipage}[b]{.24\linewidth}
\includegraphics[width=1\linewidth,bb=2 8 148 155,clip=]{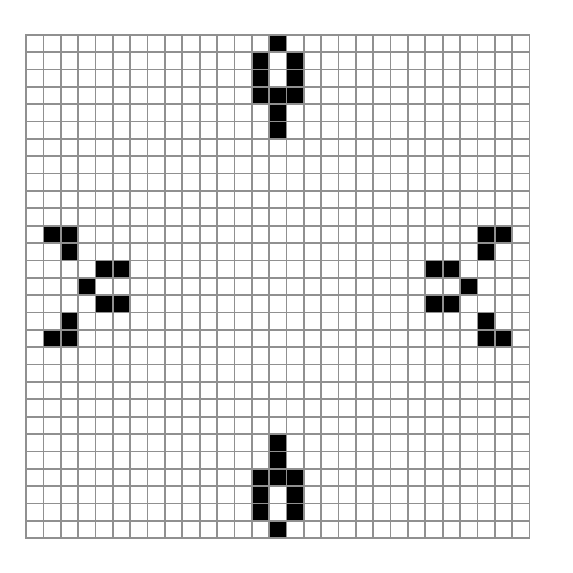}\\[-1ex]
24 sym \end{minipage}
\begin{minipage}[b]{.24\linewidth}
\includegraphics[width=1\linewidth,bb=2 8 148 155,clip=]{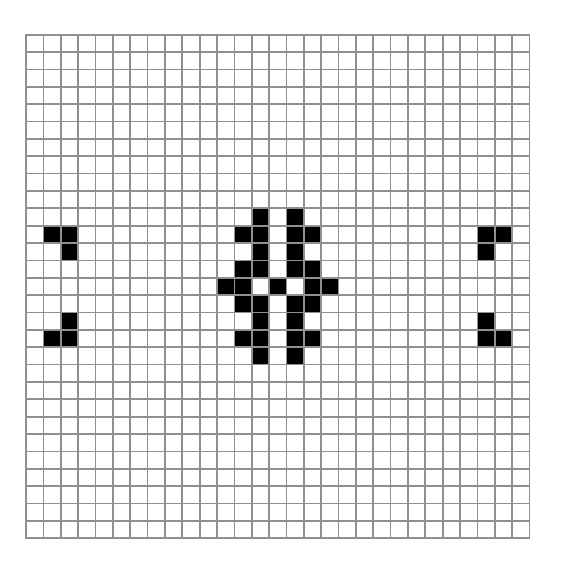}\\[-1ex]
43 sym \end{minipage}
\begin{minipage}[b]{.24\linewidth}
\includegraphics[width=1\linewidth,bb=2 8 148 155,clip=]{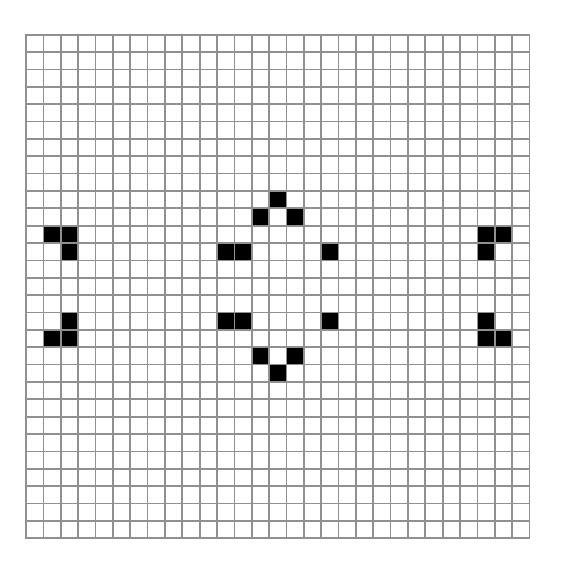}\\[-1ex]
44 asym  \end{minipage}
\begin{minipage}[b]{.24\linewidth}
\includegraphics[width=1\linewidth,bb=2 8 148 155,clip=]{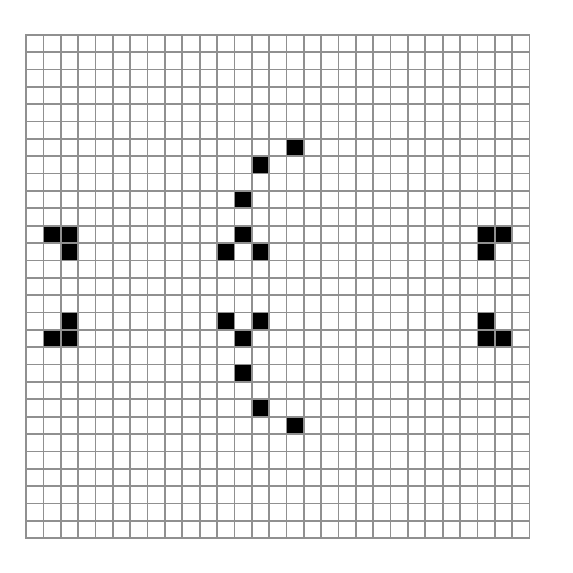}\\[-1ex]
55 asym \end{minipage}\\[2ex]
%------------------------------------------------
\begin{minipage}[b]{.24\linewidth}
\includegraphics[width=1\linewidth,bb=2 8 148 155,clip=]{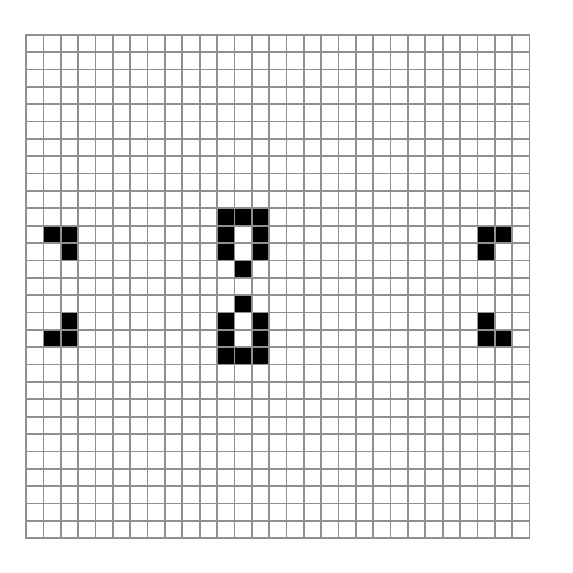}\\[-1ex]
56 dsym \end{minipage}
\begin{minipage}[b]{.24\linewidth}
\includegraphics[width=1\linewidth,bb=2 8 148 155,clip=]{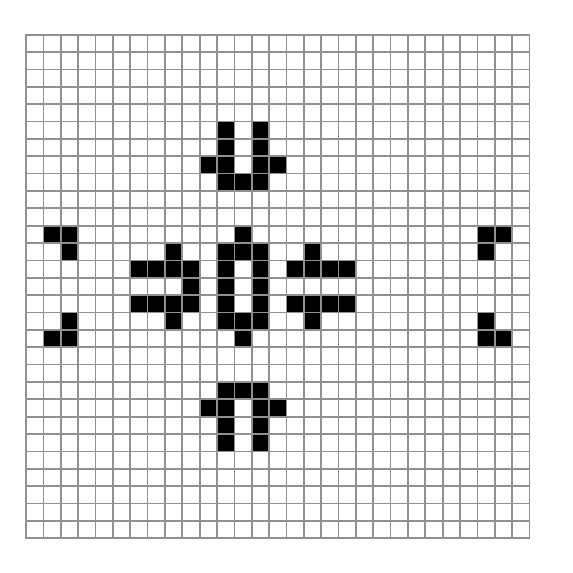}\\[-1ex]
68 dsym glider \end{minipage}
\begin{minipage}[b]{.24\linewidth}
\includegraphics[width=1\linewidth,bb=2 8 148 155,clip=]{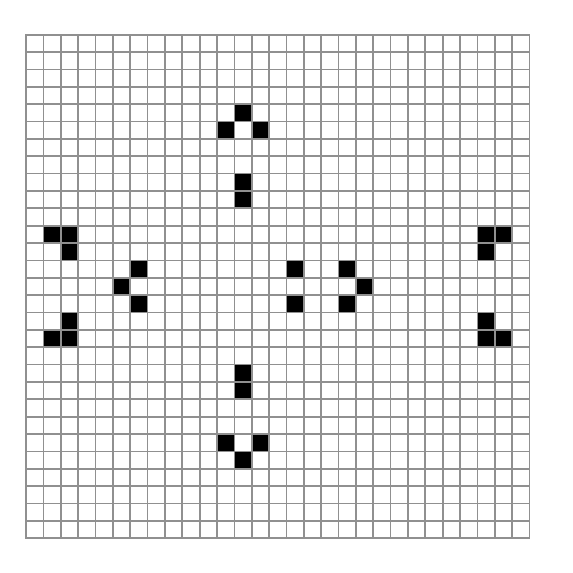}\\[-1ex]
69 asym glider \end{minipage}
\begin{minipage}[b]{.24\linewidth}
\includegraphics[width=1\linewidth,bb=2 8 148 155,clip=]{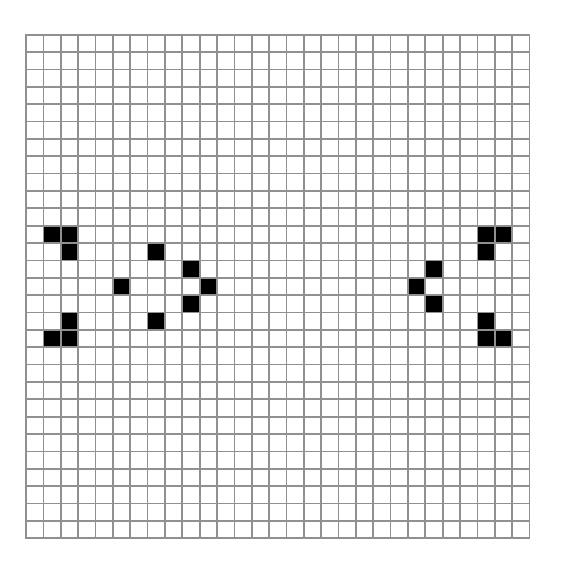}\\[-1ex]
84 asym \end{minipage}\\[2ex]
%------------------------------------------------
\begin{minipage}[b]{.24\linewidth}
\includegraphics[width=1\linewidth,bb=2 8 148 155,clip=]{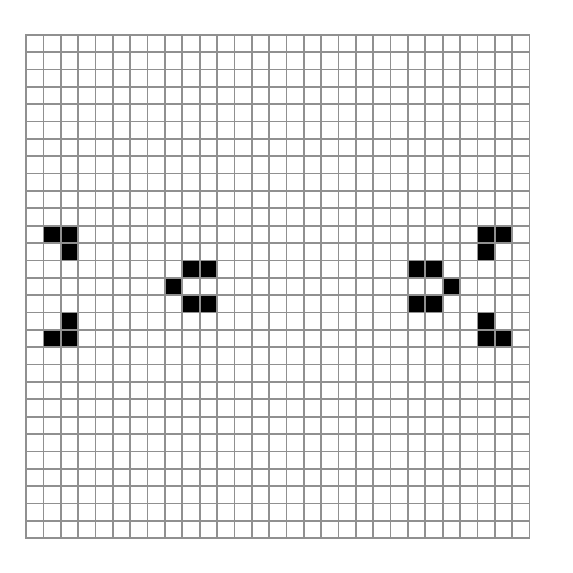}\\[-1ex]
85 dsym \end{minipage}
\begin{minipage}[b]{.24\linewidth}
\includegraphics[width=1\linewidth,bb=2 8 148 155,clip=]{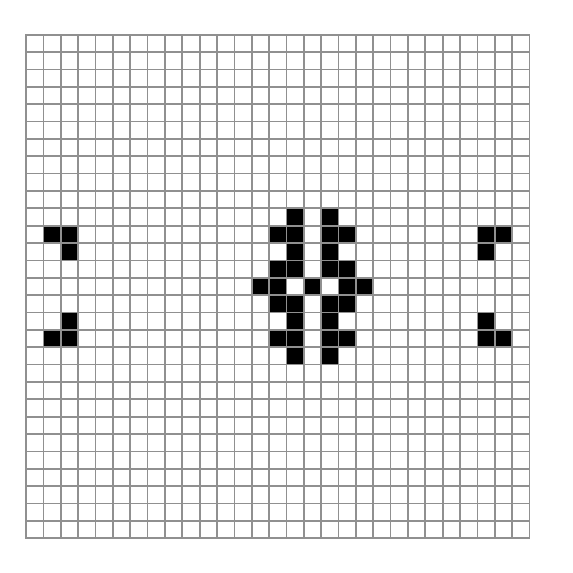}\\[-1ex]
98 dsym glider \end{minipage}
\begin{minipage}[b]{.24\linewidth}
\includegraphics[width=1\linewidth,bb=2 8 148 155,clip=]{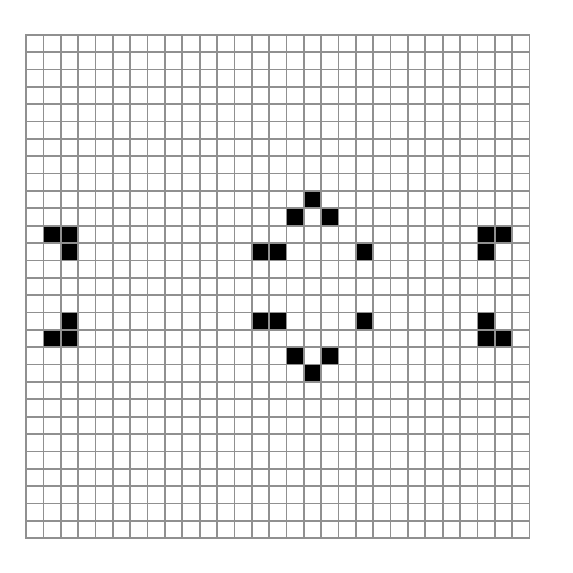}\\[-1ex]
99 asy glider \end{minipage}
\begin{minipage}[b]{.24\linewidth}
\includegraphics[width=1\linewidth,bb=2 8 148 155,clip=]{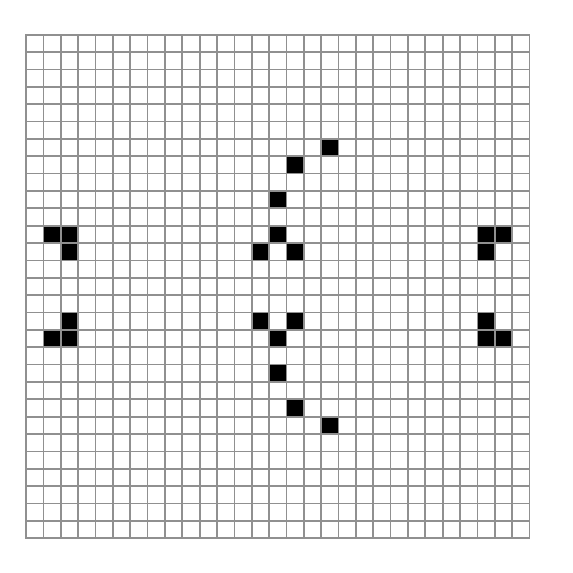}\\[-1ex]
110 asy \end{minipage}\\[-4ex]
%---------------------this grid is 29x29
}
\caption[Basic glider-guns GGb]
{\textsf{The basic orthogonal glider-gun GGb has a minimum reflector gap of 23 cells,
    an oscillator period 110 time-steps, consisting of 2 sub-periods of 55 time-steps. 
    The snapshots (on a 29$\times$29 lattice) show the span of 
    each sub-period 1--55 and 56--110, and spans of
    symmetry (sym), displaced symmetry (dsym) and asymmetry (asym) 
    of the oscillating structures. 
    In sub-period 1, symmetric structures
    and gliders lie centrally relative to the gap, at cell 12.
    Sub-period 2 is a modified copy of sub-period 1, where the axis of symmetric structures
    and gliders is initially displaced to the left, at cell 10 relative to the gap.
    Sub-period 2 has another span of displaced symmetry
    (85-98) where the axis is displaced to the right of center at cell 14 relative to the gap.
    Gliders Ga detach completely at time-steps 14 and 69 moving North and South,
    but they become apparent about 3 time-steps earlier.
    \begin{minipage}[t]{.87\linewidth} 
    Note that in the first sym/dsym span of phase 1 and 2 there is
    an isolated time-step, 9 and 64, with the same structure (right)
    which has a minor asymmetry.\end{minipage}\hspace{1.5ex}
    \begin{minipage}[t]{.09\linewidth} \color{white}----\color{black}\\[-2.5ex]
    \includegraphics[width=1\linewidth,bb=0 0 60 86,clip=]{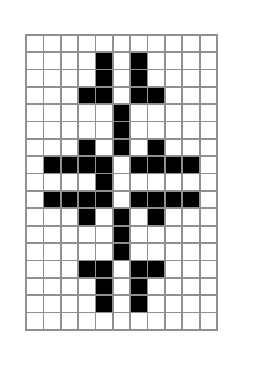}
    \end{minipage}
}}
\label{glider-gun GGb}
\end{figure}

\begin{figure}
\textsf{
\begin{minipage}[b]{1\linewidth}
\vspace{-3ex}
\hspace{-8ex}\includegraphics[width=1\linewidth,bb=320 177 630 478,clip=]{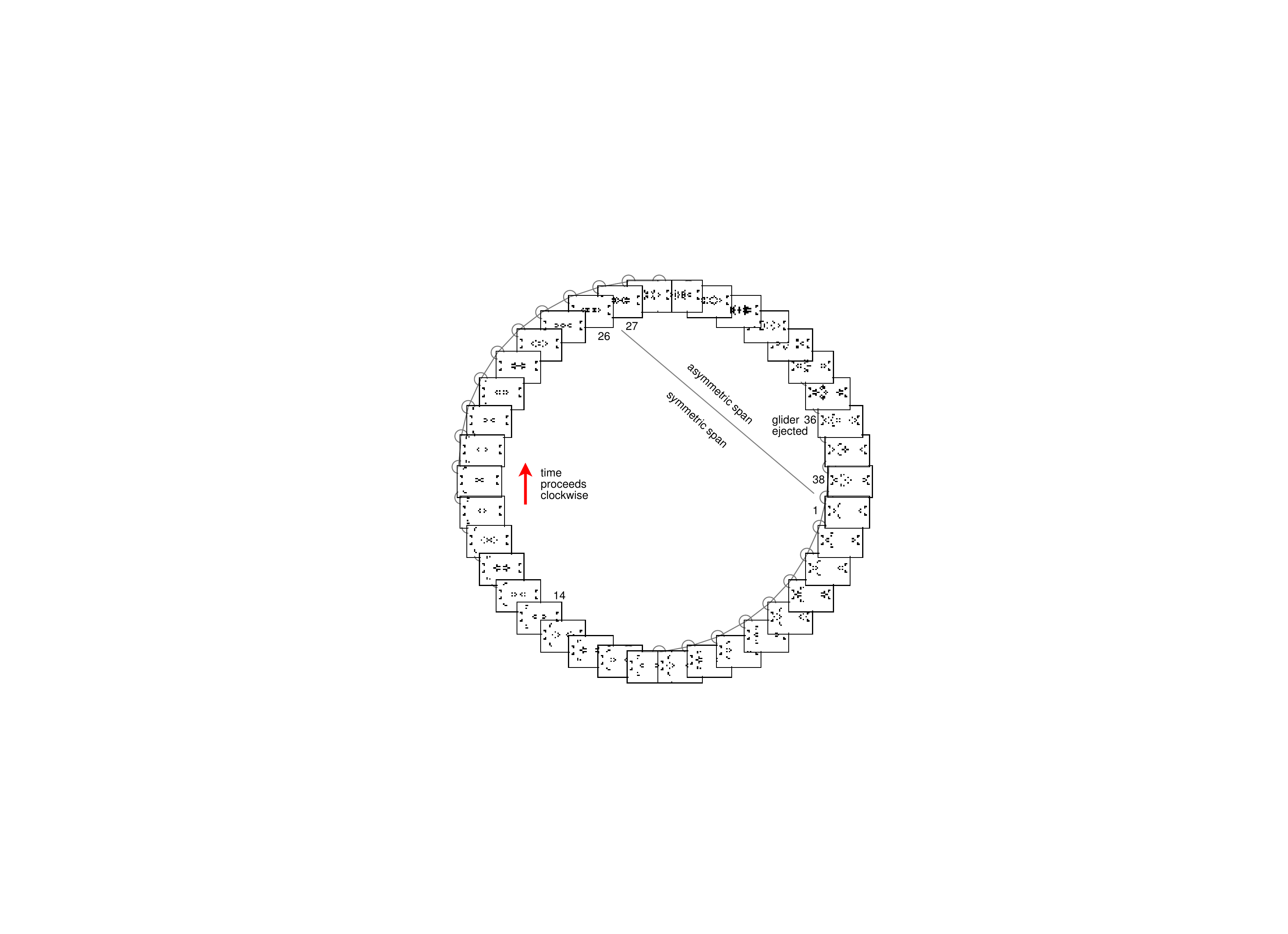}\\
\begin{minipage}[b]{.55\linewidth}
{\it (above)} GGa as in figure~\ref{glider-gun GGa}
showing all 38 attractor states -- time-step 38 is due East. Symmetric/asymmetric spans
are indicated.\\ Gliders eject at time-step 36.
\end{minipage}
\vspace{-.9\linewidth}
\end{minipage}
\begin{minipage}[b]{1\linewidth}
\includegraphics[width=1.1\linewidth,bb=142 57 546 538,clip=]{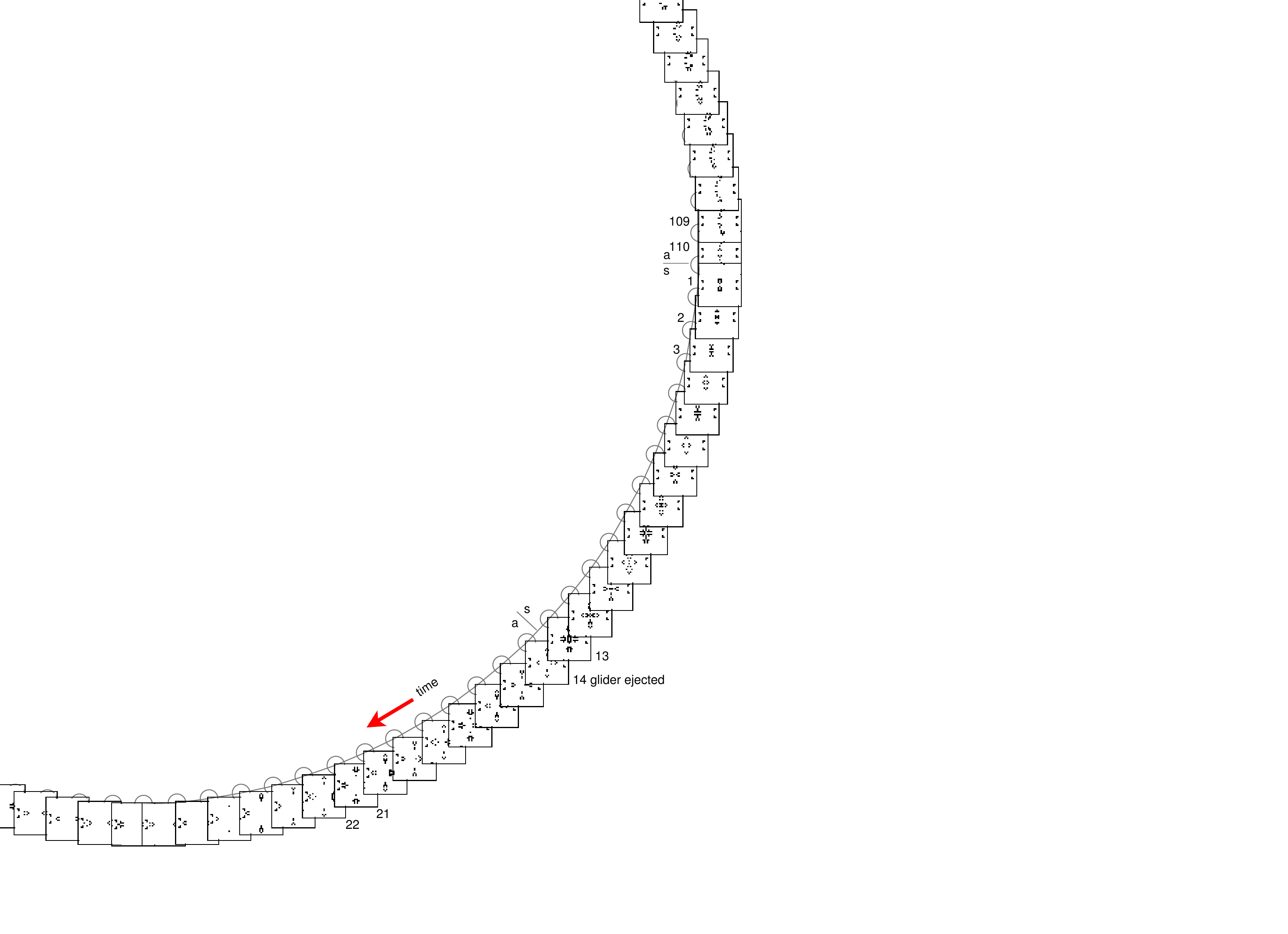}\\[-12ex]
\begin{minipage}[b]{1\linewidth}
\begin{minipage}[b]{.40\linewidth}
%xxx
\end{minipage}
\hfill
\begin{minipage}[b]{.50\linewidth}
{\it (above)} GGb as in figure~\ref{glider-gun GGb}
showing part of the attractor including time-steps 110, 1--21, and 22. Time-step 110 is due East.
Symmetric/asymmetric spans are indicated. Gliders eject at time-step 14.
\end{minipage}
\end{minipage}
\end{minipage}
}
\vspace{-3ex}
\caption[GGa and GGb attractors]
{\textsf{Glider-guns can be seen as periodic attractors on a finite toroidal lattice where
colliding gliders self-destruct.  {\it (top left)} glider-gun GGa complete attractor.
{\it (bottom right)} glider-gun GGb partial attractor. The direction of time is clockwise.
}}
\label{gga_att.pdf}
\end{figure}

\noindent
Snapshots of the X-rule's two types of glider-gun, GGa and GGb, constructed in 
section~\ref{Creating glider-guns -- the X-rule}, 
are shown in figure~\ref{glider-guns GGa and GGb} 
with greater detail in figures~\ref{glider-gun GGa} 
and \ref{glider-gun GGb}, and in the context
of attractors\cite{Wuensche92} in figure~\ref{gga_att.pdf}.
A summary of glider-gun properties are as follows, where $c$ 
is the speed of light.

Glider-gun GGa shoots 
Ga~\raisebox{-1.5ex}{\includegraphics[height=2ex,bb=0 -1 19 12 clip=]{mga1}}
\hspace{.5ex} gliders
diagonally SW and NW with speed=$c$/4.  The GGa minimum gap between
reflectors (as shown) is 24 cells with an oscillation and glider
ejection period of 38 time-steps, with a pair of gliders shot for each period.

Glider-gun GGb shoots 
%Gb~\raisebox{-1.2ex}{\epsfig{file=miniglider_b1.pdf, height=4ex,bb=0 -1 20 32, clip=}}
Gb~\raisebox{-1.5ex}{\epsfig{file=miniglider_b1.pdf, height=5ex,bb=0 -1 23 45, clip=}}\hspace{.5ex} 
gliders orthogonally South and North with speed=$c$/2.  The GGb minimum gap
between reflectors (as shown) is 23 cells and its oscillation period is 110
time-steps, but this is divided into two sub-periods -- pairs of
gliders are shot every 55 time-steps, alternating between the center
and offset by 1 cell left of center.  

Both glider-guns GGa and GGb are periodic machines developed from the
periodic oscillator of the X-rule precursor in figure~\ref{pcs}. Of
the X-rule's 512 outputs just 12 are non-isotropic but this results in
spans of continuous time-steps were the periodic oscillator is either
symmetric or asymmetric.  Glider-gun GGa starts to eject Ga gliders
towards the end of its asymmetric span, whereas glider-gun GGb starts
to eject Gb gliders at the transition from symmetry to asymmetry.
Both glider-guns can be seen as attractors on a finite periodic
(toroidal) lattice where colliding gliders self-destruct
(figure~\ref{gga_att.pdf}). Both continue to work for larger gaps
between reflectors, in steps of +4, giving
correspondingly larger periods described in 
section~\ref{Variable gap between Glider-gun reflectors}.

Although Ga gliders can move in any diagonal direction, the
glider direction from GGa is restricted to SW and NW. However, by
combining glider-guns and creating periodic collisions, any direction
can be achieved by the resulting 
compound glider-guns (section~\ref{Compound glider-guns}).

Note that the glider-guns in the game-of-Life 
(figure~\ref{Other  glider-guns}a) and in the X-rule are both artificially
constructed -- their spontaneous emergence would be highly improbable.
In contrast glider-guns in Sapin's
rule\cite{Sapin2004}(figure~\ref{Other glider-guns}c), and
the 3-value spiral rule\cite{Wuensche2006}(figure~\ref{Other glider-guns}d) 
emerge spontaneously.
\enlargethispage{4ex}

\clearpage

\clearpage

\clearpage

%----------------------------------------------------------------------
\subsection{Variable gap between Glider-gun reflectors}
\label{Variable gap between Glider-gun reflectors}

\begin{figure}[h]
\begin{minipage}[b]{1\linewidth}
\textsf{\small
\fbox{\includegraphics[height=.3\linewidth,bb=16 8 310 410,clip=]{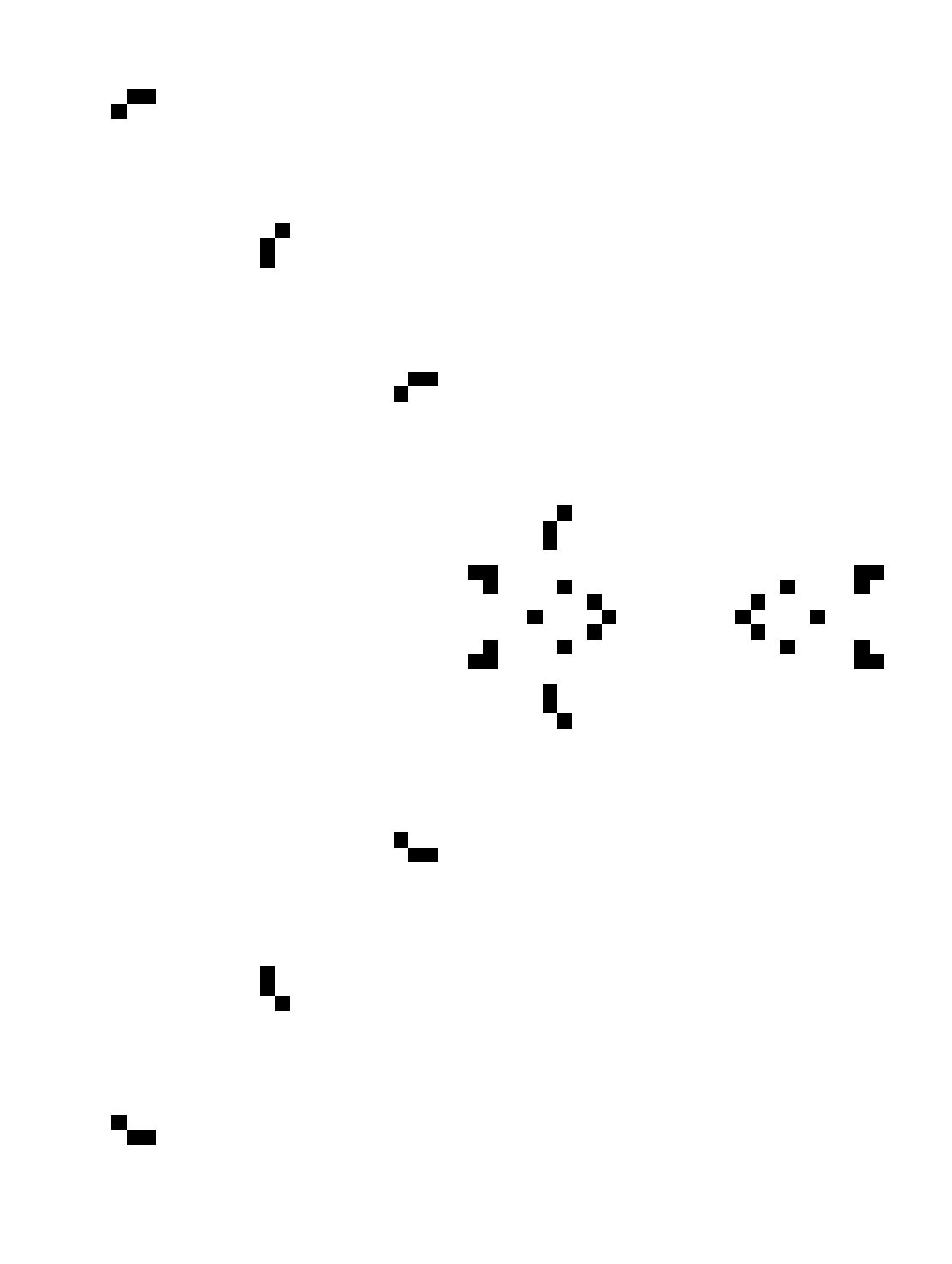}}
\hspace{-4ex}24\hspace{5ex}%64x81
\fbox{\includegraphics[height=.3\linewidth,bb=25 8 350 410,clip=]{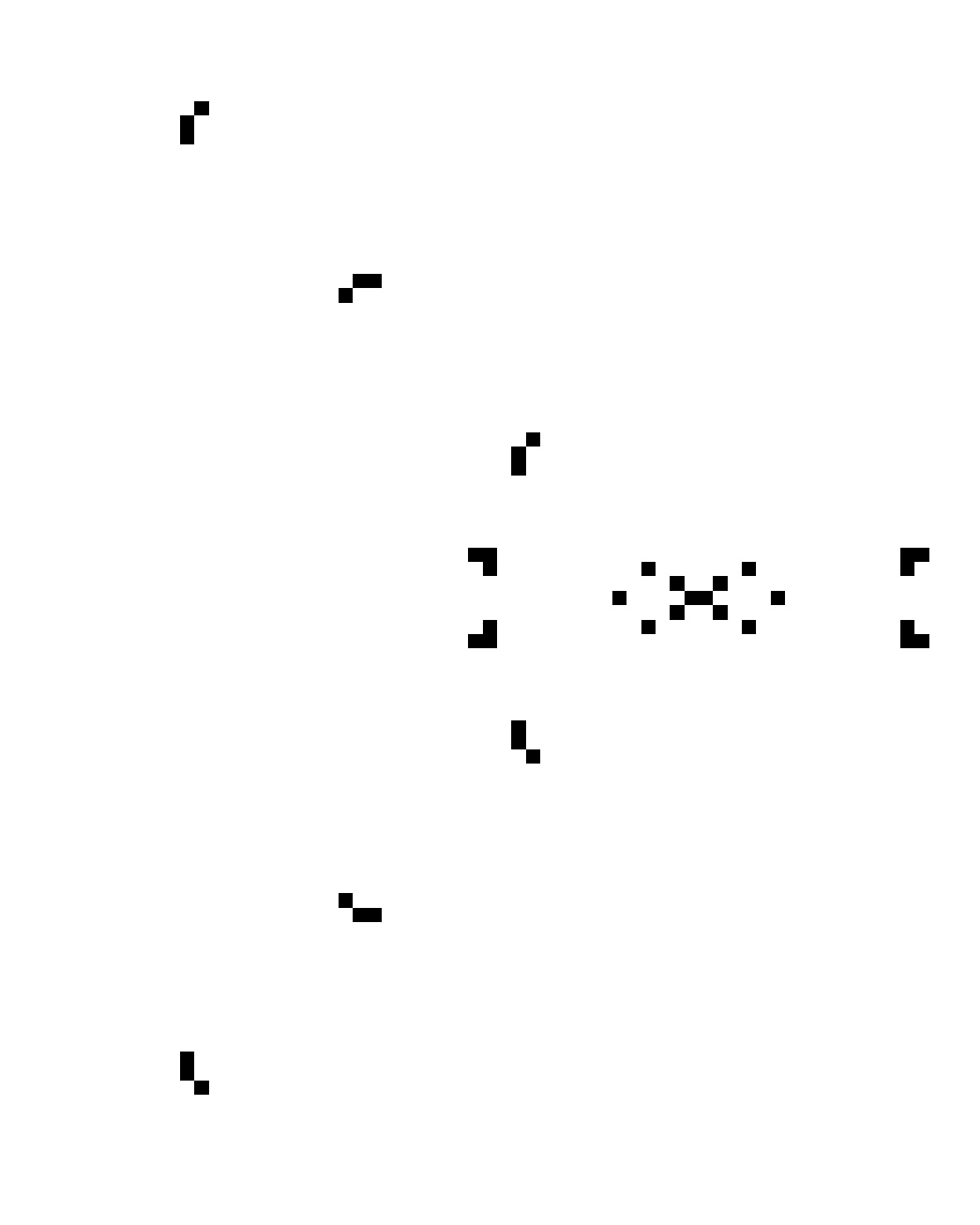}}
\hspace{-4ex}28\hspace{5ex}%60x81 
\fbox{\includegraphics[height=.3\linewidth,bb=30 8 355 410,clip=]{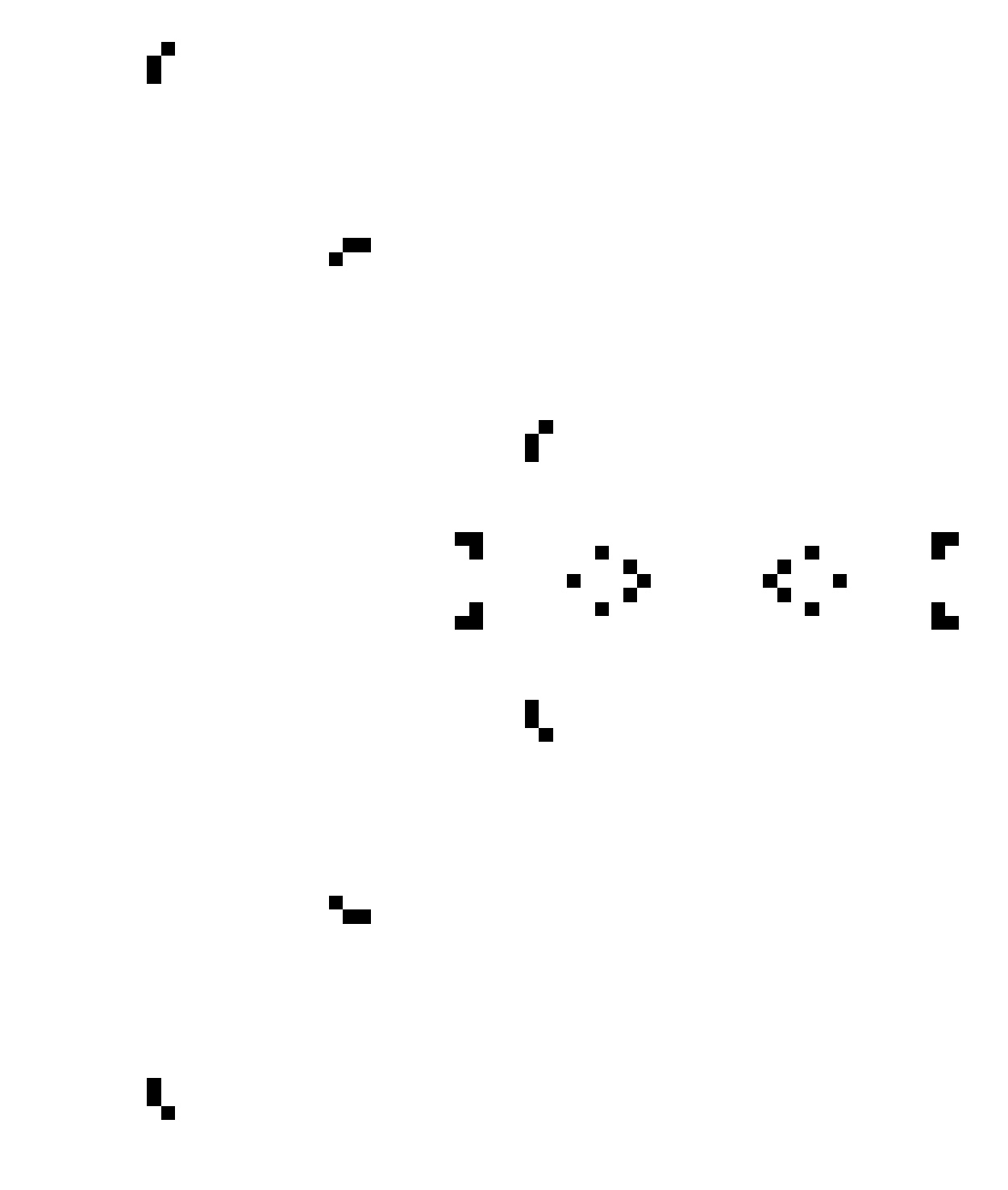}}
\hspace{-4ex}32\\[4ex]%64x81
\fbox{\includegraphics[height=.55\linewidth]{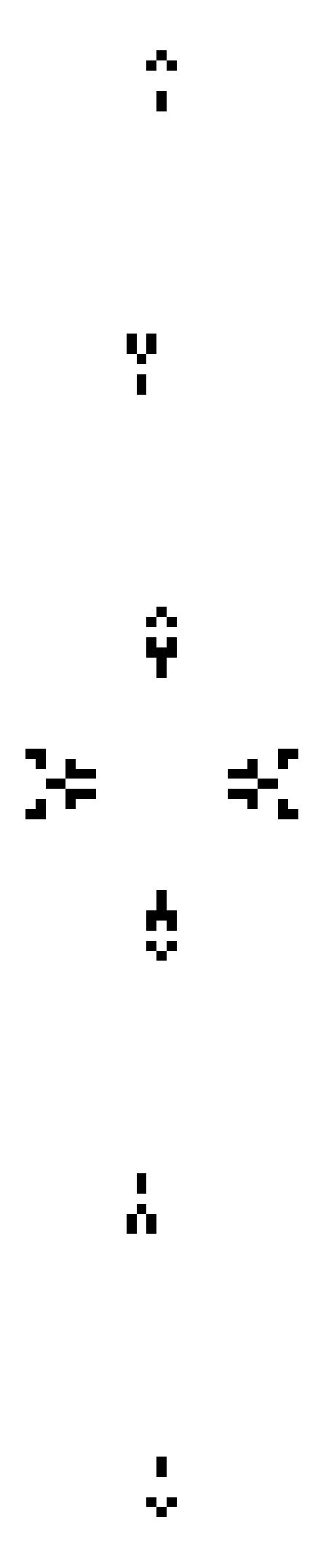}}
\hspace{-4ex}23\hspace{5ex}%29x151
\fbox{\includegraphics[height=.55\linewidth]{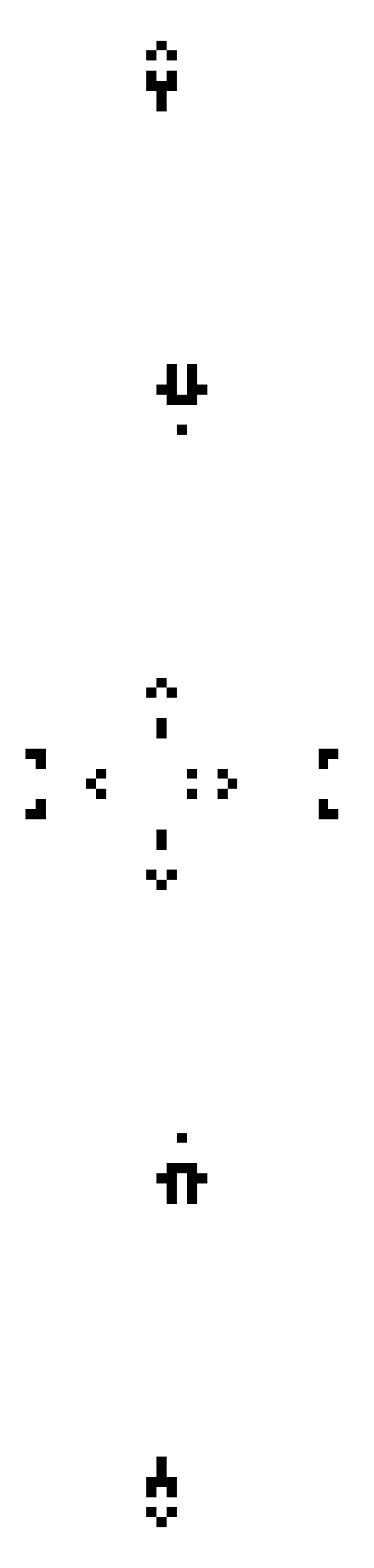}}
\hspace{-4ex}27\hspace{5ex}%33x151
\fbox{\includegraphics[height=.55\linewidth]{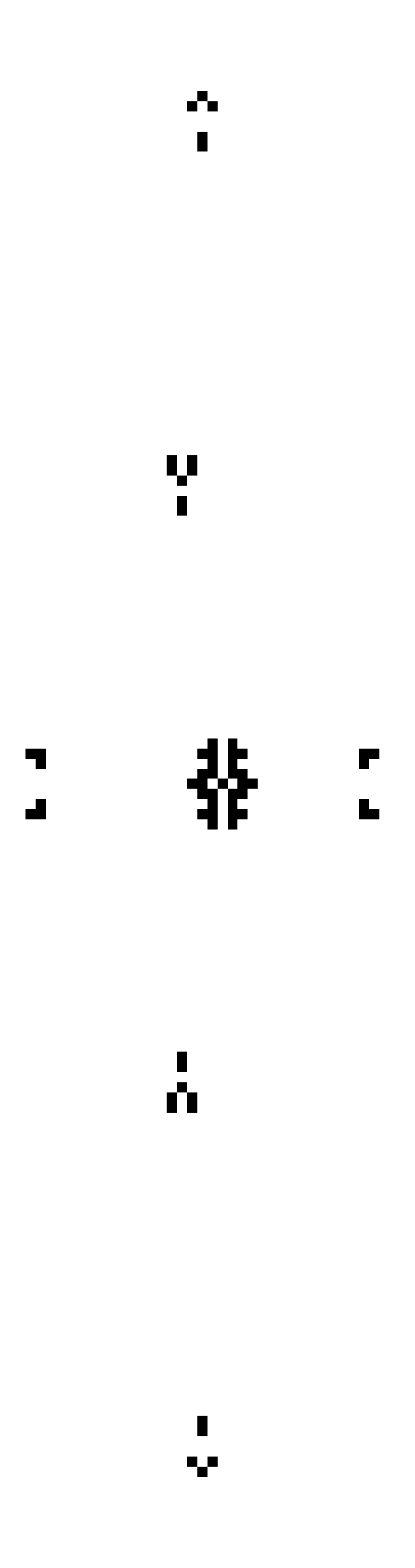}}
\hspace{-4ex}31%37x151
}
\hfill \parbox[t]{.38\linewidth}{ \vspace{-37ex}
\caption[GGa GGB with increasing gaps]
{\textsf{Example snapshots of glider-guns with increasing gaps between reflectors 
and thus lower glider frequency.\\ 
{\it (above)}\\ GGa with gaps 24, 28, 32.\\
{\it (left)}\\ GGb with gaps 23, 27, 31.}}}
\end{minipage}
%\end{center} 
\label{GGa GGB with increasing gaps}
\end{figure}

X-rule glider-guns have a special property in that the gap between reflectors
can be enlarged from the minimum -- 24 for GGa and 23 for GGb. Only increments
of +4 are valid in each case to preserve the glider-gun,
which  increases the oscillation period and thus reduces the frequency of
the resulting glider-stream. From experiment a regular pattern of gaps and periods emerge
that extend indefinitely -- listed in table~\ref{GGa gaps-periods}, and example are shown
figure~\ref{GGa GGB with increasing gaps}.

\vspace{-2ex}
\begin{table}[h]
\begin{center}
\begin{tabular}[b]{r|r|r|r|r|r|r|r|r||r}
\multicolumn{1}{r}{} & \multicolumn{9}{c}{glider-gun GGa}   +step\\\cline{2-10}
\hline
    reflector gap      &  24 & 28 & 32 & 36 & 40 & 44 &  ...&   64 &  +4\\\hline
    oscillator period  &  38 & 46 & 54 & 62 & 70 & 78 &  ...&  118 &  +8\\\hline    
\end{tabular}\\
\end{center}
\begin{center}
\begin{tabular}[b]{r|r|r|r|r|r|r|r|r||r}
\multicolumn{1}{r}{} & \multicolumn{9}{c}{glider-gun GGb}   +step\\\cline{2-10}
\hline
    reflector gap      &  23 &  27 &  31 &  35 &  39 &  43 & ...&  63 & +4\\\hline
    oscillator period  & 110 & 126 & 142 & 158 & 174 & 190 & ...& 270 & +16\\\hline    
\end{tabular}
\end{center}
\vspace{-4ex}
\caption[GGa and GGb reflector gaps and oscillator periods.]
{\textsf{GGa and GGB reflector gaps and oscillator periods.}}
\label{GGa gaps-periods}
\vspace{-1ex}
\end{table}

\enlargethispage{6ex}

\enlargethispage{4ex}
\clearpage

%^^^^^^^^^^^^^^^^^^^^^^^^^^^^^^^^^^^^^^^^^^^^^^^^^^^^^^^^^^^^^^^^^^^^^^^^^^^^
\section{Compound glider-guns}
\label{Compound glider-guns}

In the X-rule, two types of ejected glider-streams have not been 
achieved with a single glider-gun;
firstly, gliders
Gc~\raisebox{-.9ex}{\epsfig{file=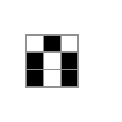, height=3.2ex,bb=2 4 23 26, clip=}}\hspace{.5ex}
in any orthogonal heading, although they are readily emergent, and
secondly, gliders 
Ga heading 
NE~\raisebox{-.9ex}{\epsfig{file=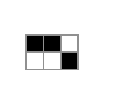, width=3.2ex,bb=4 3 25 22, clip=}}
and SE~\raisebox{-.9ex}{\epsfig{file=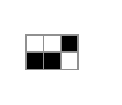, width=3.2ex,bb=4 3 25 22, clip=}},
although Ga's are readily emergent in all diagonal directions and the 
glider-gun GGa sends Ga's NW and SW.

To enhance the diversity of X-rule dynamics,
these glider-streams can be created with compound glider-guns (CGGx)
constructed from two or more basic glider-guns and eaters/reflectors,
positioned and synchronised precisely, making self-contained and sustainable
multiple oscillating colliding compound structures.
In the following examples, which are not necessarily unique solutions or the simplest,
we demonstrate CGGc shooting Gc gilders South (and North), 
where CGGcS is 
also a component within CGGaNE shooting Ga gilders NE (and SE), 
then we combine CGGcS with another GGa glider-gun to make CGGcW 
shooting Gc gilders West (and East).
  
%----------------------------------------------------------------------
\subsection{Compound glider-guns CGGcS and CGGaNE}
\label{Compound glider-guns CGGcS and CGGaNE}

\begin{figure}[htb]%93x85, file= CggaNE.eed
\includegraphics[width=.97\linewidth]{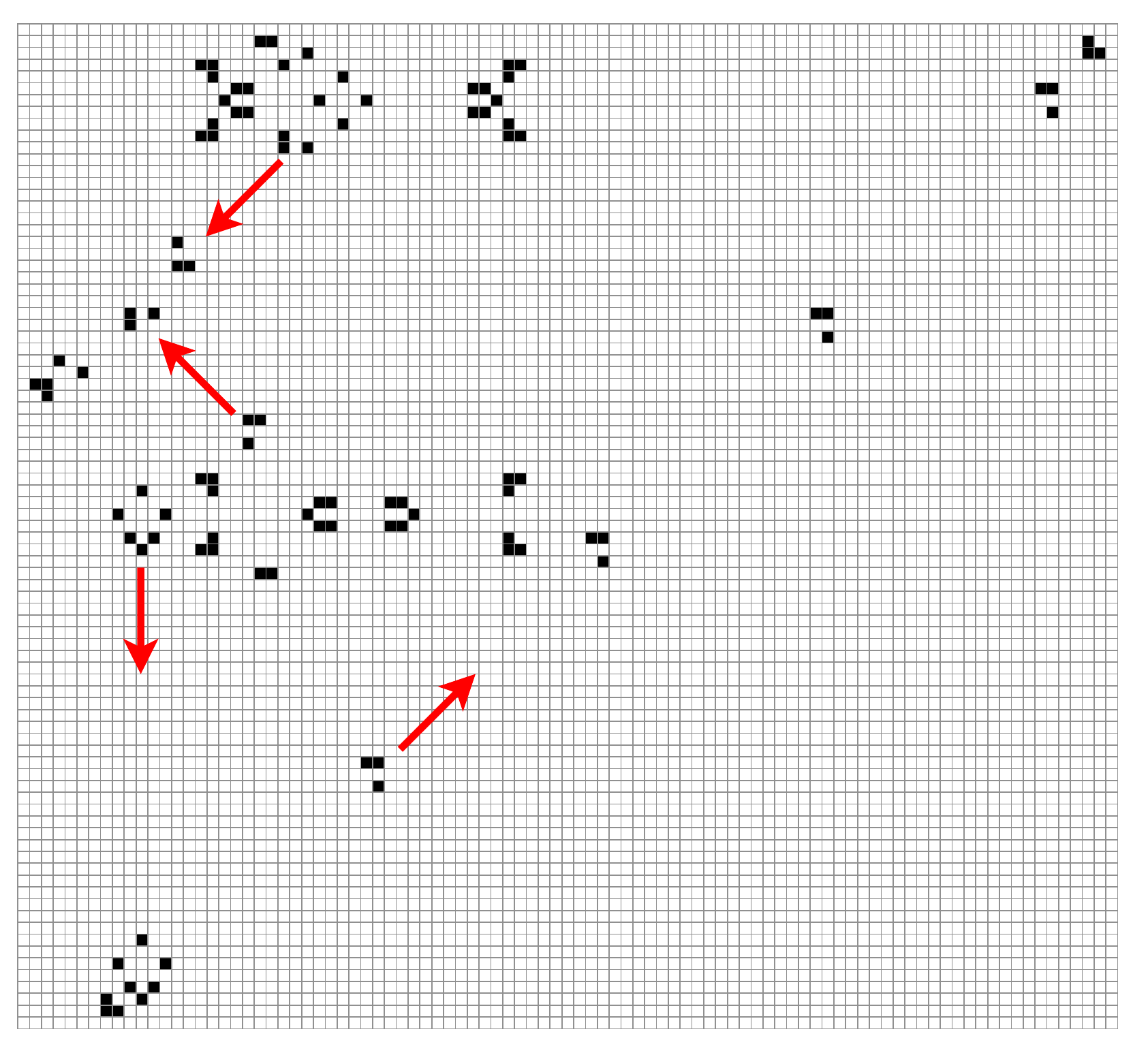}\\[-37ex]%a for arrow
\begin{flushright}
\includegraphics[width=.47\linewidth]{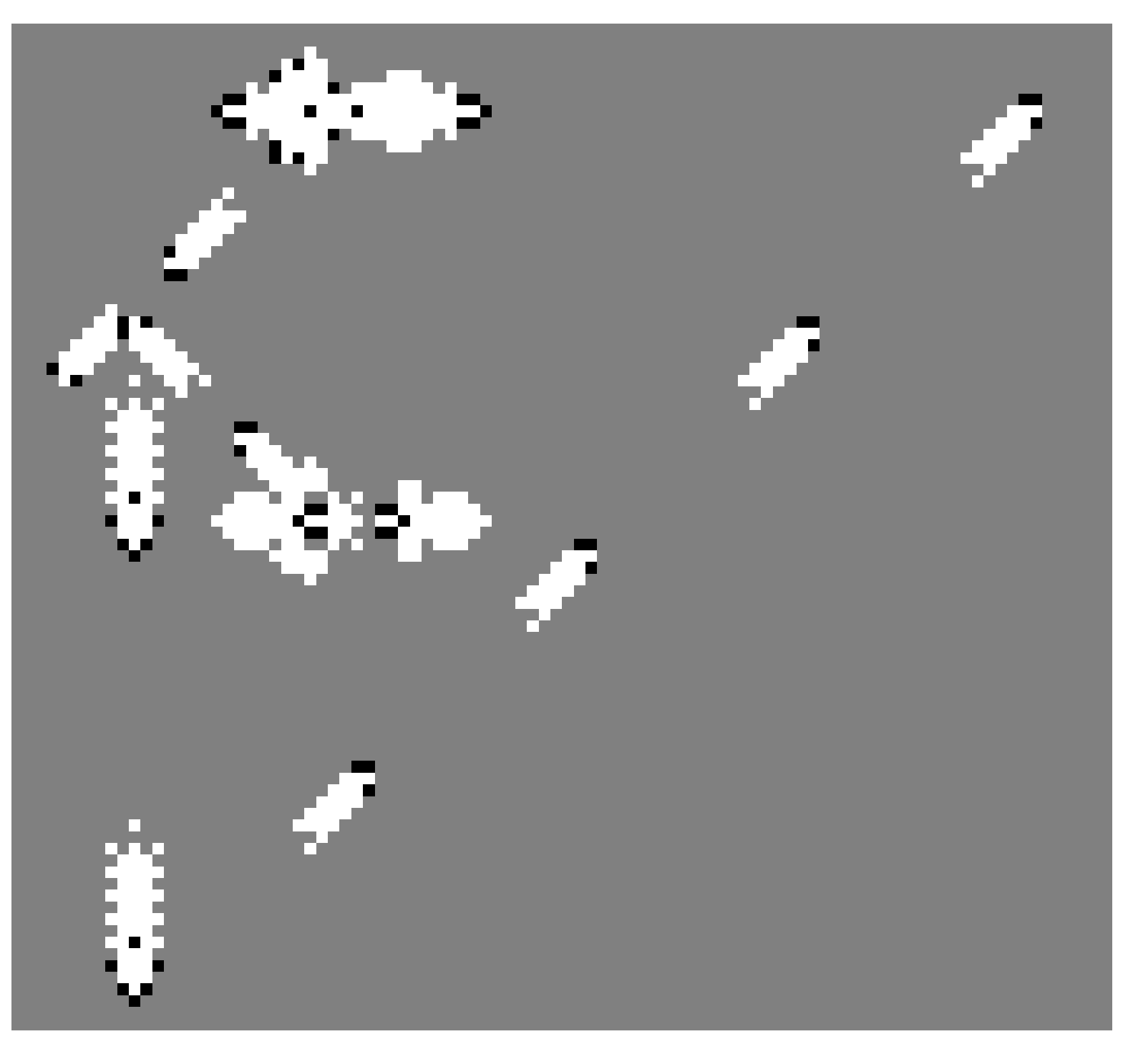}\\%
\end{flushright}
\vspace{-3ex}
\caption[CGGcS+CGGaNE]
{\textsf{Snapshot (93$\times$85) 
of the compound glider-gun CGGcS shooting Gc gliders towards the South, which
bounce of eater Ea to send Ga gliders NE, the combined system is the compound 
glider-gun CGGaNE. A vertical flip of this pattern gives the equivalent CGGaSE and CGGcN.
The inset ({\it bottom right}) shows the same snapshot with gliders leaving
white trails whose length is proportional to speed -- 
the grey background represents cells that have not changed
for the last 20 time-steps.
}}
\label{CGGcS+CGGaNE}
\end{figure}

To shoot Ga gliders NE,
we combined GGa glider-guns and employed various collision/reflection 
properties to create a compound glider-gun CGGaNE. In the process we
created a compound glider-gun CGGcS shooting
Gc~\raisebox{-.9ex}{\epsfig{file=Gc2S.pdf, height=3.2ex,bb=2 4 23 26, clip=}}\hspace{.5ex}
gliders South.  Ga gliders moving NE are generated by
bouncing glider GcS off eater 
Ea4~\raisebox{-1.3ex}{\epsfig{file=Ea4.pdf, width=3.2ex,bb=4 3 20 22, clip=}}
which survives the collision.
Figure~\ref{CGGcS+CGGaNE} shows a space-time pattern snapshot of CGGaNE which contains
CGGcS. 
A vertical flip of this pattern gives the equivalent CGGaSE and CGGcN.
   
To send a stream of Gc gliders South, two GGa glider-guns are lined up below each other
separated by a minimum of 28 vertical cells.
This compound glider-gun CGGcS will work if 
the initial state of the lower (GGa-14) 
is offset by +14 time-steps from the upper (GGa-0).
In the attractor cycle presentation in figure~\ref{gga_att.pdf},
the snapshot has GGa-0 at time-step 38 (the last in the cycle)
and GGa-14 at time-step 14. 
GGa-0 shoots gliders
Ga~\raisebox{-.9ex}{\epsfig{file=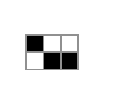, width=3.2ex,bb=4 3 25 22, clip=}}
SW and GGa-14 shoots gliders 
Ga~\raisebox{-.9ex}{\epsfig{file=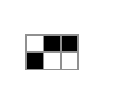, width=3.2ex,bb=4 3 25 22, clip=}}
NW, resulting in the a
Gc~\raisebox{-.9ex}{\epsfig{file=Gc2S.pdf, height=3.2ex,bb=2 4 23 26, clip=}}\hspace{.5ex}
glider heading South every alternate collision. 
One collision leaves an eater 
Ea5~\raisebox{-1.3ex}{\epsfig{file=Ea5.pdf, width=3.2ex,bb=4 3 20 22, clip=}}
and both Ga gliders are destroyed.
The next collision between GaNW and Ea5 produces a
Gc~\raisebox{-.9ex}{\epsfig{file=Gc2S.pdf, height=3.2ex,bb=2 4 23 26, clip=}}\hspace{.5ex}
glider heading South, Ea5 is destroyed,  GaSW continues and is destroyed by
an eater, as are other extraneous gliders

A continuous stream of 
GcS~~\raisebox{-.9ex}{\epsfig{file=Gc2S.pdf, height=3.2ex,bb=2 4 23 26, clip=}}\hspace{.5ex}
gliders heads South at 38 cell intervals.
Each GcS glider collides with
a appropriately positioned 
Ea4~\raisebox{-1.3ex}{\epsfig{file=Ea4.pdf, width=3.2ex,bb=4 3 20 22, clip=}}
eater (which survives), and the Gc glider is transformed (reflects) to become a 
Ga~\raisebox{-.9ex}{\epsfig{file=Ga1NE.pdf, width=3.2ex,bb=4 3 25 22, clip=}}
glider heading NE at 19 cell intervals, the closer intervals result from the
difference in speed, Gc speed=$c$/2, Ga speed=$c$/4.

%----------------------------------------------------------------------
\subsection{Compound glider-gun CGGcW}
\label{Compound glider-gun CGGcW}

To shoot Gc gliders West we combined CGGcS from 
section~\ref{Compound glider-guns CGGcS and CGGaNE} 
shooting
Gc~\raisebox{-.9ex}{\epsfig{file=Gc2S.pdf, height=3.2ex,bb=2 4 23 26, clip=}}\hspace{.5ex}
gliders South with a GGaNE basic glider-gun shooting   
Ga~\raisebox{-.9ex}{\epsfig{file=Ga1NW.pdf, width=3.2ex,bb=4 3 25 22, clip=}}
NW. Each alternate collosion produces a Gc 
glider heading West and leaves behind an
Ea5~\raisebox{-1.3ex}{\epsfig{file=Ea5.pdf, width=3.2ex,bb=4 3 20 22, clip=}}
eater. The next 
GaNW~\raisebox{-.9ex}{\epsfig{file=Ga1NW.pdf, width=3.2ex,bb=4 3 25 22, clip=}}
destroys Ea5 but reflects to become another
Gc~\raisebox{-.9ex}{\epsfig{file=Gc2S.pdf, height=3.2ex,bb=2 4 23 26, clip=}}\hspace{.5ex}
gliders heading South which is destroyed by a stable
Ea3~\raisebox{-1.3ex}{\epsfig{file=Ea3.pdf, width=3.2ex,bb=4 3 20 22, clip=}}
eater.
The result is compound glider-gun CGGcW made from three out of phase GGa's,
shooting GcW West at 38 cell intervals.

\begin{figure}[H]%%93x85, file= CGGcW-st.eed
\begin{flushright}
\includegraphics[width=.97\linewidth]{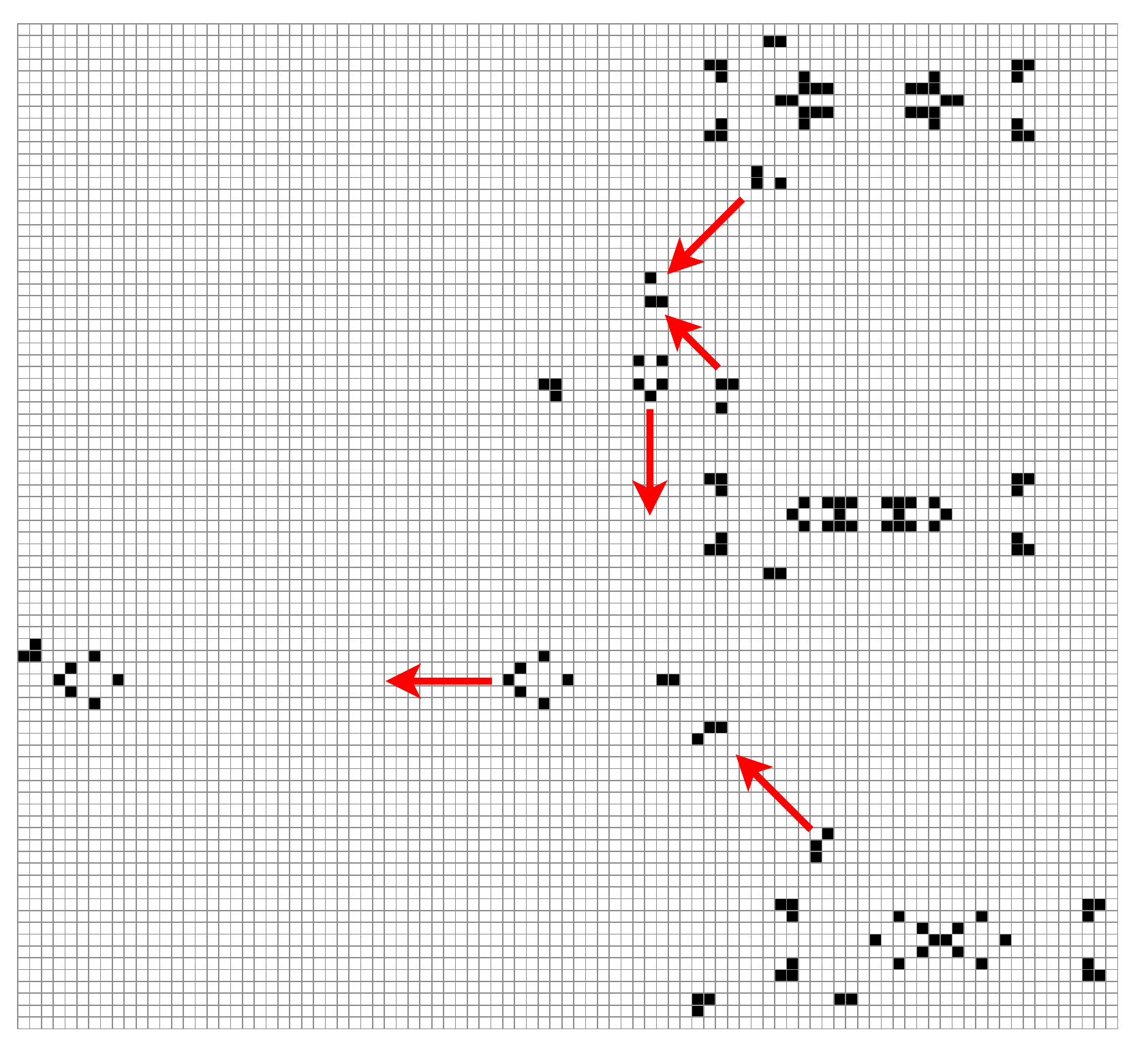}
\vspace{-70ex}
\end{flushright}
\begin{flushleft}
\includegraphics[width=.47\linewidth]{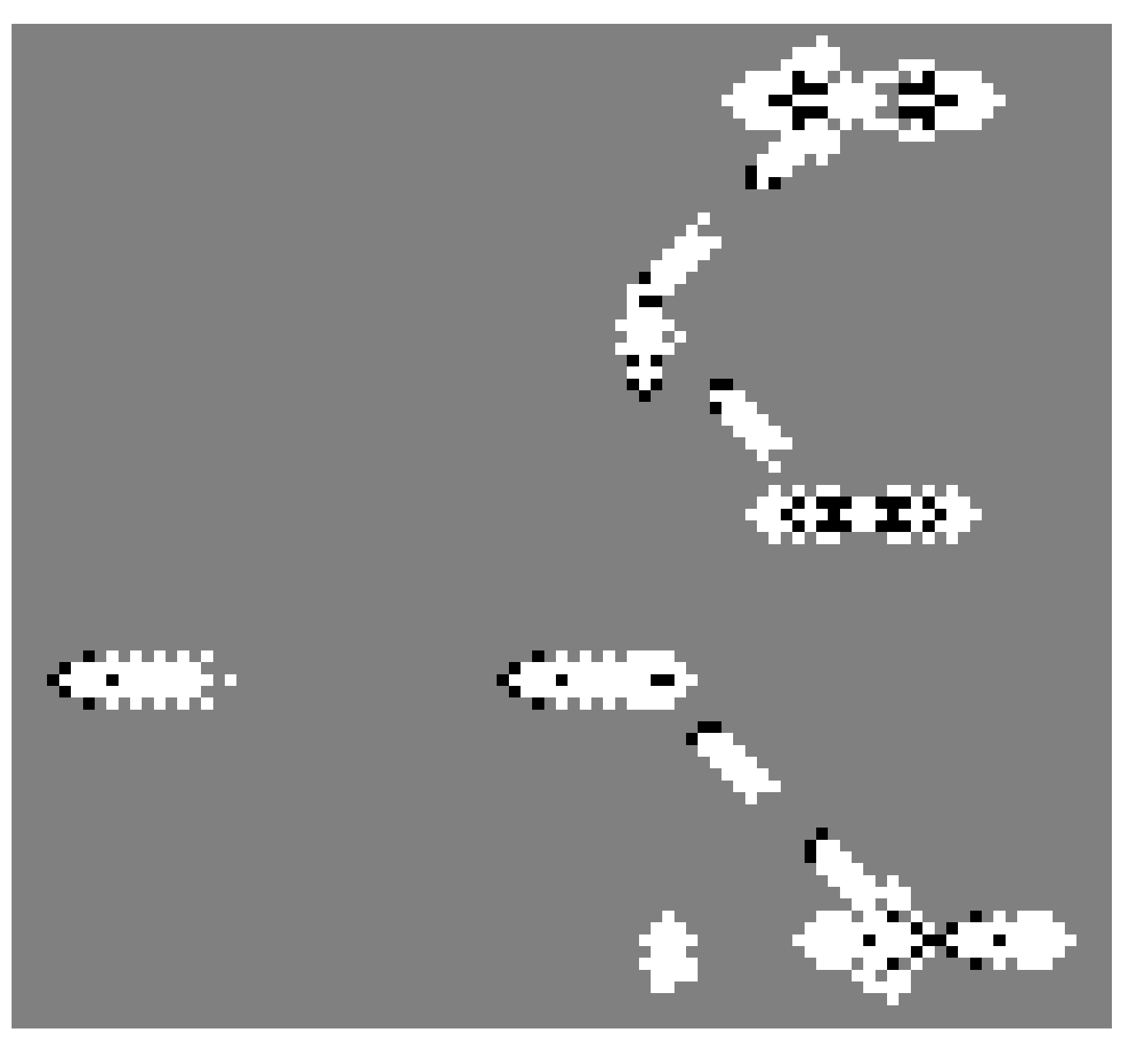}\\
\end{flushleft}
\vspace{29ex}
\caption[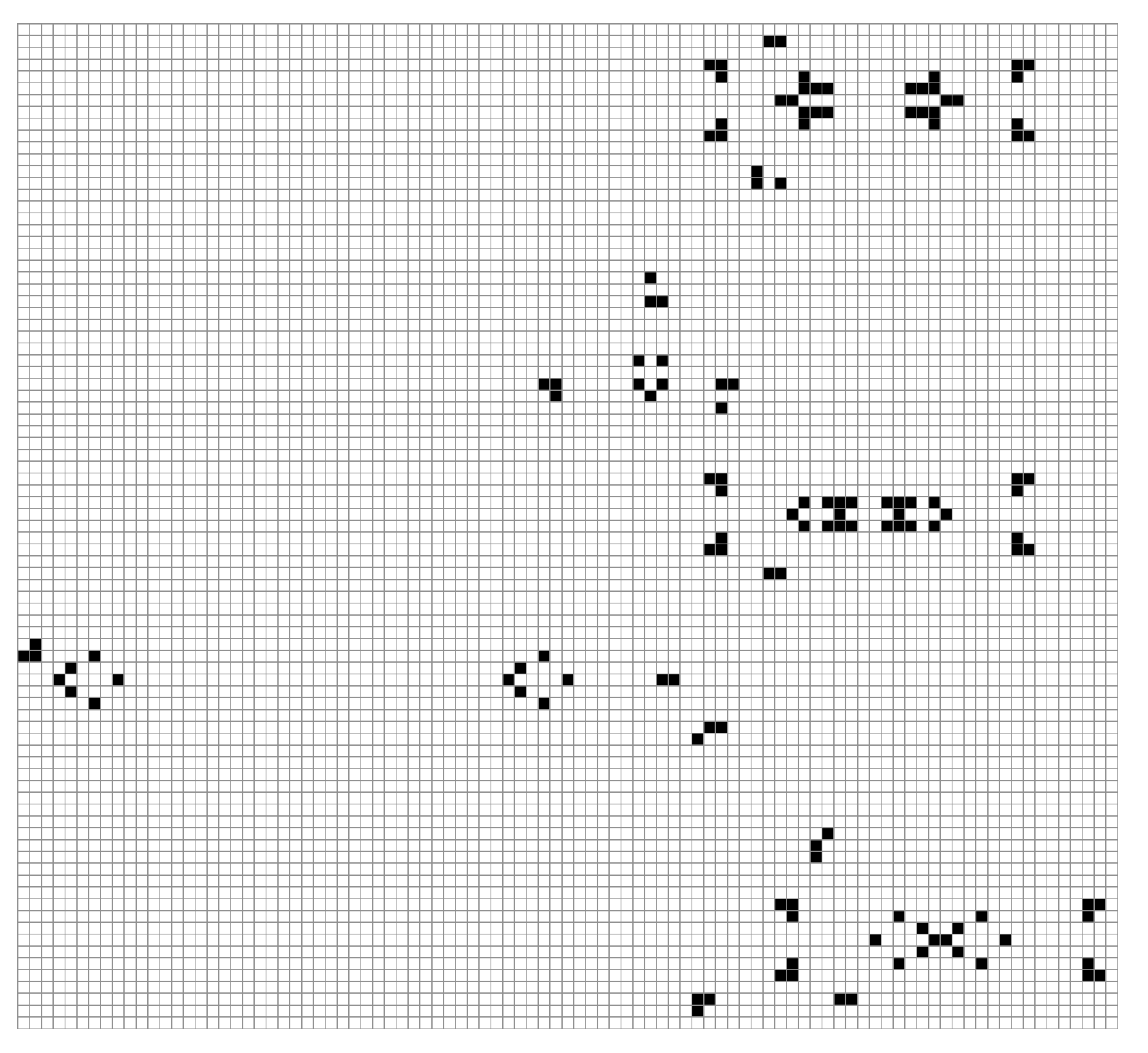]
{\textsf{Snapshot of the compound glider-gun CGGcW shooting Gc gliders towards the West
made from a CGGcS compound glider-gun shooting Gc glider South,
and a GGaNW glider-gun positioned below
shooting Ga gliders NW, so three GGa glider-guns in all.
GaNW gliders collide with alternate GcS gliders to create a GcW glider-stream.
Alternate CGcS gliders that break through are destroyed by an eater.
The inset ({\it top left}) shows the same snapshot with gliders leaving
white trails whose length is proportional to speed -- 
the grey background represents cells that have not changed
for the last 20 time-steps.
}}
\label{CGGcW.pdf}
\end{figure}

Using similar constructions to those in figures~\ref{CGGcS+CGGaNE} and \ref{CGGcW.pdf}, 
figure~\ref{CGGc North, East and West} gives examples
of CGGc shooting Gc gliders North, East, and an alternative towards the West,
demonstrating that Gc glider-streams can be projected
in any orthogonal direction, though there may be different or simpler arrangements
to achieve the same results.

\begin{figure}[H]
\textsf{
%/ggCcSs.pdf 44x84 -- not shown because CGGcS is shown as part of CGGaNE
\begin{minipage}[b]{.37\linewidth}
%44x84
\fbox{\includegraphics[width=1\linewidth]{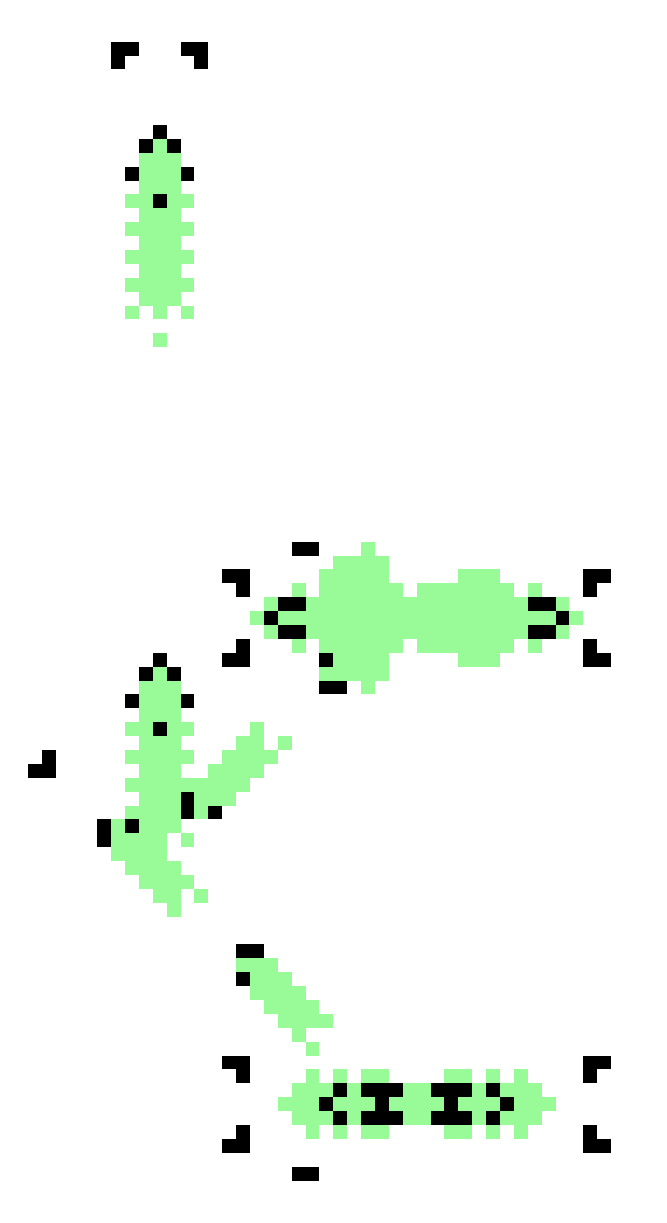}}\\[-4ex]
\begin{center}CGGcc North\end{center}
\end{minipage}
\hfill
\begin{minipage}[b]{.55\linewidth}
%91x115
\fbox{\includegraphics[width=1\linewidth]{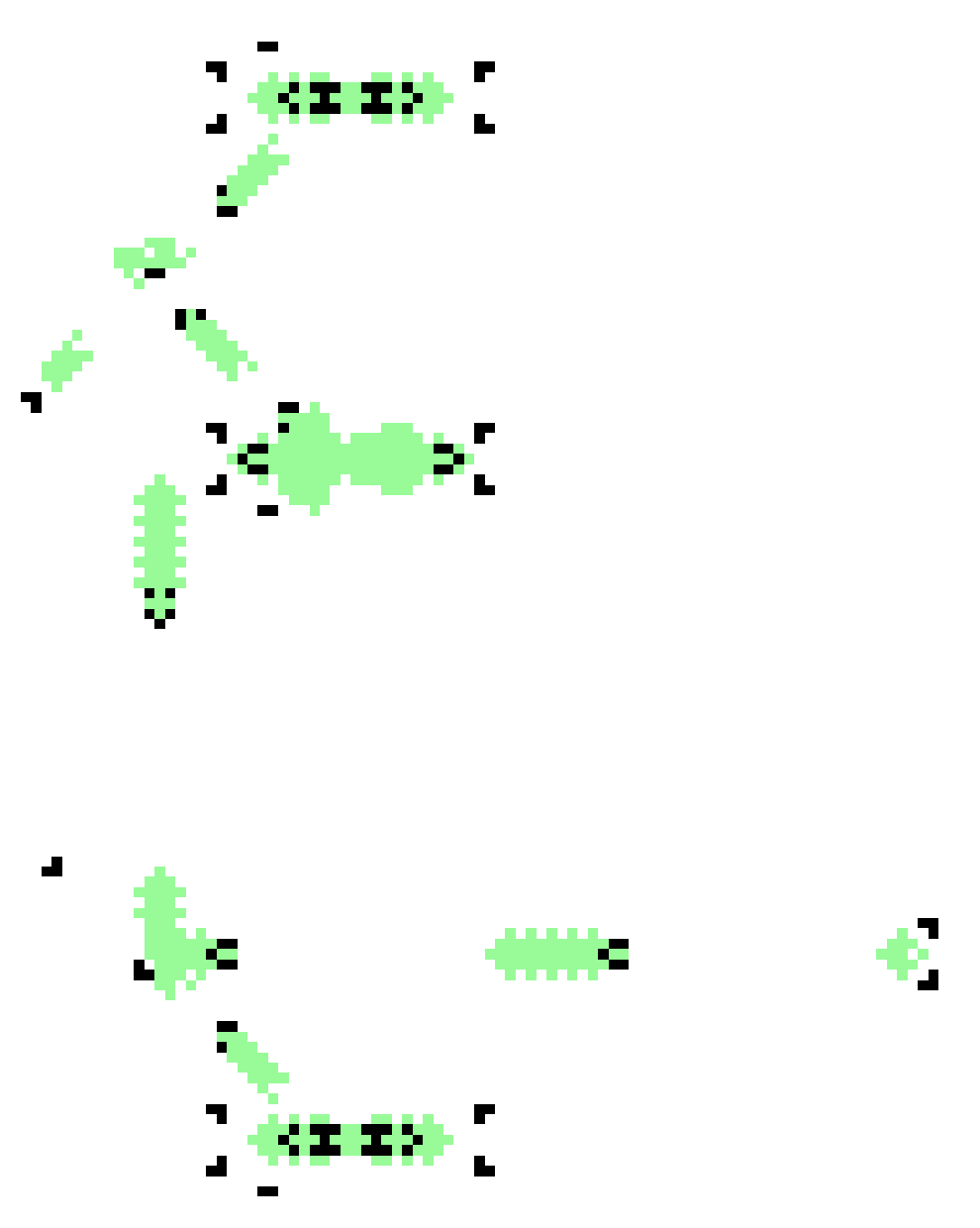}}\\[-4ex]
\begin{center}CGGc East\end{center}
\end{minipage}
\begin{minipage}[b]{1\linewidth}
\vspace{3ex}
%155x74
\fbox{\includegraphics[width=1\linewidth]{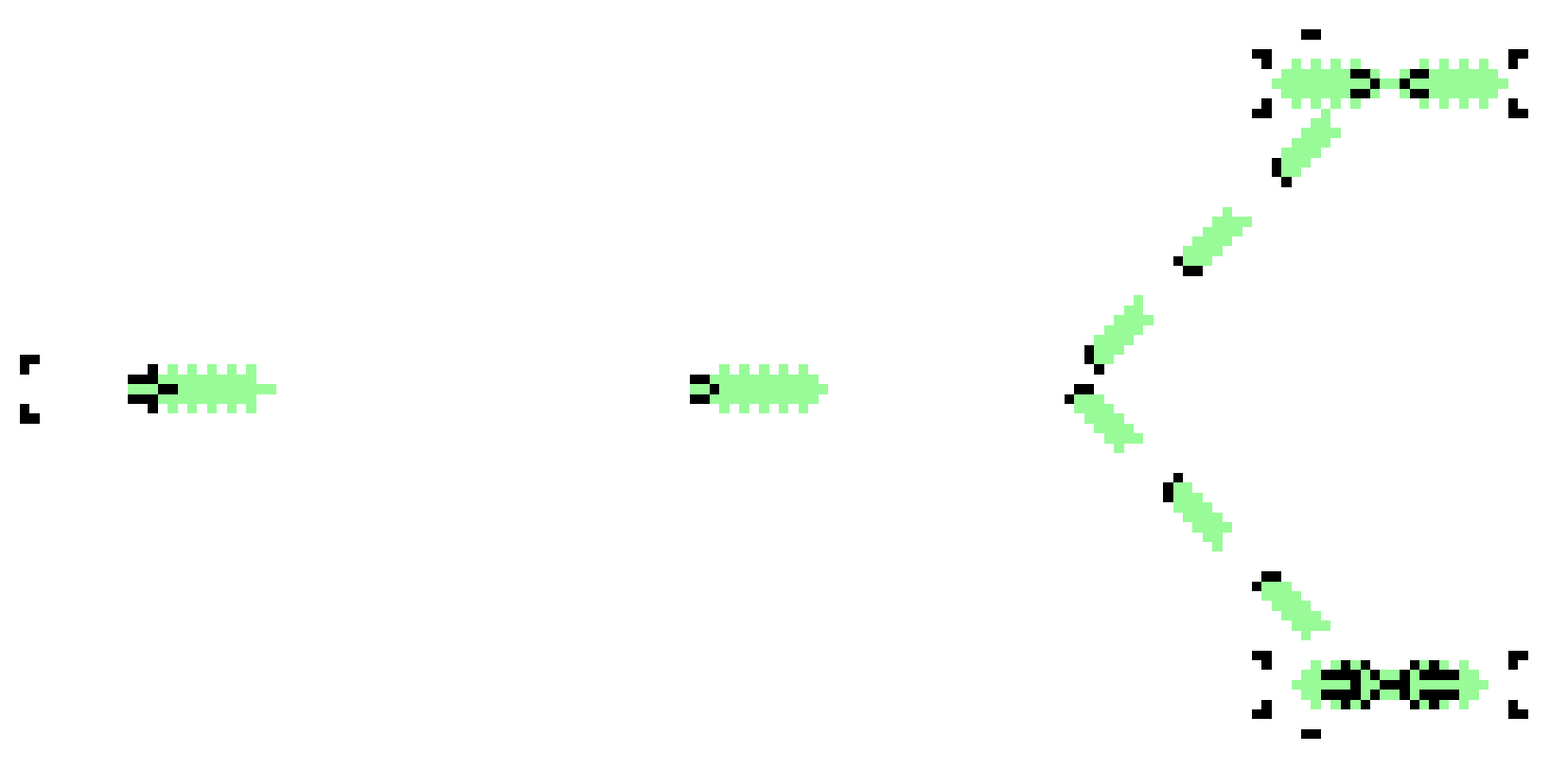}}\\[-4ex]
\begin{center}CGGc West alternative\end{center}
\end{minipage}
\vspace{-2ex}
\caption[CGGc North, East and West]
{Snapshots of CGGc glider-guns shooting Gc gliders
North and East, and an alternative towards the West 
with a wider interval of 56 cells. 
Gliders (and other mobile patterns) are shown with 
green dynamic time-trails of 20 time-steps.
}}
\label{CGGc North, East and West}
\end{figure}

%^^^^^^^^^^^^^^^^^^^^^^^^^^^^^^^^^^^^^^^^^^^^^^^^^^^^^^^^^^^^^^^^^^^^^^^^^^^^
\section{Logical gates -- logic universality}  
\label{Logical gates -- logic universality}

To demonstrate that the X-rule is universal in the logic sense, that
it can implement any logic circuit, we follow the game-of-Life method
using glider-guns as ``pulse generators'' to construct logical
gates\cite{Berlekamp1982}, starting with the simplest gate, NOT,
followed by AND and OR, and finally the functionally complete NAND gate
-- a combination of just NAND gates can implement any
logic circuit.

A proof of universality in the Turing sense is a harder proposition,
and will involve the construction of memory registers, auxiliary
storage and other components. Although we believe that the ingredients
for these structures exist in the diverse dynamics of the X-rule, we
will leave this for a later investigation.

Under the game-of-Life approach, in the glider-stream, the presence of
a glider represents the value 1, and the absence of a glider -- a gap
in the glider-stream -- represents the value 0. When two suitably
synchronised glider-streams approach each other at some angle (in
these examples at 90$^{\circ}$), gliders will either collide and
self-destruct leaving a gap, or a glider will pass through a gap and
survive.  By combining perfectly synchronised input streams
intersecting glider-streams generated by one or more glider-guns, the
logical gates can be implemented.  In the following examples, for
simplicity the input/output strings consist of 4 bits, but in
principle they could have arbitrary length.  Note that extraneous
glider-streams are stopped by strategically positioned eaters.

In the following sections 
\ref{gates towards the West} -- \ref{gates towards the East}
we gives examples of NOT, AND, OR and NAND gates
built with basic GGa diagonal glider-guns, firstly towards the West
employing basic GGa glider-guns, then towards the East which requires
compound glider-guns and a longer glider interval.
 
Taken together, these gates demonstrate that the X-rule is logic
universal in all diagonal directions.  Gates in orthogonal directions,
possibly with intersections at 45$^{\circ}$ including GGb and GGc
glider-guns, are work in progress.

%^^^^^^^^^^^^^^^^^^^^^^^^^^^^^^^^^^^^^^^^^^^^^^^^^^^^^^^^^^^^^^^^^^^^^^^^^^^^
\subsection{NOT, AND, OR and NAND gates towards the West}  
\label{gates towards the West}

Gates towards the West are built with basic GGa diagonal glider-guns,
shooting Ga gliders with a frequency of 38 time-steps.
Because Ga glider speed=$c$/4, the glider interval is 9 cells.
Each of the following examples show two snapshots separated in time. 
The initial setup state including inputs and a glider-gun or guns are
on the right (East) of a curved line --
the output at some time-steps later is on the left (West).
\clearpage

%----------------------------------------------------------------------
\subsubsection{NOT gate West}  
\label{NOT gate West}

\begin{figure}[H]%88x79, file= NOT-st.eed
\includegraphics[width=1\linewidth]{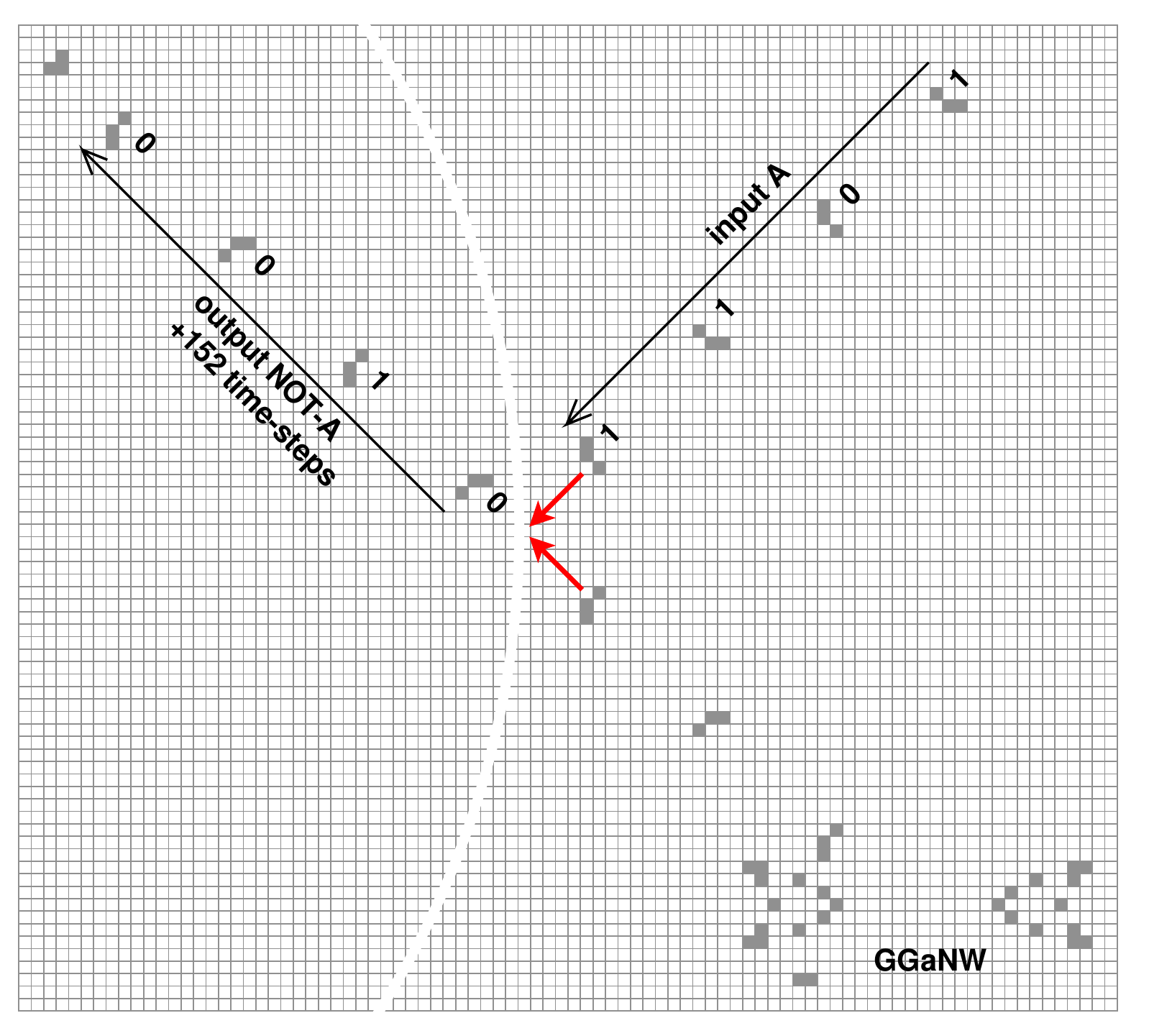}\\[-.905\linewidth]
\includegraphics[width=1\linewidth]{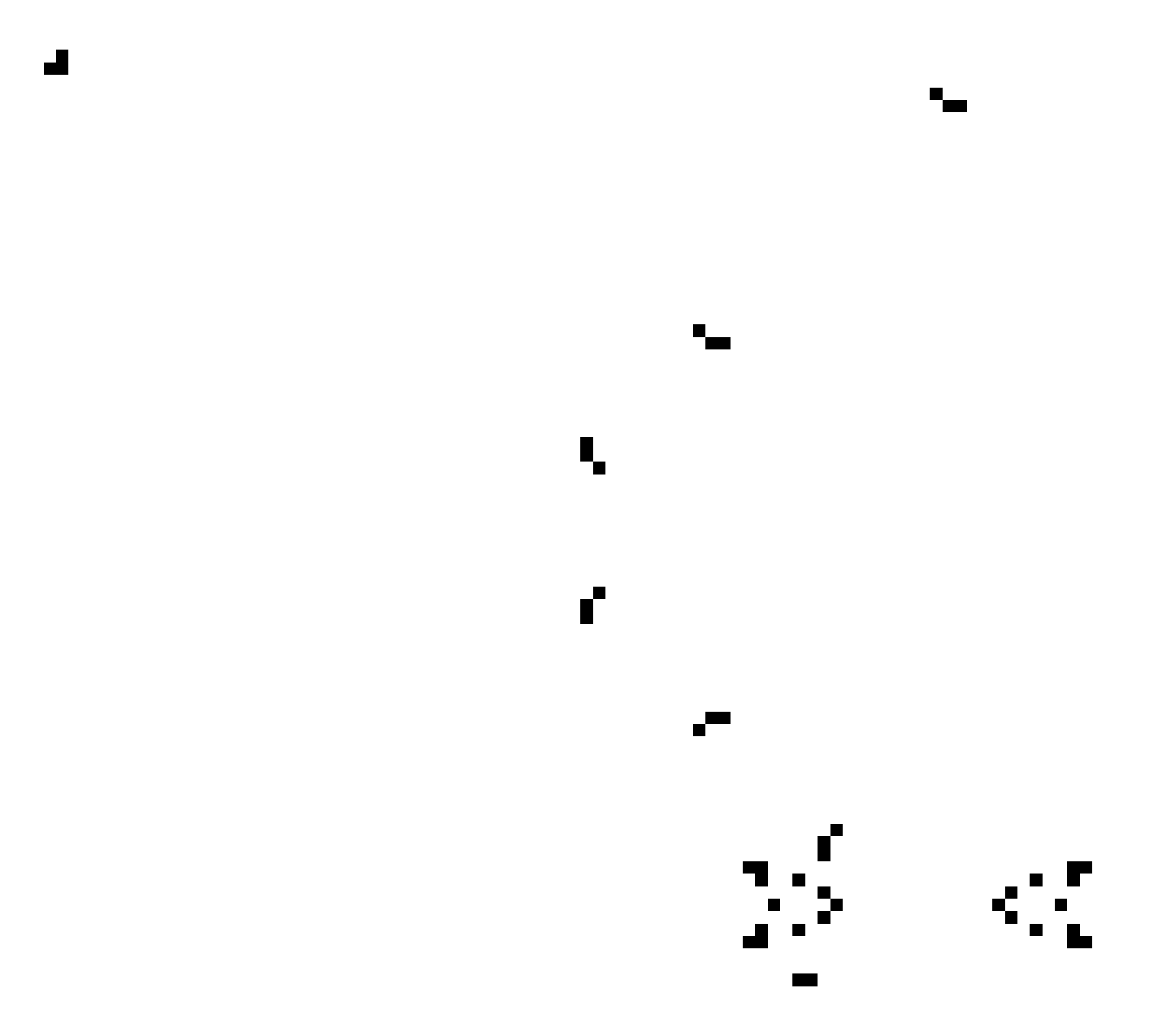}\\[-.905\linewidth]
\includegraphics[width=1\linewidth]{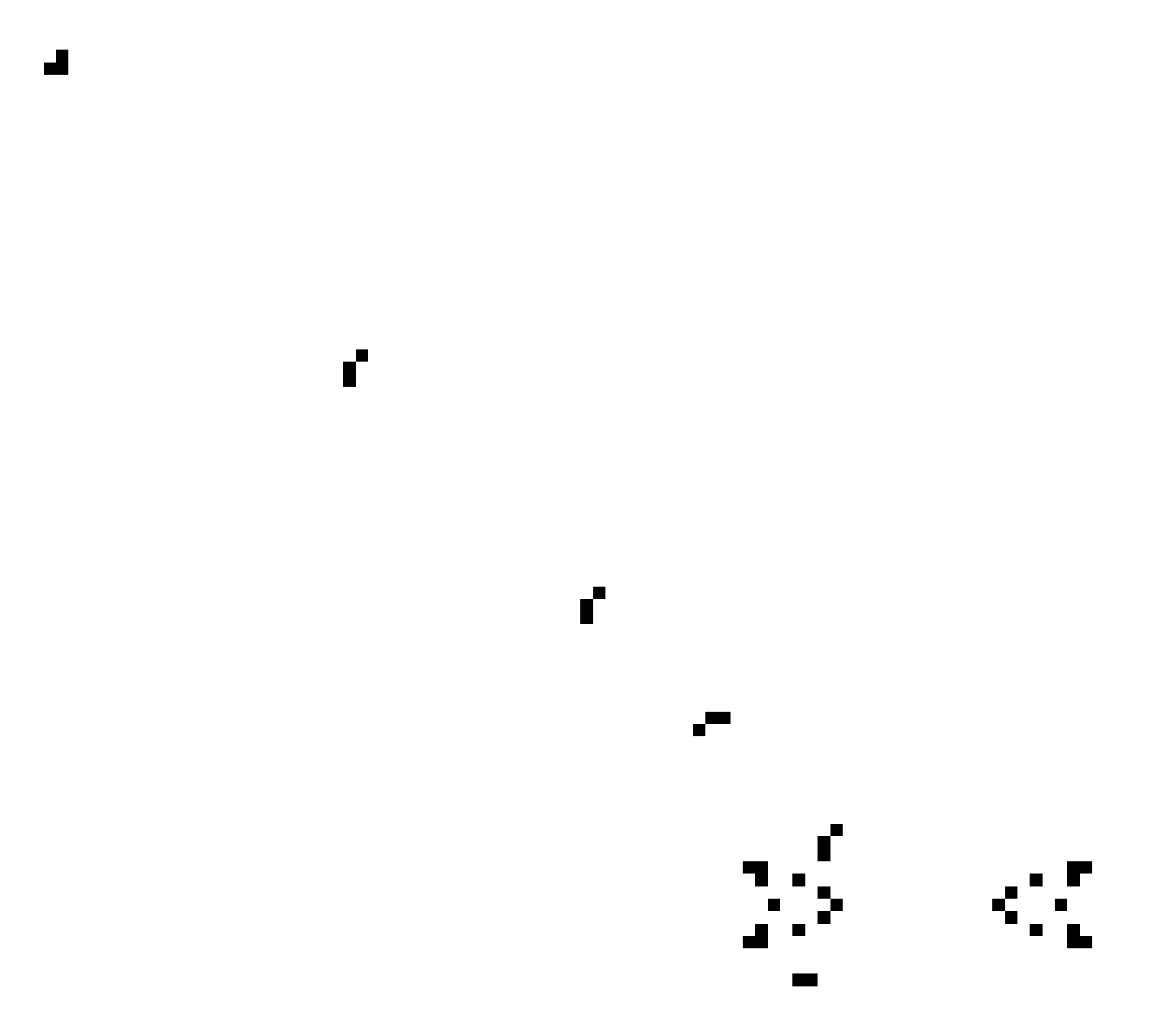}
\vspace{-2ex}
\caption[NOT gate]
{\textsf{An example of the \mbox{NOT gate} showing two snapshots separated in time.
Ghosts of absent gliders colored gray are
gaps in the glider sequence representing 0s, real gliders coloured black represent 1s.
The 4 bit input string A (1101) is transformed into the output string NOT-A (0010) 
after 152 time-steps from the 
first interaction, indicated by the red arrows,
between input A and the glider-stream from the GGaNW glider-gun.
}}
\label{fig NOT gate West} 
\end{figure}

To demonstrate the NOT-A gate (figure~\ref{fig NOT gate West}) a NW
glider-stream is generated by a GGa glider-gun.  A SW sequence of 4
gliders/gaps representing the input string A (1101), with the correct
spacing and phases, is positioned to intersect the NW glider-stream.
Gliders that collide self-destruct making a gap (value 0) in the
output, whereas gliders that pass through a gap in the input continue
NW representing value 1 in the output.  If the first interaction with
input A is at time zero, 152 time-steps later (38$\times$4), the
complete output of gliders/gaps has been built, moving NW away the
intersection point, representing NOT-A (0010).

\clearpage

%----------------------------------------------------------------------
\subsubsection{AND gate West}  
\label{AND gate West}
\begin{figure}[H]%170x140, file= AND-st.eed
\fbox{\includegraphics[width=1\linewidth]{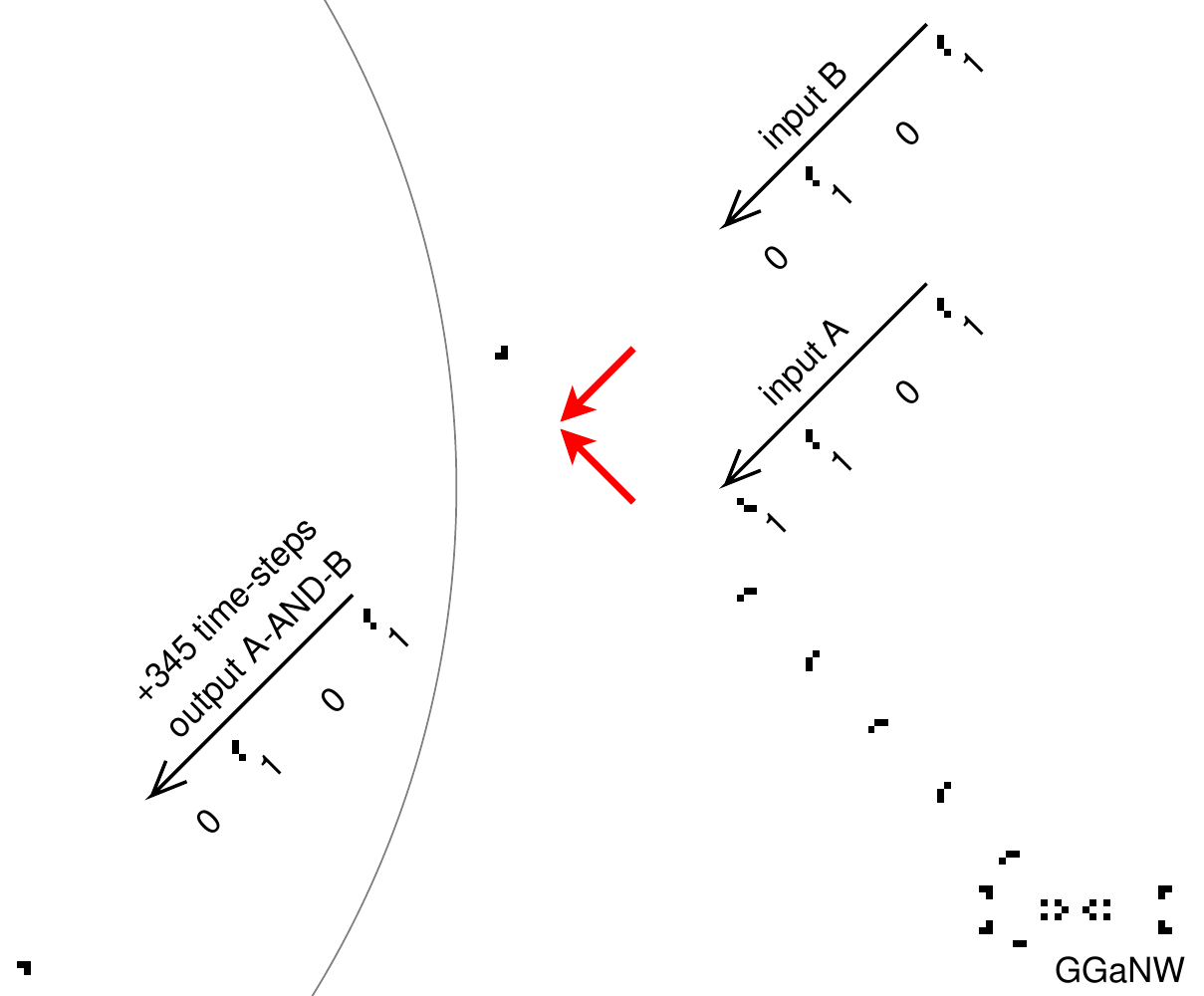}}
\vspace{-2ex}
\caption[AND gate West]
{\textsf{An example of the AND gate showing two snapshots separated in time.
Two 4 bit input strings A (1101) and B (0101) are represented by  
SW gliders+gaps. Input A intersects a NW glider-stream
generated by a GGaNW glider-gun. Gilder collisions between GaNW and A self-destruct,
allowing a glider in input B to continue. A GaNW gilder meeting a gap in A passes through
and would self-destruct if colliding with a glider in input B.
This second interaction with input B is indicated by the red arrows,
After about 214 time-steps from the first interaction with input A, 
the complete output string A-AND-B (0101) emerges moving SW.
In the figure, A-AND-B has moved further SW and is shown after about 345 time-steps.
}}
\label{fig AND gate West}
\end{figure}

The A-AND-B gate (figure~\ref{fig AND gate West}), 
builds on the initial interaction from the NOT-A gate.
Gliders that pass through a gap in input A are able to intersect
input string B (0101) which has been positioned 38 cells above string A. 
After about 214 time-steps from the first interaction with input A, 
the complete output string A-AND-B (0101) has emerged moving SW.

\clearpage

%----------------------------------------------------------------------
\subsubsection{OR gate West}  
\label{OR gate West}

\begin{figure}[H]%200x193, start file= OR-st.eed
\fbox{\includegraphics[width=1\linewidth]{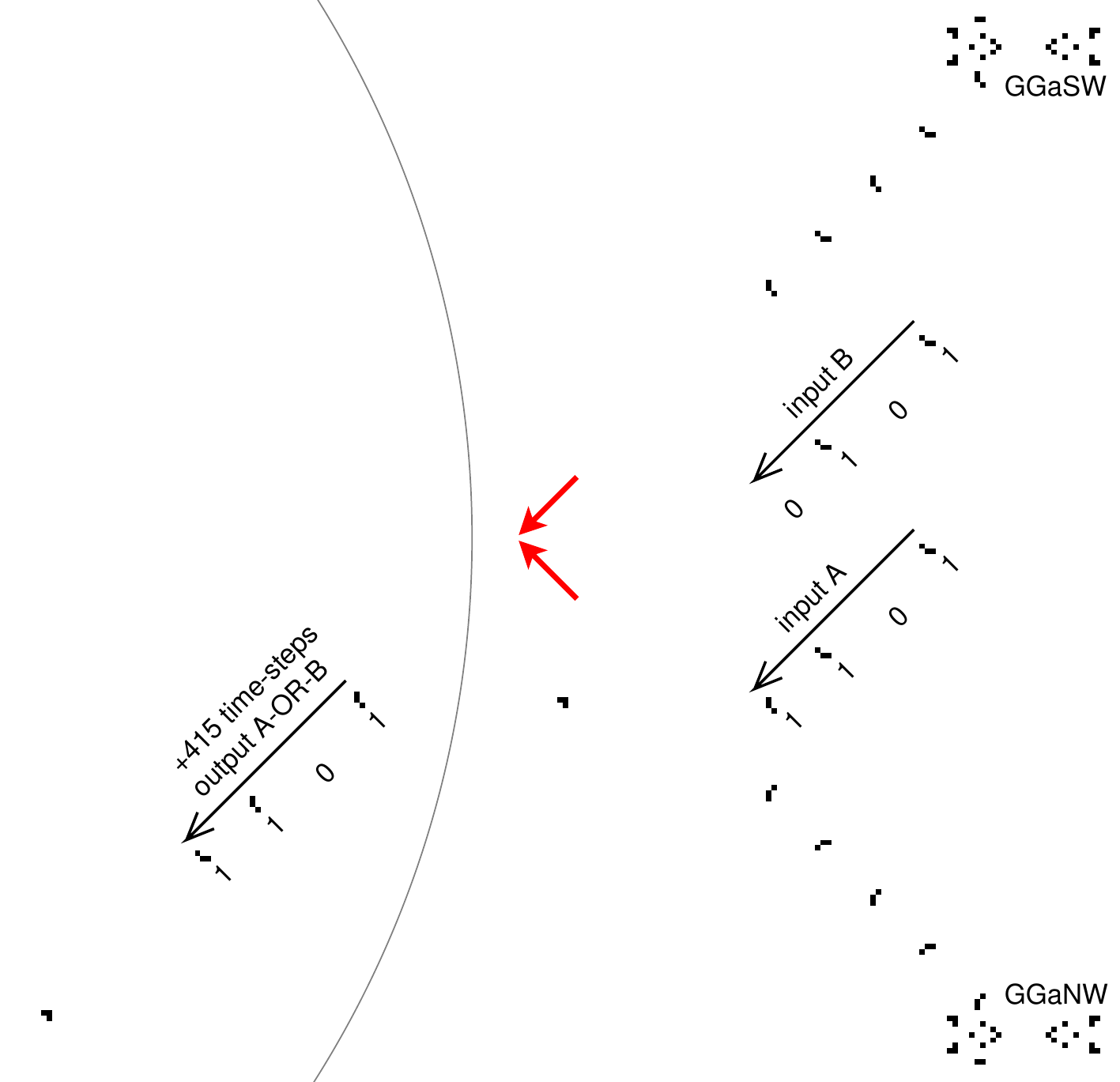}}
\vspace{-4ex}
\caption[OR gate]
{\textsf{An example of the OR gate showing two snapshots separated in time.
Two 4 bit input strings A (1101) and B (0101) are represented by  
SW gliders+gaps. Input A intersects a NW glider-stream
generated by a GGaNW glider-gun. Gilder collisions between GaNW and A self-destruct,
allowing a glider in input B to continue. A GaNW gilder meeting a gap in A passes through
and would self-destruct if colliding with a glider in input B.
NW gliders that have survived the transit of both A and B intersect a SW glider-stream
from a second GGaSW glider-gun 38 cells above B. 
This last interaction occurs at a point indicated by the red arrows.
After about 305 time-steps from the first interaction with input A,
the complete output string A-AND-B (0101) is cut off and continues SW.
This is the A-OR-B output (1101).
The glider-streams from the two glider-guns
heading NW and SW continue to self destruct.
In the figure, A-OR-B has moved further SW and is shown after about 415 time-steps.
}}
\label{fig OR gate West}
\end{figure}

The A-OR-B gate (figure~\ref{fig OR gate West}) includes
procedures from the AND gate combined with a second GGaSW glider-gun --
commentary is in the caption. 
\enlargethispage{4ex}

%----------------------------------------------------------------------
\subsubsection{NAND gate West}  
\label{NAND gate West}

\begin{figure}[H]%141x174, start file= NAND-st.eed
\fbox{\includegraphics[width=1\linewidth]{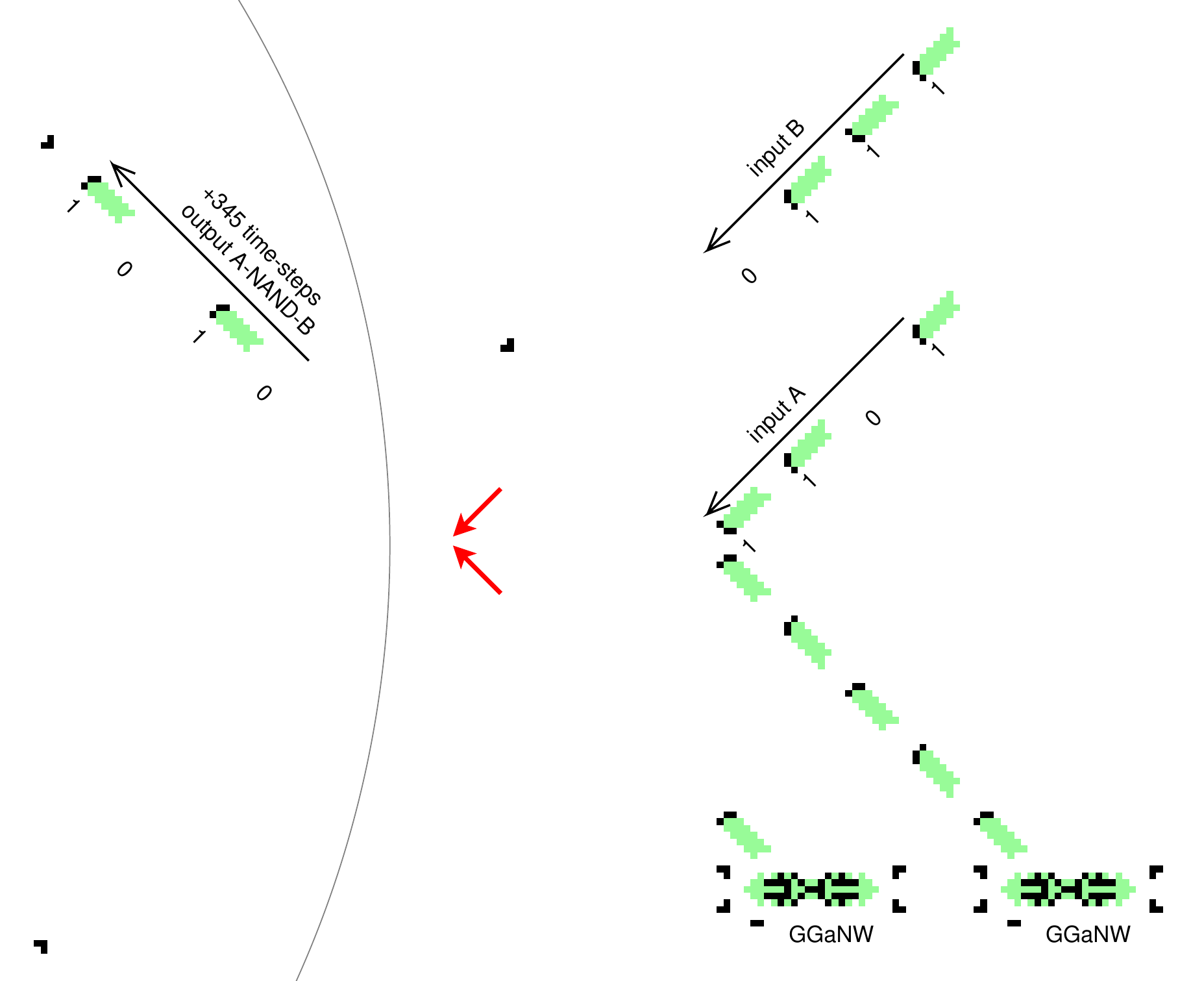}}
\vspace{-3ex}
\caption[NAND gate West]
{\textsf{An example of the NAND gate showing two snapshots separated in time.
Gliders (and other mobile patterns) are shown with 
green dynamic time-trails of 20 time-steps.
Two 4 bit input strings A (1101) and B (0101) are represented by  
SW gliders+gaps. The first set of interactions with a NW glider-stream
generated by a GGaNW glider-gun follows the AND gate procedure similar to
figure~\ref{fig AND gate West}. The output A-AND-B then intersects a second
NW glider-stream at a point indicated by the red arrows,
generated from a second GGaNW glider-gun. This implements
the NOT gate procedure similar to figure~\ref{fig NOT gate West}.
The final A-NAND-B output (1010) heads NW, shown after about 374 time-steps
from the first interaction and 163 from the second.
}}
\label{fig NAND gate West}
\end{figure}

The A-NAND-B gate (figure~\ref{fig NAND gate West}) is a combination of
the AND and NOT gates, and includes two identical GGaNW glider-guns
-- commentary is in the caption. 

%^^^^^^^^^^^^^^^^^^^^^^^^^^^^^^^^^^^^^^^^^^^^^^^^^^^^^^^^^^^^^^^^^^^^^^^^^^^^
\subsection{NOT, AND, OR and NAND gates towards the East}  
\label{gates towards the East}
Gates towards the East are built with CGGa compound diagonal glider-guns,
shooting Ga gliders with a frequency one every 76 time-steps.
Because Ga glider speed=$c$/4, the glider interval is 19 cells,
wider than the basic GGa glider interval of 9 cells.
Each of the following examples show two snapshots separated in time. 
The initial setup state including inputs and glider-guns are
on the left (West) of a curved line --
the output at some time-steps later is on the right (East).

%----------------------------------------------------------------------
\subsubsection{NOT gate East}  
\label{NOT gate East}

\begin{figure}[H]%164x150, file= cNOT-st.eed
\fbox{\includegraphics[width=1\linewidth]{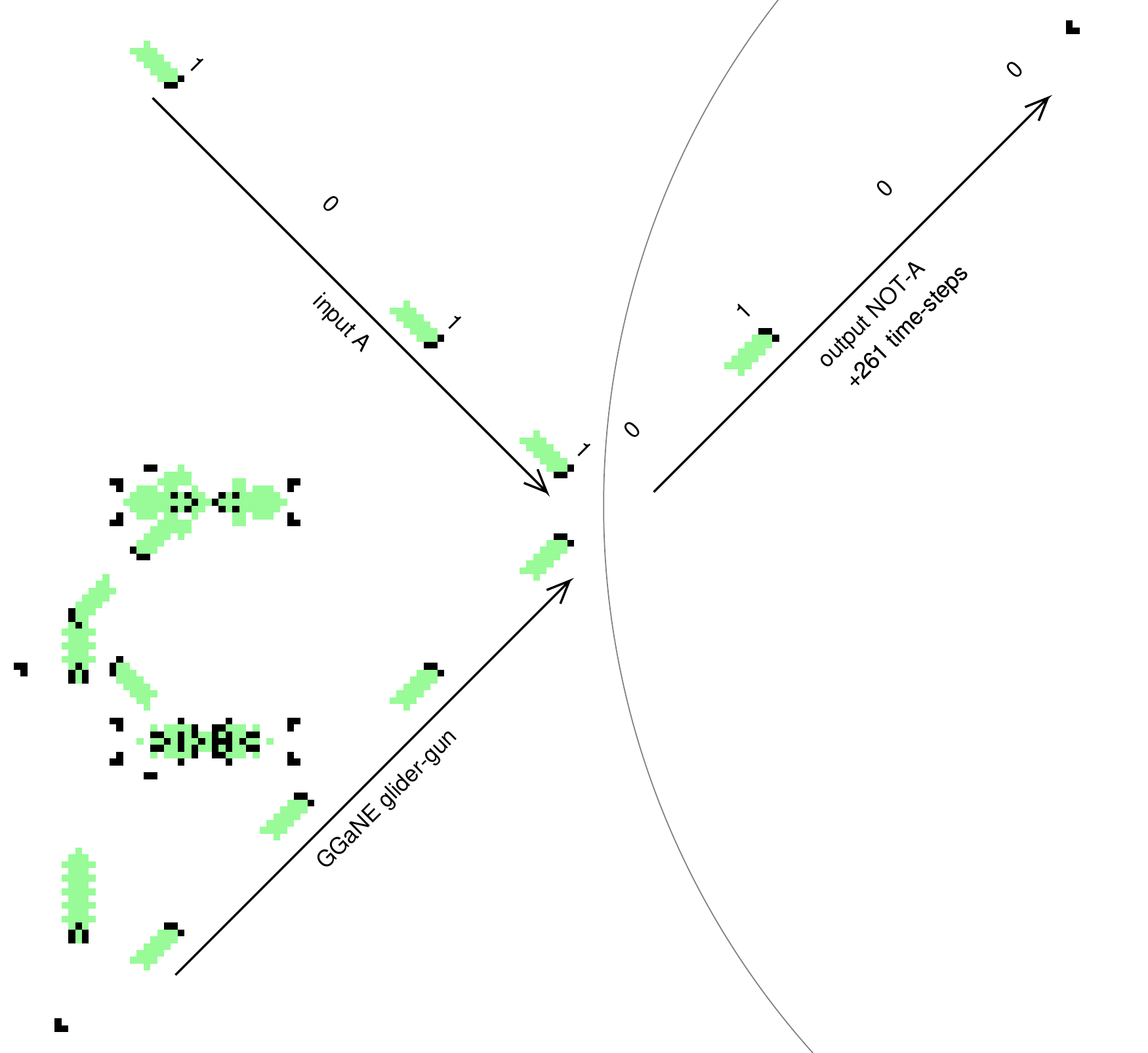}}
\vspace{-2ex}
\caption[Compound NOT gate towards the East]
{\textsf{An example of the NOT towards the East showing two snapshots separated in time.
Input A (1101) moving SE interacts with a NE glider-stream
generated by the compound glider-gun CGGaNE, and transforms
the glider stream into the output NOT-A (0010) which continues NE,
show about 261 time-steps from the first interaction.
}}
\label{fig Compound NOT gate East} 
\end{figure}

To demonstrate the NOT-A gate towards the East 
(figure~\ref{fig Compound NOT gate East}) we harness the
CGGaNE compound glider-gun shooting gliders NE (figure~\ref{CGGcS+CGGaNE}),
and construct the gate following similar interactions as in 
figure~\ref{fig NOT gate West}.

\clearpage

%----------------------------------------------------------------------
\subsubsection{AND gate East}  
\label{AND gate East}

\begin{figure}[H]%228x228, file= cAND-in.eed
\fbox{\includegraphics[width=1\linewidth]{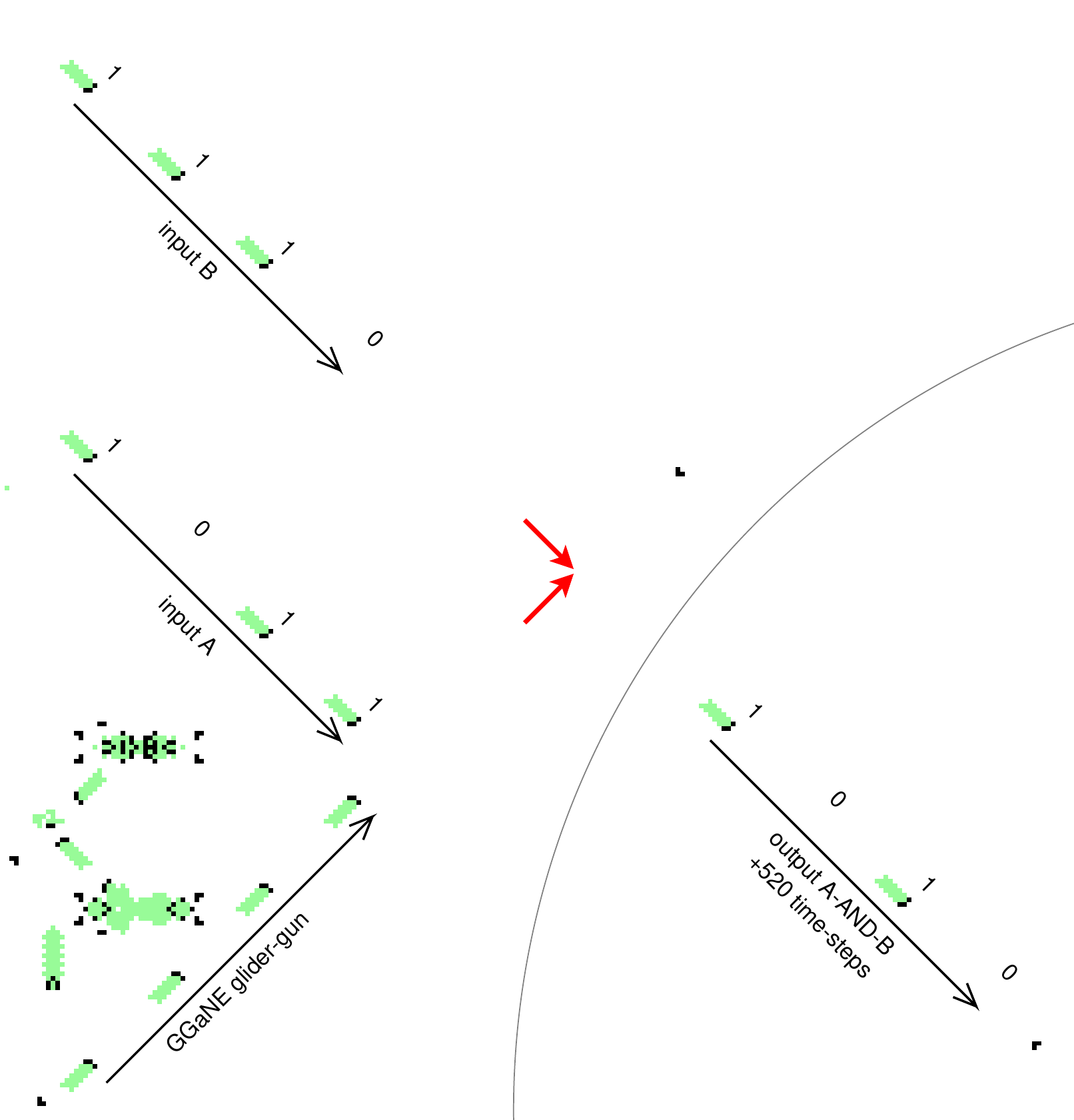}}
\vspace{-2ex}
\caption[Compound AND East]
{\textsf{An example of the AND gate towards the East showing two snapshots separated in time.
Inputs A (1101) and B (0111) moving SW interact in turn with a NE glider-stream
generated by the compound glider-gun CGGaNE. The surviving glider-stream, perturbed by A,
then interacts with B at the point indicated by the red arrows, and B is transformed into
the output A-AND-B (0101) which continues SE, and is shown about 520 time-steps
from the first interaction.
}}
\label{fig AND gate East} 
\end{figure}

To demonstrate the A-AND-B gate towards the East 
(figure~\ref{fig AND gate East}) we harness the
CGGaNE compound glider-gun shooting gliders NE (figure~\ref{CGGcS+CGGaNE}),
and construct the gate following similar interactions as in 
figure~\ref{fig AND gate West}.

%----------------------------------------------------------------------
\subsubsection{OR gate East}  
\label{OR gate East}

\begin{figure}[H]%300x330, file= cOR-st.eed
\fbox{\includegraphics[width=1\linewidth]{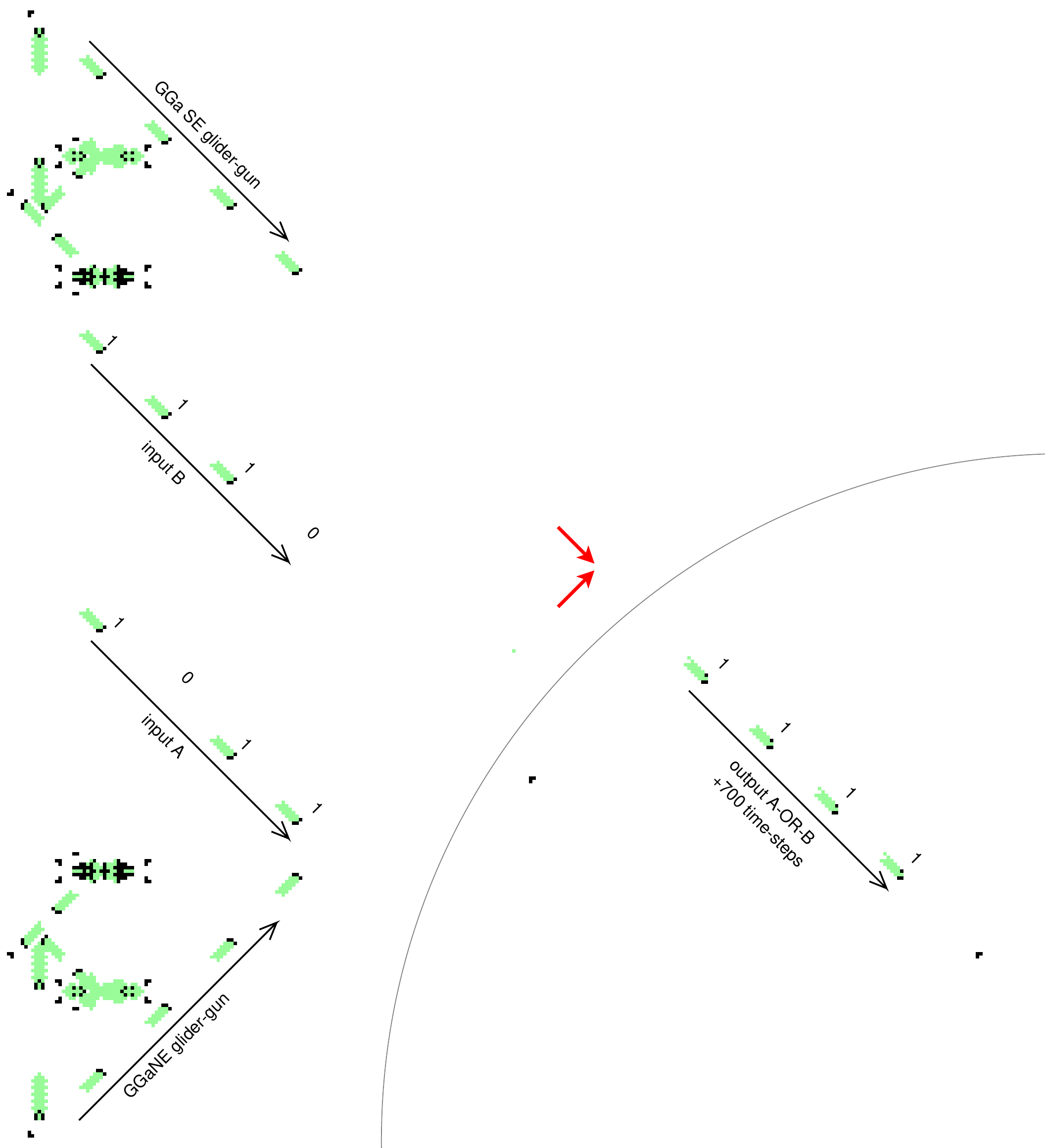}}
\vspace{-4ex}
\caption[Compound OR gate towards the East]
{\textsf{An example of the OR gate towards the East showing two snapshots separated in time.
Inputs A (1010) and B (0111) moving SE interact in turn with a NE glider-stream
from compound glider-gun CGGaNE. The surviving glider-stream
then intersects a second glider-stream from the compound glider-gun CGGaSE at the
point indicated by the red arrows.
After about 700 time-steps from the first interaction
the complete output string A-OR-B (1111) is cut off and continues SE.
}}
\label{fig OR gate East} 
\end{figure}

To demonstrate the A-OR-B gate towards the East 
(figure~\ref{fig OR gate East}) we harness two compound glider-guns,
CGGaNE and CGGaSE (figure~\ref{CGGcS+CGGaNE}),
and construct the OR gate following similar interactions as in 
figure~\ref{fig OR gate West}.

%----------------------------------------------------------------------
\subsubsection{NAND gate East}  
\label{NAND gate East}

\begin{figure}[H]%260x350, file= cNAND-st.eed
\begin{center}
\fbox{\includegraphics[width=.9\linewidth,bb=4 5 261 339,clip=]{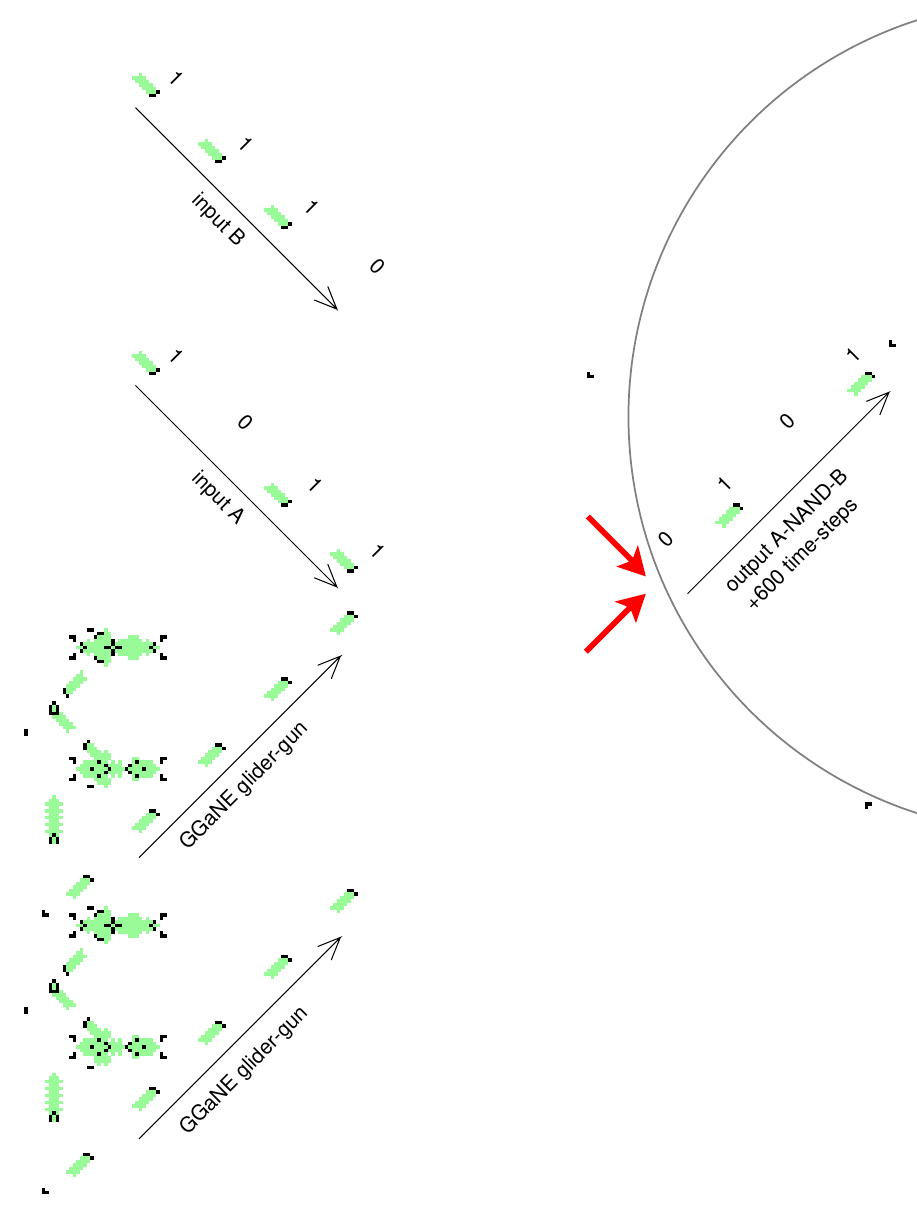}}
\end{center}
\vspace{-2ex}
\caption[Compound NAND gate towards the East]
{\textsf{
An example of the NAND gate towards the East showing two snapshots
separated in time.  A glider-stream from the upper GGaNE compound
glider-gun intersects input strings A (1101) and B (0111) in turn,
following the AND gate procedure. The output A-AND-B then intersects a
second NE glider-stream at a point indicated by the red arrows. This
implements the NOT gate procedure giving the final A-NAND-B output
(1010) heading NE, shown after about 600 time-steps from the first
interaction.
}}
\label{fig NAND gate East} 
\end{figure}

To demonstrate the A-NAND-B gate towards the East 
(figure~\ref{fig NAND gate East}) we use two compound glider-guns
CGGaNE (figure~\ref{CGGcS+CGGaNE}),
and construct the NAND gate following similar interactions as in 
figure~\ref{fig NAND gate West}.
\enlargethispage{6ex}

%^^^^^^^^^^^^^^^^^^^^^^^^^^^^^^^^^^^^^^^^^^^^^^^^^^^^^^^^^^^^^^^^^^^^^^^^^^^^
\section{Concluding remarks}  
\label{Concluding remarks}

\noindent The X-rule is a novel 2D binary CA with a diversity of
emergent structures -- gliders and eaters/reflectors -- from which
glider-guns and logical gates have been constructed and demonstrated,
showing that the X-rule is universal in the logic sense in that it can
implement any logic circuit.  The structures, experiments and results
made so far and documented in this paper are an initial exploration of
what can be done, which suggests to us that memory functions required
for universality in the Turing sense could be also feasible.

The key component in a rule that has the potential for universality is
the periodic glider-gun, either naturally emergent or constructed.
Whereas Gosper's glider-gun in the game-of-Life is an asymmetric
periodic structure within an isometric rule, the glider-guns in the
X-rule have an underlying symmetry within a marginally non-isotropic
rule.  Just a few outputs, selected by an automatic method, were changed
in the isotropic precursor's rule-table, with the effect that periodic
oscillators based on reflecting/bouncing behaviour temporarily break
symmetry to eject gliders.

The X-rule's glider-guns and compound gilder-guns, shooting a variety
of glider types at various frequencies and in various directions, are
constructed from a kit of parts that can be put together in many
combinations, but just a few of the available ingredients have been
included. A next step would be to complete a full catalogue of basic
glider/glider and glider/eater collisions.

The search method for X-rule precursors suggests there are many binary
isotropic rules in rule-space with periodic reflecting/bouncing
properties, and that these rules would be susceptible to analogous
non-isotropic adjustments to make glider-guns.  Alternatively,
glider-guns based on non-symmetric periodic oscillators in isotropic
rules could be designed in ways analogous to Gosper's glider-gun, and
we intend to try this approach for the X-rule precursor.

In the scatter plot of entropy variability against mean entropy, the
X-rule is found in a different place from both the game-of-Life and the
Sapin rule. So logic universality occurs in unexpected places -- 
we do not yet know the diversity of condition for logic universality.

Like the game-of-Life, the X-rule is open-ended.  As in nature, and
given a quasi-infinite space-time, it would be impossible to pin down
a complete description of behaviour.

\section{Acknowledgements}  
\label{Acknowledgements}

\enlargethispage{3ex}
\noindent Figures were made with Discrete Dynamics Lab (www.ddlab.org).
Experiments were done with DDLab and Mathematica.
The research was done during a collaboration at June
workshops in 2013 and 2014 at the DDLab Complex Systems Institute in
Ariege, France, and also at the Universidad Aut\'onoma de Zacatecas,
M\'exico, and in London, UK.  
J.M. G\'omez Soto also acknowledges his residency at the DDLab Complex Systems
Institute and financial support from the Research
Council of Zacatecas (COZCyT).
The website www.ddlab.org/Xrule/ is
available for more information and for DDLab files to duplicate these
results.
\clearpage

%^^^^^^^^^^^^^^^^^^^^^^^^^^^^^^^^^^^^^^^^^^^^^^^^^^^^^^^^^^^^^^^^^^^^^^^^^^^^

\end{document}